\documentclass[nolinenum]{jfp}
\newif\iftr\trtrue
\usepackage{styles/pl-syntax/pl-syntax}
\usepackage{styles/LC}
\usepackage{styles/comments}
\usepackage{amsmath}
\usepackage{amsthm}
\usepackage{bbm}
\usepackage{stmaryrd}
\usepackage{mathpartir}
\usepackage{latexsym}
\usepackage{amssymb}
\usepackage{placeins}

\date{}
\theoremstyle{theorem}
\newtheorem{theorem}{Theorem}
\newtheorem{lemma}{Lemma}
\newtheorem{corollary}{Corollary}
\theoremstyle{definition}
\newtheorem{definition}{Definition}
\newtheorem{example}{Example}

\NewDocumentCommand{\rulen}{m}{\hyperlink{rule:#1}{[\textsf{#1}]}}

\begin{document}

\journaltitle{JFP}
\cpr{Cambridge University Press}
\doival{10.1017/xxxxx}

\totalpg{\pageref{lastpage01}}
\jnlDoiYr{2023}
  
\title{Alice or Bob? Process Polymorphism in Choreographies}
\begin{authgrp}
  \author{Eva Graversen}
  \affiliation{Department of Mathematics and Computer Science, University of Southern Denmark
    \email{efgraversen@imada.sdu.dk}}

  \author{Andrew K. Hirsch}
  \affiliation{Department of Computer Science and Engineering, University at Buffalo, SUNY
    \email{akhirsch@buffalo.edu}}

  \author{Fabrizio Montesi}
  \affiliation{Department of Mathematics and Computer Science, University of Southern Denmark
    \email{fmontesi@imada.sdu.dk}}
\end{authgrp}
\begin{abstract}
  We present \chorlam, a language for higher-order functional \emph{choreographic programming}---an emerging paradigm by which programmers write the desired cooperative behaviour of a system of communicating processes and then compile it into distributed implementations for each process, a translation called \emph{endpoint projection}.\fabrizio{there was some repetition here (programming... programming), does it read ok now?}
Unlike its predecessor, Chor$\lambda$, \chorlam has both type and \emph{process} polymorphism inspired by System F$_\omega$.
That is, \chorlam is the first (higher-order) functional choreographic language which gives programmers the ability to write generic choreographies and determine the participants at runtime.
This novel combination of features also allows \chorlam processes to communicate \emph{distributed values}, leading to a new and intuitive way to write delegation.
While some of the functional features of \chorlam give it a weaker correspondence between the semantics of choreographies and their endpoint-projected concurrent systems than some other choreographic languages, we still get the hallmark end result of choreographic programming: projected programs are deadlock-free by design.

\end{abstract}

\maketitle

\section{Introduction}
\label{sec:introduction}

Concurrent programs involve interacting processes.
Usually, programmers write one program per process, and then compose those programs in parallel.
These programs contain \emph{send} and \emph{receive} expressions which transmit data between processes.
This makes it easy to write code that \emph{deadlocks}, or gets stuck because patterns of sends and receives do not match.
\emph{Session types}~\citep{Honda93,HondaVK98} can be used to describe the patterns of sends and receives in a program, offering a foundation for static analyses aimed at preventing communication mismatches and deadlocks~\citep{ScalasY19,CairesP10,Wadler12,DeYoungCPT12,HondaYC16,DardhaGS12}.\fabrizio{Sends and receives had multiple different fonts, with no good reason imho. I made them normal.}
Working with session types enables the programmer to ensure the communications in their system follow compatible send/receive patterns.

Alternatively, programmers can use a \emph{global} point of view~\citep[see e.g.,][]{PredaGGLM16,GiallorenzoMPRS21,CarboneM13,CruzFilipeGLMP21,HirschG22,RichterKWSFM22,JongmansB22,LopezNN16}.
A programmer using the global point of view writes one program, which is then compiled to a program for each process.
\emph{Choreographic programming}~\citep{Montesi13} is a global programming paradigm with particularly well-explored foundations~\citep{Cruz-FilipeM17,Montesi22}.
In choreographies, \textsf{send} and \textsf{receive} commands are written together, usually using a notation inspired by security protocols~\citep{NeedhamS78}. This has two key advantages.
First, it gives programmers the power to express the desired communication flow among processes, but without the burden of manually coding send and receive actions.
Second, it ensures that there is no mismatch which can cause deadlock, a property that has become known as \emph{deadlock-freedom by design}~\citep{CarboneM13}.

To see the power of this, consider the (in)famous bookseller example---a recurring example in the literature of choreographic programming and session types~\citep{CarboneM13,HondaYC16,Montesi22}.
\LCProcBuyer wants to buy a book from \LCProcSeller.
To this end, \LCProcBuyer sends the title of the book---say, ``The Importance of Being Earnest''---to \LCProcSeller, who then sends back the price.
\LCProcBuyer then can compare the price with its budget and based on the result informs \LCProcSeller that they want to buy the book if it is within their budget, or informs them that they do not want to buy the book otherwise.
We can describe this via the following choreography:
$$
\NewDocumentCommand{\pricelookup}{}{\textsf{price\_lookup}}
\LClet*{x}{
  \LCcom{}{\LCProcBuyer}{\LCProcSeller}{(\LCLocalVal{\text{``The Importance of Being Earnest''}}[\LCProcBuyer])}
}
{
  \LClet*{y}{
    \LCcom{}{\LCProcSeller}{\LCProcBuyer}{(\LCapp{\pricelookup}{x})}
  }
  {
    \LCif*{y < \textsf{budget}}
    {\LCselect{\LCProcBuyer}{\LCProcSeller}{\textsf{Buy}}{\left(\LCUnitVal[\LCProcSeller]\right)}}
    {\LCselect{\LCProcBuyer}{\LCProcSeller}{\textsf{Quit}}{\left(\LCUnitVal[\LCProcSeller]\right)}}
  }
}
$$
In this choreography, as in all choreographic programs, computation takes place among multiple \emph{processes} communicating via message passing.
Values are located at processes; for example, in the first line of the choreography, the title of the book is initially located at~\LCProcBuyer.
The function $\LCcomV{}{\LCProcP}{\LCProcQ}$ communicates a value from the process~\LCProcP to the process~\LCProcQ.
It takes a local value at \LCProcP and returns a local value at \LCProcQ.\footnote{When we formalize our language in Section~\ref{sec:chorlambda-language}, we require a type annotation on $\LCcomV{}{\LCProcP}{\LCProcQ}$, but in the introduction we leave it unannotated for the clarity of our examples.}
Thus, $x$ represents the string ``The Importance of Being Earnest'' at the process~\LCProcSeller, while $y$ represents the price at the process~\LCProcBuyer.
Finally, we check locally if the book's price is in \LCProcBuyer's budget.
Either way, we use function $\LCKWselect$ to send a label from \LCProcBuyer to \LCProcSeller representing \LCProcBuyer's choice to either proceed with the purchase or not.
Either way, the choreography returns the dummy value $()$ at \LCProcSeller.

While most of the early work on choreographies focused on simple lower-order imperative programming like in the example above, recent work has shown how to develop higher-order choreographic programming languages.
These languages allow a programmer to write deadlock-free code using the usual abstractions of higher-order programming, such as objects~\citep{GiallorenzoMP20} and higher-order functions~\citep{HirschG22,CruzFilipeGLMP21}.

For instance, the program above bakes in the title and the value of the book.
However, we may want to use this code whenever \LCProcBuyer wants to buy any book, and let \LCProcBuyer use any local function to decide whether to buy the book at a price.
$$
\NewDocumentCommand{\titlevar}{}{\textsf{title}}
\NewDocumentCommand{\pricelookup}{}{\textsf{price\_lookup}}
\NewDocumentCommand{\inbudget}{}{\textsf{buyAtPrice?}}
\LCfun*{\titlevar}[\LCString{\LCProcBuyer}]{
    \LCfun*{\inbudget}[\LCArr{\LCNat[\LCProcBuyer]}{}{\LCBool{\LCProcBuyer}}]
    {
      \LClet*{x}{
        \LCcom{}{\LCProcBuyer}{\LCProcSeller}{\titlevar}
      }{
        \LClet*{y}{
          \LCcom{}{\LCProcSeller}{\LCProcBuyer}{(\LCapp{\pricelookup}{x})}
        }{
          \LCif*{\LCapp{\inbudget}{y}}
          {\LCselect{\LCProcBuyer}{\LCProcSeller}{\textsf{Buy}}{\left(\LCUnitVal[\LCProcSeller]}\right)}
          {\LCselect{\LCProcBuyer}{\LCProcSeller}{\textsf{Quit}}{\left(\LCUnitVal[\LCProcSeller]}\right)}
        }
      }
    }
}
$$

A programmer using a higher-order choreographic language, like a programmer using any higher-order programming language, can write a program once and use it in a large number of situations.
For instance, by supplying different values of \textsf{title} and \textsf{buyAtPrice?}, this same code can be used to buy several different titles and \LCProcBuyer can determine if they are willing to buy the book at the price using any method they desire.

While the move from first-order programming to higher-order programming is significant, previous work on the theoretical foundations of higher-order choreographic programming still did not account for other forms of abstraction~\citep{HirschG22,CruzFilipeGLMP21}.
In particular, they did not allow for \emph{polymorphism}, where programs can abstract over types as well as data, allowing them to operate in many more settings; nor did they allow for \emph{delegation}, where one process can ask another process to act in its stead.

These forms of abstraction are relatively standard: delegation is an important operation in concurrent calculi, and polymorphism is vital to modern programming.
In choreographic programming, however, another form of abstraction becomes natural: abstraction over processes.
Current higher-order choreographic languages require that code mention concrete process names.
However, we often want to write more-generic code, allowing the same code to run on many processes.
For example, the above program allows \LCProcBuyer to decide whether to buy a book from \LCProcSeller using any local function~\mbox{\textsf{buyAtPrice?}}.
It would be more natural to write \LCProcSeller as a book-selling \emph{service} which different clients could interact with in the same way to buy a book.

In this paper, we tackle three styles of abstraction.
Firstly, we show that abstraction over processes is a type of polymorphism, which we refer to as \emph{process polymorphism}.
Secondly, we extend Chor$\lambda$---a simply-typed functional choreographic language---with polymorphism, including process polymorphism, and call this new language \chorlam.
Thirdly, we add the ability to communicate distributed values such as functions.
Surprisingly, this addition also allows us to write code with delegation, giving a clean language to study all three forms of abstraction.

Let us examine the bookseller \emph{service} in our extended language:
$$
\NewDocumentCommand{\titlevar}{}{\textsf{title}}
\NewDocumentCommand{\pricelookup}{}{\textsf{price\_lookup}}
\NewDocumentCommand{\inbudget}{}{\textsf{buyAtPrice?}}
\LCFun*{B}[\LCProc]{
  \LCfun*{\titlevar}[\LCString{B}]{
    \LCfun*{\inbudget}[\LCArr{\LCNat[B]}{\LCEmptyProcSet}{\LCBool{B}}]
      {
        \LClet*{x}{
          \LCcom{}{B}{\LCProcSeller}{\titlevar}
        }{
          \LClet*{y}{
            \LCcom{}{\LCProcSeller}{B}{(\LCapp{\pricelookup}{x})}
          }{
            \LCif*{\LCapp{\inbudget}{y}}
            {\LCselect{B}{\LCProcSeller}{\textsf{Buy}}{\left(\LCUnitVal[\LCProcSeller]}\right)}
            {\LCselect{B}{\LCProcSeller}{\textsf{Quit}}{\left(\LCUnitVal[\LCProcSeller]}\right)}
          }
        }
    }
  }
}
$$
This program allows a process named~$B$ to connect with \LCProcSeller to buy a book.
$B$ then provides a string~\textsf{title} and a decision function~$\textsf{buyAtPrice?}$.
Thus, we no longer have to write a separate function for every process which may want to buy a book from \LCProcSeller.

While this addition may appear simple, it poses some unique theoretical challenges.
First, the goal of a choreographic language is to compile a global program to one local program per process.
However, since $B$ does not represent any particular process, it is unclear how to compile the polymorphic code above.
We solve this problem via a simple principle: each process knows its identity.
With this principle in place, we can compile the code to a conditional in each process: one option to run if they take the role of $B$, and the other to run if they do not.

Notably, each process chooses dynamically which interpretation of the code to run.
This flexibility is important, since we may want to allow different processes to occupy $B$'s place dynamically.
For instance, we can imagine a situation where \LCProcBuyer[1] and \LCProcBuyer[2] work together to buy a particularly expensive book: perhaps they compare bank accounts, and whoever has more money buys the book for them to share.
This can be achieved in our system with the following code, where \textsf{seller\_service} is the name of the function above:\fabrizio{We have lots of examples, so instead of saying `above', `previous', `next', etc. I would rather put the code that we need to refer to in equation envs and then refer to them using eqref}
$$
\NewDocumentCommand{\titlevar}{}{\textsf{title}}
\LCfun*{\titlevar}[\LCString{\LCProcBuyer[1]}]{
  \LClet*{x}{
    \LCcom{}{\LCProcBuyer[1]}{\LCProcBuyer[2]}{\textsf{bank\_balance}_1}
  }{
    \LCif*{x < \textsf{bank\_balance}_2}{
      \LCselect{\LCProcBuyer[2]}{\LCProcBuyer[1]}{\textsf{Me}}{\LCselect*{\LCProcBuyer[2]}{\LCProcSeller}{\textsf{Me}}{\left(\begin{array}{ll}\LCapp{\LCApp{\textsf{seller\_service}&}{\LCProcBuyer[2]}}{(\LCcom{}{\LCProcBuyer[1]}{\LCProcBuyer[2]}{\titlevar})}\\&(\LCfun{z}{z < \textsf{bank\_balance}_2})\end{array}\right)}}
    }
    {
      \LCselect{\LCProcBuyer[2]}{\LCProcBuyer[1]}{\textsf{You}}{\LCselect*{\LCProcBuyer[2]}{\LCProcSeller}{\textsf{Them}}{(\LCapp{\LCapp{\LCApp{\textsf{seller\_service}}{\LCProcBuyer[1]}}{\titlevar}}{(\LCfun{z}{z < x})})}}
    }
  }
}
$$

A related challenge shows up in the operational semantics of our extended language.
Languages like \chorlam generally have operational semantics which match the semantics of the compiled code by allowing \emph{out-of-order execution}: redices in different processes might be reduced in any order.
However, care must be taken with process polymorphism, since it may not be clear whether two redices are in the same or different processes.

In addition to type and process polymorphism, \chorlam is the first choreographic language to allow the communication of distributed values: values not located entirely at the sender. These values include full choreographies described by distributed functions, which can be used to model delegation.
To see how process polymorphism and communication of distributed values enables delegation, consider the following scenario:
when a buyer asks for a book, the seller first checks whether it is in stock.
If it is, the sale continues as normal.
If not, the seller delegates to a second seller, which may sell the book to the buyer.
We can define this code as follows:

\makeatletter
\begin{samepage}
$$
\NewDocumentCommand{\titlevar}{}{\textsf{title}}
\NewDocumentCommand{\pricelookup}{}{\textsf{price\_lookup}}
\NewDocumentCommand{\inbudget}{}{\textsf{buyAtPrice?}}
\LCFun*{B}[\LCProc]{
  \LCfun*{\titlevar}[\LCString{B}]{
    \LCfun*{\inbudget}[\LCArr{\LCNat[B]}{\varnothing}{\LCBool{B}}]
    {
      \LClet*{x}{
        \LCcom{}{B}{\LCProcSeller}{\titlevar}
      }{
        \LCif*{\LCapp{\textsf{found}{(\LCapp{\pricelookup}{x}}})}{
          \LCselect*{\LCProcSeller}{B}{\textsf{Continue}}{
            \LCselect*{\LCProcSeller}{\LCProcSeller[2]}{\textsf{Disconnect}}{
              \LClet*{y}{
                \LCcom{}{\LCProcSeller}{B}{(\LCapp{\textsf{price}{(\LCapp{\pricelookup}{x})}})}
              }{
                \LCif*{\LCapp{\inbudget}{y}}
                {\LCselect{B}{\LCProcSeller}{\textsf{Buy}}{\left(\LCUnitVal[B]\right)}}
                {\LCselect{B}{\LCProcSeller}{\textsf{Quit}}{\left(\LCUnitVal[B]\right)}}
              }
            }
          }
        }{
          \LCselect*{\LCProcSeller}{B}{\textsf{Delegate}}{
            \LCselect*{\LCProcSeller}{\LCProcSeller[2]}{\textsf{Delegate}}{
              \LClet*{F}{
                \LCcom{}{\LCProcSeller}{\LCProcSeller[2]}{
                  \left(
                    \begin{array}[t]{l}
                      \vspace{-5.5em}\\
                      \LClambda\mkern3mu\titlevar_2 \mathrel{\@LC@KeywordColor{:}} \LCString{\LCProcSeller}\mkern-1mu\@LC@KeywordColor{.}\\\mkern10mu
                      \LCif*{\LCapp{\textsf{found}{(\LCapp{\pricelookup_2}{\titlevar_2}}})}{
                      \LCselect*{\LCProcSeller}{B}{\textsf{Continue}}{
                      \LClet*{y'}{
                      \LCapp{\textsf{price}{(\LCapp{\pricelookup_2}{\titlevar_2})}}
                      }{
                      \LClet*{y}{
                      \LCcom{}{\LCProcSeller}{B}{y'}
                      }{
                      \LCif*{\LCapp{\inbudget}{y}}
                      {\LCselect{B}{\LCProcSeller}{\textsf{Buy}}{\left(\LCUnitVal[B]\right)}}
                      {\LCselect{B}{\LCProcSeller}{\textsf{Quit}}{\left(\LCUnitVal[B]\right)}}
                      }
                      }
                      }
                      }{
                      \LCselect{\LCProcSeller}{B}{\textsf{Quit}}{\left(\LCUnitVal[B]\right)}
                      }
                    \end{array}
                  \right) 
                }}
              {
                      \LClet*{\titlevar_2}{\LCcom{}{\LCProcSeller}{\LCProcSeller[2]}{x}}
                      {
                      \LCapp{F}{\titlevar_2}
                      }
                    }
            }
          }
        }
      }
    }
  }
}
$$
\end{samepage}
\makeatother

\paragraph*{Structure of the Paper.}
We begin in Section~\ref{sec:system-model} by examining the system model of \chorlam.
We then proceed with the following contributions:
\begin{itemize}
\item In Section~\ref{sec:chorlambda-language}, we describe the \chorlam language in detail.
  This language includes both type polymorphism and process polymorphism.
  We develop both a type system and kind system and an operational semantics for \chorlam.
\item In Section~\ref{sec:epp}, we describe the local \emph{network language} used to describe the distributed implementation.
  We also detail how to obtain this implementation via \emph{endpoint projection}, which compiles \chorlam programs to a program for each process.
\item In Section~\ref{sec:epp-correct} we describe the main theorem of this paper, the correctness of endpoint projection with respect to our operational semantics.
  Because of the dynamic nature of process polymorphism, this requires significant reasoning compared to previous works on choreographies.
\end{itemize}
Finally, we discuss related work in Section~\ref{sec:related} and conclude in Section~\ref{sec:conclusion}.


\section{System Model}
\label{sec:system-model}

We begin by discussing the assumptions we make about how \chorlam programs will be run.
These assumptions are as light as possible, allowing for \chorlam to be run in many different scenarios.
In particular, we assume that we have a fixed set of processes, which can communicate via messages.
These processes can each be described by a polymorphic $\lambda$-calculus, similar to System~F$\omega$, but with the addition of communication primitives.

\subsection{Processes}
\label{sec:processes}

We assume that there is a fixed set~\ProcNames of process names~\LCProcP, \LCProcQ, \LCProcAlice, et cetera.
These processes can represent nodes in a distributed system, system processes, threads, or more.
Process polymorphism allows us to refer to processes using type variables, which may go in or out of scope.
Despite this, the set of physically-running processes remains the same.

We assume every process knows its identity.
Thus, every process can choose what code to run on the basis of its identity.
This assumption is reasonable for many practical settings,\fabrizio{We originally said it's unusual but it is actually matched a lot.. Maybe we meant `unusual in theoretical studies'? But isn't it usual in many theories of distributed protocols anyway?} for instance it is common for nodes in distributed systems to know their identity.\fabrizio{It is also common for processes and threads in operating systems, is it relevant?}
This capability is essential to our strategy for enabling process polymorphism.

\subsection{Communication}
\label{sec:communication}

We assume that process communicate via synchronous message passing.
Thus, if \LCProcP sends a message to \LCProcQ, then \LCProcP does not continue until \LCProcQ has received the message.
Moreover, we assume that message passing is instantaneous and certain, so messages do not get lost.

Processes can receive two kinds of messages: values of local programs (described below) and \emph{labels} describing choices made during a computation.
These are used to ensure that different processes stay in lock-step with each other.

\subsection{Local Programs}
\label{sec:local-programs}

We assume that processes run a \emph{local} language described in Section~\ref{sec:epp}.
This is a functional language extended with communication features, similar to the language~GV~\citep{GayV10,Wadler12}.
However, unlike GV or its derivatives, our local language includes type and process polymorphism.

Endpoint projection translates \chorlam into this ``Network Process'' language.
We have thus further extended GV with features required for our endpoint-projection mechanism.
For instance, we provide an \LCKWAmI expression form, which allows a process to choose which code to run based on its identity.
Despite these extensions, the language should feel familiar to any reader familiar with polymorphic $\lambda$-calculi.


\section{The Polymorphic Chor$\lambda$ Language}
\label{sec:chorlambda-language}

We now turn to our first major contribution: the design of the polymorphic, choreographic $\lambda$-calculus, \chorlam.
This calculus extends the choreographic $\lambda$-calculus Chor$\lambda$ of~\citet{CruzFilipeGLMP21} with both type and, more importantly, process polymorphism.
We begin by describing the features that \chorlam shares with the base~Chor$\lambda$ before describing the new features.
The syntax of \chorlam can be found in Figure~\ref{fig:syntax}.

\begin{figure}
  \begin{syntax}
    \abstractCategory[Variables]{x,y,\ldots}
    \abstractCategory[Type Variables]{X,Y,\ldots}
    \abstractCategory[Integers]{n}
    \abstractCategory[Labels]{\ell}
    \abstractCategory[Process Names]{\LCProcP}
    \categoryFromSet[Process-Name Sets]{\LCProcrho}{2^{\text{Type Values}}}
    \category[Kinds]{\LCKindK}
      \alternative{\LCKindAst}
      \alternative{\LCKindArrow{\LCKindK[1]}{\LCKindK[2]}}
      \alternative{\LCProc}
      \alternative{\LCKindWithout{\LCKindK}{\LCProcrho}}
    \category[Types]{\LCTypeT}
      \alternative{\LCTypeV}
      \alternative{\LCapp{\LCTypeT[1]}{\LCTypeT[2]}}
      \alternative{\LCArr{\LCTypeT[1]}{\LCProcrho}{\LCTypeT[2]}}\\
      \alternative{\LCSum{\LCTypeT[1]}{\LCTypeT[2]}}
      \alternative{\LCProd{\LCTypeT[1]}{\LCTypeT[2]}}
      \alternative{\LCForall{X}[\LCKindK]{\LCTypeT}}
      \alternative{\LCTypefun{X}[\LCKindK]{\LCTypeT}}
    \category[Type Values]{\LCTypeV}
      \alternative{X}
      \alternative{\LCUnit[\LCTypeV]}
      \alternative{\LCNat[\LCTypeV]}
      \alternative{\LCArr{\LCTypeV[1]}{\LCProcrho}{\LCTypeV[2]}}
      \alternative{\LCProcP}\\
      \alternative{\LCSum{\LCTypeV[1]}{\LCTypeV[2]}}
      \alternative{\LCProd{\LCTypeV[1]}{\LCTypeV[2]}}
      \alternative{\LCForall{X}[\LCKindK]{\LCTypeV}}
      \alternative{\LCTypefun{X}[\LCKindK]{\LCTypeV}}
    \category[Expressions]{M,N,\ldots}
      \alternative{x}
      \alternative{\LCUnitVal[\LCTypeV]}
      \alternative{\LCLocalVal{n}[\LCTypeV]}
      \alternative{\LCfun{x}[\LCTypeT]{M}}
      \alternative{\LCFun{X}[\LCKindK]{M}}\\
      \alternative{\LCapp{M}{N}}
      \alternative{\LCApp{M}{\LCTypeT}}
      \alternative{\LCinl{\LCTypeT}{M}}
      \alternative{\LCinr{\LCTypeT}{M}}\\
      \alternative{\LCcase{M}{x}{N_1}{y}{N_2}}\\
      \alternative{\LCpair{M}{N}}
      \alternative{\LCfst{M}}
      \alternative{\LCsnd{M}}\\
      \alternative{\LCcomV{\LCTypeT}{\LCTypeV[1]}{\LCTypeV[2]}}
      \alternative{\LCselect{\LCTypeV[1]}{\LCTypeV[2]}{\ell}{M}}
      \alternative{f}
    \category[Values]{V}
      \alternative{x}
      \alternative{\LCUnitVal[\LCTypeV]}
      \alternative{\LCLocalVal{n}[\LCTypeV]}
      \alternative{\LCfun{x}[\LCTypeT]{M}}
      \alternative{\LCFun{x}[\LCKindK]{M}}\\
      \alternative{\LCinl{\LCTypeT}{V}}
      \alternative{\LCinr{\LCTypeT}{V}}
      \alternative{\LCpair{V_1}{V_2}}\\
      \alternative{\LCcomV{\LCTypeT}{\LCTypeV[1]}{\LCTypeV[2]}}
  \end{syntax}
  
  \caption{\chorlam Syntax}
  \label{fig:syntax}
\end{figure}

\vspace{0.5em}\noindent\textbf{Syntax Inherited from Chor$\lambda$.}
Since choreographic programs describe the behavior of an entire communicating network of processes, we need to reason about where terms are located.
In other words, we need to know which processes store the data denoted by a term.
Terms of base type, like integers, are stored by exactly one process.
This is represented in our type system by matching base types with a process name.
For example, integers stored by the process \LCProcAlice are represented by the type $\LCNat[\LCProcAlice]$.
Values of this type also mark the process which stores them, so a value $\LCLocalVal{5}[\LCProcAlice]$ ( read ``the integer $5$ at \LCProcAlice'') has type $\LCNat[\LCProcAlice]$.
In Figure~\ref{fig:syntax}, the only base types are $\LCUnit[\LCProcP]$ and $\LCNat[\LCProcP]$, but it is easy to extend the language with other base types, such as the types~$\LCString{\LCProcP}$ or~$\LCBool{\LCProcP}$ used in the introduction.
We will continue to freely use other base types in our examples.

While base types are located on just one process, data of more-complex types may involve multiple processes.
For instance, the term $\LCpair{\LCLocalVal{5}[\LCProcAlice]}{\LCLocalVal{42}[\LCProcBob]}$ involves both data stored by \LCProcAlice and~\LCProcBob.
This is still recorded in the type: the term above has type $\LCProd{\LCNat[\LCProcAlice]}{\LCNat[\LCProcBob]}$.
In addition to base types and product types, \chorlam also has sum types (written~$\LCSum{\LCTypeT[1]}{\LCTypeT[2]}$), along with their normal introduction and elimination forms.

Functions are treated more unusually: while we have standard $\LClambda$ and application forms, we also allow functions to be defined mutually-recursively with each other.
In order to do so, any \chorlam choreography is associated with a list, $D$, of bindings of functions to \emph{function variables}~$f$, which are also expressions. A function variable can then during execution be instantiated with its definition according to this list.
As we will see in Section~\ref{sec:semantics}, \chorlam terms are evaluated in a context which associates each function variable with a term.
Note that, while in the original Chor$\lambda$ types were mutually recursive in a similar way, in \chorlam we do not support recursive types.
To see why, note that we syntactically restrict many types to \emph{type~values}.
This prevents us having to reason about processes denoted by arbitrary terms---e.g., we cannot send to the ``process''~$\LCapp{(\LCTypefun{X}[\LCProc]{X})}{\LCProcP}$ but we can write $\LCapp{(\LCFun{Y}[\LCProc]{\LCcomV{\LCTypeT}{\LCProcQ}{Y}})}{(\LCapp{(\LCTypefun{X}[\LCProc]{X})}{\LCProcP})}$ which, due to our call-by-value semantics, will force the type to reduce to $\LCProcP$ before $Y$ gets instantiated.
As we will see in Section~\ref{sec:epp}, allowing communication between arbitrary types would make endpoint projection difficult.
However, since recursive types cannot necessarily reduce to a type value, they cannot be used in many parts of the type system.

Function types are also more specific than their usual construction in $\lambda$-calculus: they are written $\LCArr{\LCTypeT[1]}{\LCProcrho}{\LCTypeT[2]}$.
Here, \LCProcrho is the set of processes which describes which processes, other than those in $\LCTypeT[1]$ and $\LCTypeT[2]$, may participate in the function body (note that $\LCProcrho$ may contain type variables).
Thus, if \LCProcAlice wants to communicate an integer to \LCProcBob directly (without intermediaries), then she should use a function of type $\LCArr{\LCNat[\LCProcAlice]}{\LCEmptyProcSet}{\LCNat[\LCProcBob]}$.
However, if she is willing to use the process \LCProcProxy as an intermediary, then she should use a function of type \mbox{$\LCArr{\LCNat[\LCProcAlice]}{\LCProcSet{\LCProcProxy}}{\LCNat[\LCProcBob]}$}.

In order to allow values to be communicated between processes, we provide the primitive communication function~$\LCcomV{\LCTypeT}{\LCProcP}{\LCProcQ}$.
This function takes a value of type \LCTypeT at \LCProcP and returns the corresponding type at \LCProcQ.
As mentioned in the introduction, most choreographic languages provide a communication term modeled after the ``Alice-and-Bob'' notation of cryptographic protocols.
For instance, $\LCProcAlice \mathbin{\texttt{->}} \LCProcBob\colon 5$ might represent \LCProcAlice sending $5$ to \LCProcBob.
This is easily recovered by applying the function~$\LCcomV{\LCTypeT}{\LCProcAlice}{\LCProcBob}$.
For example, the term $\LCcom{\LCNat[\LCProcAlice]}{\LCProcAlice}{\LCProcBob}{(\LCLocalVal{5}[\LCProcAlice])}$ represents \LCProcAlice sending a message containing $5$ to \LCProcBob:
it evaluates to $\LCLocalVal{5}[\LCProcBob]$ and has type $\LCNat[\LCProcBob]$.

Finally, consider the following, where $M$ has type $\LCSum{\LCNat[\LCProcAlice]}{\LCNat[\LCProcAlice]}$: $$\LCcase*{M}{x}{\LCLocalVal{3}[\LCProcBob]}{y}{\LCLocalVal{4}[\LCProcBob]}$$
Clearly, \LCProcBob needs to know which branch is taken, since he needs to store a different return value in each branch.
However, only \LCProcAlice knows which whether $M$ evaluates to $\LCinr{\LCNat[\LCProcAlice]}{V}$ or $\LCinl{\LCNat[\LCProcAlice]}{V}$ (here $\LCKWinl$ and $\LCKWinr$ are used to denote that a value is either the right or left part of a sum and annotated with the type of the other part of the sum to ensure type principality).
Thus, this choreography cannot correspond to any network program. Using the terminology found in the literature of choreographic languages, we might say that the choreography is \emph{unrealisable} because there is insufficient \emph{knowledge of choice}~\citep{CDP12,Montesi22}.

In order to enable programs where a process's behaviour differs depending on other processes data, such as how \LCProcBob behaved differently depending on \LCProcAlice's data, we provide \LCKWselect terms.
These allow one process to tell another which branch has been taken, preventing knowledge from ``appearing out of nowhere.''
For instance, we can extend the program above to:
$$\LCcase*{M}{x}{\LCselect{\LCProcAlice}{\LCProcBob}{\textsf{Left}}{(\LCLocalVal{3}[\LCProcBob])}}{y}{\LCselect{\LCProcAlice}{\LCProcBob}{\textsf{Right}}{(\LCLocalVal{4}[\LCProcBob])}}$$
This represents the same program as above, except \LCProcAlice tells \LCProcBob whether the left or the right branch has been taken.
Unlike the previous version of this example, it \emph{does} represent a (deadlock-free) network program.
In general, we allow arbitrary labels to be sent by \LCKWselect terms, so semantically-meaningful labels can be chosen.

While $\LCKWcom$ and $\LCKWselect$ both transfer information between two processes, they differ in what information they transfer. $\LCKWcom$ moves a value, e.g., as an integer or a function, from the sender to the receiver. $\LCKWselect$ on the other hand uses a label to inform the receiver of a choice made by the sender. Some choreographic languages combine the two, so both a label and a value is communicated at the same time, but like most choreographic languages \chorlam keeps the two separate.

\vspace{0.5em}\noindent\textbf{Syntax Additions over Chor$\lambda$.}\fabrizio{You can just use paragraph* (see at the end of the intro `Structure of the Paper').}
In order to achieve (both type and process) polymorphism in \chorlam, we add several features based on System~F$\omega$~\citep{Girard72}.
In particular, we add kinds and universal types~$\LCForall{X}[\LCKindK]{\LCTypeT}$ along with type abstraction and application.
From System~F$\omega$ we inherit the kind $\LCKindAst$, which is the kind of types.
We additionally inherit the kind $\LCKindArrow{\LCKindK[1]}{\LCKindK[2]}$ which represents functions from types to types.

Moreover, we inherit type-level functions~$\LCTypefun{X}[\LCKindK]{\LCTypeT}$ from System~F$\omega$.
These represent the definition of type constructors.
We additionally have type-level function application~$\LCapp{\LCTypeT[1]}{\LCTypeT[2]}$.
Since types contain computation, we also define type \emph{values}, which are simply types without application.

Note that the base types~$\LCUnit[\LCTypeV]$ and~$\LCNat[\LCTypeV]$, like local values, are \emph{syntactically} restricted to only allow type values as subterms.
This allows us to use a type variable to compute the location of a value dynamically, but not arbitrary terms, which would make it much harder to tell at time of projection where the value is located.
Thus, we can write $\LCapp{(\LCTypefun{X}[\LCProc]{\LCNat[X]})}{(\LCapp{Y}{\LCProcP})}$ to compute the location of an integer dynamically ($\LCapp{Y}{\LCProcP}$ has to reduce to a type value before $X$ can be instantiated), but we cannot write $\LCNat[(\LCapp{Y}{\LCProcP})]$ directly.
This way, our projected calculus can tell when instantiating $X$ (at runtime) whether it gets instantiated as $\LCProcP$.
It would be more complicated to create runtime checks for whether $Y$ gets instantiated as a function type that outputs $\LCProcP$ or not.

In addition to the kinds $\LCKindAst$ and $\LCKindArrow{\LCKindK[1]}{\LCKindK[2]}$ of System~F$\omega$, we also have the kind $\LCProc$ of \emph{process names}.
Thus, process names are types, but they cannot be used to type any terms.
Additionally, we have kinds $\LCKindWithout{\LCKindK}{\LCProcrho}$, which represents types of kind~\LCKindK which do not mention any of the processes in the set $\LCProcrho$.
Since we  restrict the types that can be communicated based on which processes they contain, as we will see soon, this restricted kind can be used to define polymorphic functions which contain communication.
For instance, the term $$\LCFun{X}[\LCProc]{\LCFun{Y}[\LCKindWithout{\LCProc}{\{X\}}]{\LCcom{\LCNat[X]}{X}{Y}{(\LCLocalVal{5}[X])}}}$$ defines a function which, given \emph{distinct} processes~$X$ and~$Y$, causes $X$ to send $5$ to $Y$.

In the rest of this section, we explore the semantics of \chorlam.
First, we look at its static semantics, both in the form of typing and kinding.
Second, we describe its operational semantics.
Throughout, we will continue to give intuitions based on the concurrent interpretation of \chorlam, though the semantics we give here does not correspond directly to that interpretation.

\subsection{Typing}
\label{sec:typing}

We now turn to the type system for \chorlam.
As before, our type system builds on that for Chor$\lambda$.
Here, we focus on the rules that are new in this work.
Thus, we focus on rules related to polymorphism, and those that have had to change due to polymorphism.

Typing judgements for \chorlam have the form $\Theta;\Gamma \vdash M : \LCTypeT$, where $\Theta$ is the set of process names---either names in \ProcNames or type variables with kind \LCProc---used in $M$ or the type of $M$.
The \emph{typing environment}~$\Gamma$ is a list associating variables and function names to their types and type variables and process names to their kinds.
We sometimes refer to the pair $\Theta;\Gamma$ as a \emph{typing context}.

\begin{figure}
  \begin{mathpar}
    \inferrule*[left=Tunit]{
      \Theta;\Gamma\vdash \LCTypeV:: \LCProc
    }{
      \Theta;\Gamma\vdash \LCUnitVal[\LCTypeV] : \LCUnit[\LCTypeV]
    }\and
    \inferrule*[left=Tint]{
      \Theta;\Gamma\vdash \LCTypeV :: \LCProc
    }{
      \Theta;\Gamma\vdash \LCLocalVal{n}[\LCTypeV]:\LCNat[\LCTypeV]
    } \and
    \inferrule*[left=Tapp]{
      \Theta;\Gamma\vdash N:\LCArr{\LCTypeT[1]}{\LCProcrho}{\LCTypeT[2]}\\
      \Theta;\Gamma\vdash M:\LCTypeT[1]
    }{
      \Theta;\Gamma\vdash N~M:\LCTypeT[2]
    } \and
    \inferrule*[left=Tabs]{
      \Theta;\Gamma\vdash \LCTypeT[1]::\LCKindAst \\
      \Theta;\Gamma'\vdash \LCTypeV:: \LCProc \text{ for all } \LCTypeV\in\rho\\
      \Theta\cap(\LCProcrho\cup \roles(\LCTypeT[1]) \cup \roles(\LCTypeT[2])\cup\ftv(\LCTypeT[1])\cup\ftv(\LCTypeT[2]));\Gamma,x:\LCTypeT[1]\vdash M:\LCTypeT[2] \\
    }{
      \Theta;\Gamma\vdash\LCfun{x}[\LCTypeT[1]]{M}:\LCArr{\LCTypeT[1]}{\LCProcrho}{\LCTypeT[2]}
    } \and
    \inferrule*[left=Tsel]{
      \Theta;\Gamma\vdash \LCTypeV[1] :: \LCProc\\
      \Theta;\Gamma\vdash \LCTypeV[2] :: \LCProc\\
      \Theta;\Gamma\vdash M : \LCTypeT
    }{
      \Theta;\Gamma\vdash \LCselect{\LCTypeV[1]}{\LCTypeV[2]}{\ell}{M}:\LCTypeT
    } \and
    \inferrule*[left=Tcom]{
      \Theta;\Gamma\vdash \LCTypeT:: \LCKindArrow{\LCProc}{\LCKindAst}\\
      \Theta;\Gamma\vdash \LCTypeV[1]:: \LCKindWithout{\LCProc}{(\mn(\LCTypeT)\cup \ftv(\LCTypeT)})\\
      \Theta;\Gamma\vdash \LCTypeV[2]::\LCKindWithout{\LCProc}{(\mn(\LCTypeT)\cup \ftv(\LCTypeT)})
    }{
      \Theta;\Gamma \vdash \LCcomV{\LCTypeT}{\LCTypeV[1]}{\LCTypeV[2]}:(\LCArr{\LCTypeT~\LCTypeV[1]}{\LCEmptyProcSet}{\LCTypeT~\LCTypeV[2]})
    } \and
    \inferrule*[left=TappT]{
      \Theta;\Gamma\vdash M:\LCForall{X}[\LCKindK]{\LCTypeT[1]}\\
      \Theta;\Gamma \vdash \LCTypeT[2]::\LCKindK
    }{
      \Theta;\Gamma\vdash \LCApp{M}{\LCTypeT[2]} : \SingleSubst{\LCTypeT[]}{X}{\LCTypeT[2]}
    } \and
    \inferrule*[left=TabsT]{
      \Theta';\Gamma', X :: \LCKindK \vdash M : \LCTypeT\\
      \text{if}\;\exists \LCKindKPrime, \LCProcrho.\,\LCKindK = \LCKindWithout{\LCKindKPrime}{\LCProcrho}
      \mathrel{\text{then}}\Gamma' = \addkind{(\Gamma + X)}{\LCProcrho}{X}
      \mathrel{\text{else}}\Gamma' = \Gamma + X\\
      \text{if}\;\LCKindK = \LCProc \mathrel{\text{or}} \exists \LCProcrho.\, \LCKindK = \LCKindWithout{\LCProc}{\LCProcrho}
      \mathrel{\text{then}} \Theta' = \Theta\cup \{X\}
      \mathrel{\text{else}} \Theta' = \Theta
    }{
      \Theta;\Gamma\vdash\LCFun{X}[\LCKindK]{M}:\LCForall{X}[\LCKindK]{\LCTypeT}
    }\and
    \inferrule*[left=Teq]{
      \Theta;\Gamma\vdash M:\LCTypeT[1]\\
      \LCTypeT[1] \equiv \LCTypeT[2]\\
      \Theta;\Gamma \vdash \LCTypeT[2]::\LCKindAst
    }{
      \Theta;\Gamma\vdash M:\LCTypeT[2]
    }    
  \end{mathpar}

\caption{Typing Rules (Selected)}\label{fig:type}
\end{figure}

Selected rules for our type system can be found in Figure~\ref{fig:type}.
\iftr The full collection of rules are given in Appendix~\ref{sec:full-typing-rules}. \fi
Again, many of the rules are inherited directly from Chor$\lambda$~\citep{CruzFilipeGLMP21}; we thus focus on the rules that have changed due to our additions.
Many, if not most, of these rules are inspired by System~F$\omega$.
However, the addition of the kind of processes and Without kinds---i.e., kinds of the form $\LCKindWithout{\LCKindK}{\LCProcrho}$---also lead to some changes.

The rules~\rulen{Tunit} and~\rulen{Tint} give types to values of base types.
Here, we have to ensure that the location of the term is a process.
Intuitively, then, we want the location to have kind~\LCProc.
However, it might have restrictions---that is, it might be of the form $\LCKindWithout{\LCProc}{\LCProcrho}$.
In this case, our subkinding system (which you can find details about in Section~\ref{sec:kinding}) still allows us to apply the rule.

We express function application and abstraction via the \rulen{Tapp} and \rulen{Tabs} rules, respectively.
The application rule~\rulen{Tapp} is largely standard---the only addition is the addition of a set~\LCProcrho on the function type, as discussed earlier.
The abstraction rule~\rulen{Tabs}, on the other hand, is more complicated.
First, it ensures that the argument type,~\LCTypeT[1], has kind~\LCKindAst.
Then, it ensures that every element in the set decorating the arrow is a process name---i.e., that it has kind \LCProc.
Finally, it checks that, in an extended environment, the body of the function has the output type~\LCTypeT[2].
As is usual, this extended environment gives a type to the argument.
However, it restricts the available process names to those in the set~\LCProcrho and those mentioned in the types~\LCTypeT[1] and~\LCTypeT[2].

There are two ways that a type~\LCTypeT can mention a process: it can either name it directly, or it can name it via a type variable.
Thus, in the rule \rulen{Tabs} we allow the free variables of \LCTypeT[1] and \LCTypeT[2] to remain in the process context, computing them using the (standard) free-type-variable function where $\LCForall{X}[\LCKindK]{M}$ and $\LCTypefun{X}[\LCKindK]{M}$ both bind $X$.
However, we must also identify the \emph{involved processes} in a type, which we write $\roles(\LCTypeT)$ and compute as follows:
  \begin{mathpar}
    \roles(X) = \LCEmptyProcSet \and
    \roles(\LCProcP) = \LCProcP \and
    \roles(\LCUnit[\LCTypeV]) = \roles(\LCNat[\LCTypeV]) = \roles(\LCTypeV) \and
    \roles(\LCForall{X}[\LCKindWithout{\LCKindK}{\LCProcrho}]{\LCTypeT}) = \roles(\LCTypefun{X}[\LCKindWithout{\LCKindK}{\LCProcrho}]{\LCTypeT}) = \roles(\LCTypeT) \cup (\ProcNames \setminus \LCProcrho) \and
    \roles(\LCForall{X}[\LCKindK]{\LCTypeT}) = \roles(\LCTypefun{X}[\LCKindK]{\LCTypeT}) = \ProcNames~\text{if}~\nexists \LCKindK',\LCProcrho.\,\LCKindK=\LCKindWithout{\LCKindK'}{\LCProcrho}
  \end{mathpar}
The involved processes of other types are defined homomorphically.

The communication primitives~\LCKWselect and~\LCKWcom are typed with \rulen{Tsel} and \rulen{Tcom}, respectively.
A term $\LCselect{\LCTypeV[1]}{\LCTypeV[2]}{\ell}{M}$ behaves as $M$, where the process~$\LCTypeV[1]$ informs the process~$\LCTypeV[2]$ that the $\ell$ branch has been taken, as we saw earlier.
Thus, the entire term has type \LCTypeT if $M$ does.
Moreover, \LCTypeV[1] and \LCTypeV[2] must be processes.

The rule~\rulen{Tcom} types \LCKWcom terms. So far we have been simplifying the type used in $\LCcomV{\LCTypeT}{\LCProcP}{\LCProcQ}$ for readability. We have been using $\LCTypeT$ to denote the input type, but as it turns out to type $\LCcomV{\LCTypeT}{\LCProcP}{\LCProcQ}$ correctly, we have to complicate things a little.
Intuitively, a term $\LCcom{\LCTypeT}{\LCTypeV[1]}{\LCTypeV[2]}{M}$ represents \LCTypeV[1] communicating the parts of $M$ on \LCTypeV[1] to \LCTypeV[2].
Thus, we require that \LCTypeT be a type \emph{transformer} requiring a process.
Moreover, \LCTypeV[1] and \LCTypeV[2] cannot be mentioned in \LCTypeT; otherwise not every part of the type of $M$ on \LCTypeV[1] in our example above would transfer to \LCTypeV[2].
For this we use the following notion of \emph{mentioned processes}:
  \begin{mathpar}
    \mn(X) = \LCEmptyProcSet \and
    \mn(\LCProcP) = \LCProcP \and
    \mn(\LCUnit[\LCTypeV]) = \mn(\LCNat[\LCTypeV]) = \mn(\LCTypeV) \and
    \mn(\LCForall{X}[\LCKindWithout{\LCKindK}{\LCProcrho}]{\LCTypeT}) = \mn(\LCTypefun{X}[\LCKindWithout{\LCKindK}{\LCProcrho}]{\LCTypeT}) = \mn(\LCTypeT) \cup \LCProcrho \and
    \mn(\LCForall{X}[\LCKindK]{\LCTypeT}) = \mn(\LCTypefun{X}[\LCKindK]{\LCTypeT}) = \mn(\LCTypeT)~\text{if}~\nexists \LCKindK',\LCProcrho.\,\LCKindK=\LCKindWithout{\LCKindK'}{\LCProcrho}
  \end{mathpar}
  Again, with other types being defined homomorphically.
The difference between involved and mentioned processes is subtle. If there is no polymorphism, they are the same, but when dealing with polymorphism with restriction they are opposites: involved processes includes every process not in the restriction (the variable could be instantiated as something involving those processes and thus they may be involved), while mentioned names includes the processes mentioned in the restriction. Mentioned names is used only when typing $\LCKWcom$.
If we have such a type-level function, \LCTypeT, and two type values $\LCTypeV[1]$ and $\LCTypeV[2]$ which are not and will not be instantiated to anything mentioned in $\LCTypeT$ then we can type $\LCcom{\LCTypeT}{\LCTypeV[1]}{\LCTypeV[2]}$ as a function from $\LCapp{\LCTypeT}{\LCTypeV[1]}$ to $\LCapp{\LCTypeT}{\LCTypeV[2]}$.
Since this is direct communication, no intermediaries are necessary and we can associate this arrow with the empty set~$\LCEmptyProcSet$.

It is worth noting at this point that the communication rule inspired our use of System~F$\omega$ rather than plain System~F, which lacks type-level computation.
In Chor$\lambda$ and other previous choreographic languages, communicated values must be local to the sender.
In \chorlam, this would mean not allowing the communicated type to include type variables or processes other than the sender.
Since we are introducing the idea of using communication as a means of delegation, we have slackened that restriction.
This means that \chorlam programs can communicate larger choreographies whose type may involve other processes, and importantly other type variables.
We see this in the delegation example at the end of Section~\ref{sec:introduction}, where we have the communication $\LCcomV{}{\LCProcSeller}{\LCProcSeller[2]}$.
Adding in the required type annotation (which we had suppressed in the introduction), this becomes $\LCcomV{\LCTypefun{X}[\LCProc]{\LCArr{\LCString{X}}{\varnothing}{\LCUnit[B]}}}{\LCProcSeller}{\LCProcSeller[2]}$.
Note that this still leaves us with a free type variable $B$, representing the unknown process that $\LCProcSeller$ is telling $\LCProcSeller[2]$ to interact with!
Since we cannot ban free type variables in communicated types, we must create a typing system that can handle them, and this requires type level computation.

To see why this led us to type-level computation, consider the alternative.
In Chor$\lambda$ and other choreographic works, we would have a type communication using process \emph{substitution} instead of communication.
The annotated program would then be $\LCcomV{\LCArr{\LCString{\LCProcSeller}}{\varnothing}{\LCUnit[B]}}{\LCProcSeller}{\LCProcSeller[2]}$.
When applied to a program of appropriate type, the result would have type $$(\LCArr{\LCString{\LCProcSeller}}{\varnothing}{\LCUnit[B]})[\LCProcSeller \mapsto \LCProcSeller[2]] = \LCArr{\LCString{\LCProcSeller[2]}}{\varnothing}{\LCUnit[B]}$$
Note that, because $B$ is a type variable, it was ignored by the substitution.
If $B$ is later substituted with something which contains \LCProcSeller, then we have lost the information that we need to change this to $\LCProcSeller[2]$.
Thus, we need some mechanism to delay this substitution; rather than use a mechanism like explicit substitutions, we instead reached for the standard tool of System~F$\omega$.
This seemed more elegant and less ad-hoc; moreover, it adds features which a real-world implementation of \chorlam would want anyway.

Returning now to the typing rules of Figure~\ref{fig:type}, we next have the \rulen{TappT} and \rulen{TabsT} rules, which type universal quantification.
The \rulen{TappT} rule is completely standard, while the \rulen{TabsT} rule is our most-complicated rule. In this rule, we have 4 different definitions for the typing context of $M$, depending on the kind of $X$.
As is standard, we check if the body of the function has the right type when the parameter $X$ has kind~\LCKindK.
But first, if $X$ is a process, then we need to extend $\Theta$ with $X$.
In addition, we must further manipulate the context in order to ensure that the types whose kinds are restricted of $X$ correspond to the restriction on the kind of $X$.

First, the new type variable $X$ may shadow a previously-defined $X$.
Thus, we need to remove $X$ from any restrictions already in the context.
We do this using the following operation~$\LCKindK + \LCTypeV$:
$$(\LCKindWithout{\LCKindK}{\LCProcrho}) + \LCTypeV = \LCKindWithout{(\LCKindK + \LCTypeV)}{(\LCProcrho \setminus \LCProcSet{\LCTypeV})}$$
We define $+$ on other kinds homomorphically, and extend this to contexts as usual:
$$\Gamma + \LCTypeV = \{x : \LCTypeT \mid x : \LCTypeT \in \Gamma\} \cup \{X : \LCKindK + \LCTypeV \mid X : \LCKindK \in \Gamma\}$$

Furthermore, if $X$ itself has a Without kind---that is, $X$'s kind tells us it cannot be any of the processes in $\LCProcrho$---then we need to symmetrically add a restriction on $X$ to every type in \LCProcrho.
Otherwise, we would not be able to use the roles in \LCProcrho in any place where we cannot use $X$, even though we know $X$ will not be instantiated with them.
We do this with the operation $\addkind{\Gamma}{\LCProcrho}{X}$, which we define as follows:
$$
\addkind{\Gamma}{\LCProcrho}{X} =
\begin{array}[t]{l}
  \{x:\LCTypeT\mid x:\LCTypeT\in\Gamma\}\cup\{\LCTypeT::\LCKindK\mid \LCTypeT::\LCKindK\in\Gamma \text{ and }\LCTypeT\notin \LCProcrho\} \\
  {} \cup \{\LCTypeT::\LCKindWithout{\LCKindK}{(\LCProcrho[2]\cup\LCProcSet{X})}\mid \LCTypeT::\LCKindWithout{\LCKindK}{\LCProcrho[2]} \in\Gamma \text{ and }\LCTypeT\in \LCProcrho\}  \\
  {} \cup \{\LCTypeT::\LCKindWithout{\LCKindK}{\LCProcSet{X}}\mid \LCTypeT::\LCKindK\in\Gamma\text{, } \LCKindK\neq \LCKindWithout{\LCKindK[2]}{\LCProcrho[2]} \text{, and }\LCTypeT\in \LCProcrho\}
\end{array}
$$

With these operations in place, we can now fully understand \rulen{TabsT}.
When $\LCKindK$ is actually a Without kind, then we must handle both shadowing and symmetrical restrictions.
However, when it is not a Without kind, we must only handle shadowing.

The final addition to our type system is the rule \rulen{Teq}.
This is another standard rule from System~F$\omega$; it tells us that we are allowed to compute in types.
More specifically, it tells us that we can replace a type with an equivalent type, using the following equivalence:
\begin{mathpar}
  \inferrule*{ }{
    \LCTypeT \equiv \LCTypeT
  }\and
  \inferrule*{
    \LCTypeT[1] \equiv \LCTypeT[2]
  }{
    \LCTypeT[2] \equiv \LCTypeT[1]
  }\and
  \inferrule*{
    \LCTypeT[1] \equiv \LCTypeT[2]\\
    \LCTypeT[2] \equiv \LCTypeT[3]
  }{
    \LCTypeT[1] \equiv \LCTypeT[3]
  }\\
  \inferrule*{
    \LCTypeT[1]\equiv\LCTypeTPrime[1]\\
    \LCTypeT[2]\equiv\LCTypeTPrime[2]
  }{
    \LCArr{\LCTypeT[1]}{\LCProcrho}{\LCTypeT[2]} \equiv \LCArr{\LCTypeTPrime[1]}{\LCProcrho}{\LCTypeTPrime[2]}
  }\and
  \inferrule*{
    \LCTypeT[1]\equiv\LCTypeTPrime[1]\\
    \LCTypeT[2]\equiv\LCTypeTPrime[2]
  }{
    \LCSum{\LCTypeT[1]}{\LCTypeT[2]}\equiv\LCSum{\LCTypeTPrime[1]}{\LCTypeTPrime[2]}
  } \and
  \inferrule*{
    \LCTypeT[1]\equiv\LCTypeTPrime[1]\\
    \LCTypeT[2]\equiv\LCTypeTPrime[2]
  }{
    \LCProd{\LCTypeT[1]}{\LCTypeT[2]}\equiv\LCProd{\LCTypeTPrime[1]}{\LCTypeTPrime[2]}
  } \and
  \inferrule*{
    \LCTypeT \equiv \LCTypeTPrime
  }{
    \LCTypefun{X}[\LCKindK]{\LCTypeT}\equiv \LCTypefun{X}[\LCKindK]{\LCTypeTPrime}
  } \and
  \inferrule*{ }
  {
    \LCapp{(\LCTypefun{X}[\LCKindK]{\LCTypeT[1]})}{\LCTypeT[2]} \equiv \SingleSubst{\LCTypeT[1]}{X}{\LCTypeT[2]}
  } \and
  \inferrule*{
    \LCTypeT[1]\equiv\LCTypeTPrime[1]\\
    \LCTypeT[2]\equiv\LCTypeTPrime[2]
  }{
    \LCapp{\LCTypeT[1]}{\LCTypeT[2]} \equiv\LCapp{\LCTypeTPrime[1]}{\LCTypeTPrime[2]}
  }\and
  \inferrule*{
    \LCTypeT \equiv \LCTypeTPrime
  }{
    \LCForall{X}[\LCKindK]{\LCTypeT}\equiv \LCForall{X}[\LCKindK]{\LCTypeTPrime}
  }
\end{mathpar}

While \rulen{Teq} is the last rule we added to our type system, it is not the last rule in Figure~\ref{fig:type}.
We also add an extra judgement of the form $\Theta; \Gamma \vdash D$ where $\Theta; \Gamma$ is a typing context as before, and $D$ is a set of definitions for function variables---i.e., $D = \{f_1 = M_1, \ldots f_n = M_n\}$.
We write $D(f)$ for the term associated with $f$ in $D$.
The only rule for this judgement is \rulen{Tdefs}, which says that a set of definitions is well-formed if every variable in $D$ is associated with a type \LCTypeT in $\Gamma$, and the body of $f$ in $D$ can be given be given type~\LCTypeT in the context $\emptyset; \Gamma$.
We require that the body of $f$ can be typed with an empty set of roles because they are global predefined functions, and as such they should not be local to any one process.

$$\inferrule*[left=Tdefs]{
      \forall f\in \mathsf{domain}(D).\, f \mathrel{:} \LCTypeT \in \Gamma \land \emptyset;\Gamma\vdash D(f) \mathrel{:} \LCTypeT
    }{
      \Theta;\Gamma\vdash D
    }$$

\subsection{Kinding}
\label{sec:kinding}

We finish our discussion of the static semantics of \chorlam by looking at our kinding system.
Our kinding system uses only one judgement, $\Theta; \Gamma \vdash \LCTypeT :: \LCKindK$, which says that in the typing context $\Theta; \Gamma$, the type~\LCTypeT has kind~\LCKindK.
You can find the rules of our kinding system in Figure~\ref{fig:kind}.
These are mostly directly inherited from System~F$\omega$.
However, we must account for \LCProc and Without kinds.

\begin{figure}
  \begin{mathpar}
    \inferrule*[left=Kvar]{
      X::\LCKindK\in \Gamma \\
      \text{if}\; \LCKindK \in \{\LCProc,\LCKindWithout{\LCProc}{\LCProcrho}\} \mathrel{\text{then}} X \in \Theta
    }{
      \Theta;\Gamma \vdash X::\LCKindK
    }\and
    \inferrule*[left=Krole]{
      \LCProcP :: \LCKindK \in \Gamma\\
      \LCKindK\in\{\LCProc,\LCKindWithout{\LCProc}{\LCProcrho}\}\\
      \LCProcP \in \Theta \\
      \text{if}\;\LCKindK=\LCKindWithout{\LCProc}{\LCProcrho} \mathrel{\text{then}} \LCProcP\notin\LCProcrho
    }{
      \Theta;\Gamma \vdash \LCProcP:: \LCKindK
    }\and
    \inferrule*[left=Kunit]{
      \Theta;\Gamma\vdash \LCTypeT::\LCKindWithout{\LCProc}{\LCProcrho}
    }{
      \Theta;\Gamma \vdash \LCUnit[\LCTypeT]::\LCKindWithout{\LCKindAst}{\LCProcrho}
    }\and
    \inferrule*[left=Kint]{
      \Theta;\Gamma\vdash \LCTypeT::\LCKindWithout{\LCProc}{\LCProcrho}
    }{
      \Theta;\Gamma \vdash \LCNat[\LCTypeT]::\LCKindWithout{\LCKindAst}{\LCProcrho}
    }\and
    \inferrule*[left=Kfun]{
      \Theta;\Gamma\vdash \LCTypeT[1]::\LCKindWithout{\LCKindAst}{\LCProcrho[2]}\\
      \Theta;\Gamma\vdash \LCTypeT[2]::\LCKindWithout{\LCKindAst}{\LCProcrho[2]}\\
      \forall \LCTypeV \in \LCProcrho[1].\,\Theta;\Gamma\vdash \LCTypeV::\LCKindWithout{\LCProc}{\LCProcrho[2]}
    }{
      \Theta;\Gamma \vdash \LCArr{\LCTypeT[1]}{\LCProcrho[1]}{\LCTypeT[2]}::\LCKindWithout{\LCKindAst}{\LCProcrho[2]}
    }\and \inferrule*[left=Kabs]{
      \Theta';\Gamma,X::\LCKindK[1] \vdash \LCTypeT::\LCKindK[2]
    }{
      \Theta;\Gamma \vdash \LCTypefun{X}[\LCKindK[1]]{\LCTypeT}::\LCKindArrow{\LCKindK[1]}{\LCKindK[2]} 
    } \and
    \inferrule*[left=Kall]{
      \Theta';\Gamma,X::\LCKindK \vdash \LCTypeT::\LCKindWithout{\LCKindAst}{\LCProcrho} 
    }{
      \Theta;\Gamma \vdash \LCForall{X}[\LCKindK]{\LCTypeT}::\LCKindWithout{\LCKindAst}{\LCProcrho}
    }\and
    \inferrule*[left=Karr]{
      \Theta;\Gamma\vdash \LCTypeT::\LCKindArrow{(\LCKindWithout{\LCKindK[1]}{\LCProcrho})}{ (\LCKindWithout{\LCKindK[2]}{\LCProcrho})}
    }{
      \Theta;\Gamma \vdash \LCTypeT::\LCKindWithout{(\LCKindArrow{\LCKindK[1]}{\LCKindK[2]})}{\LCProcrho}
    }\and
    \inferrule*[left=Ksub]{
      \Theta;\Gamma \vdash \LCTypeT::\LCKindK[1]\\
      \subkind{\LCKindK[1]}{\LCKindK[2]}
    }{
      \Theta;\Gamma \vdash \LCTypeT::\LCKindK[2]
    }\and
    \inferrule*[left=Ksum]{
      \Theta;\Gamma \vdash {\LCTypeT[1]}:: \LCKindWithout{\LCKindAst}{\LCProcrho} \\
      \Theta;\Gamma \vdash {\LCTypeT[2]}:: \LCKindWithout{\LCKindAst}{\LCProcrho}
    }{
      \Theta;\Gamma \vdash \LCSum{\LCTypeT[1]}{\LCTypeT[2]}:: \LCKindWithout{\LCKindAst}{\LCProcrho}
    }\and
    \inferrule*[left=Kprod]{
      \Theta;\Gamma \vdash \LCTypeT[1] :: \LCKindWithout{\LCKindAst}{\LCProcrho}\\
      \Theta;\Gamma \vdash \LCTypeT[2] :: \LCKindWithout{\LCKindAst}{\LCProcrho}
    }{
      \Theta;\Gamma \vdash \LCProd{\LCTypeT[1]}{\LCTypeT[2]}:: \LCKindWithout{\LCKindAst}{\LCProcrho}
    }
  \end{mathpar}
  \caption{Kinding Rules}
  \label{fig:kind}
\end{figure}

For instance, the rules~\rulen{Kunit} and~\rulen{Kint} check that the type representing which process is storing the data indeed has the kind~\LCProc.
Similarly, \rulen{Kfun} ensures that all of the types in the set of possible intermediaries are processes.
The rule for type variables, \rulen{Kvar}, ensures that if a type variable $X$ is assigned kind~\LCProc, then $X$ must also be in $\Theta$.

One of the biggest differences between our kinding system and that of System~F$\omega$, however, is the rule~\rulen{Ksub} which tells us that our system enjoys \emph{subkinding}.
The subkinding rules come from the subset ordering on Without kinds.
The rules for subkinding are as follows:
\begin{mathpar}
  \inferrule*[left=SKRefl]{ }{\subkind{\LCKindK}{\LCKindK}}\and
  \inferrule*[left=SKTrans]{
    \subkind{\LCKindK[1]}{\LCKindK[2]}\\
    \subkind{\LCKindK[2]}{\LCKindK[3]}
  }{
    \subkind{\LCKindK[1]}{\LCKindK[3]}
  }\and
  \inferrule*[left=SKArr]{
    \subkind{\LCKindK[1]}{\LCKindKPrime[1]}\\
    \subkind{\LCKindK[2]}{\LCKindKPrime[2]}\\
  }{
    \subkind{\LCKindArrow{\LCKindK[1]}{\LCKindK[2]}}{\LCKindArrow{\LCKindKPrime[1]}{\LCKindKPrime[2]}}
  }\and
  \inferrule*[left=SKEmpty]{ }{\subkind{\LCKindK}{\LCKindWithout{\LCKindK}{\LCEmptyProcSet}}}\and
  \inferrule*[left=SKWithoutL]{
    \subkind{\LCKindK[1]}{\LCKindK[2]}
  }{
    \subkind{\LCKindWithout{\LCKindK[1]}{\LCProcrho}}{\LCKindK[2]}
  }\and
  \inferrule*[left=SKWithoutUnion]{
    \subkind{\LCKindK[1]}{\LCKindK[2]}
  }{
    \subkind{\LCKindWithout{\LCKindK[1]}{(\LCProcrho[1] \cup \LCProcrho[2])}}
            {\LCKindWithout{\LCKindK[2]}{\LCProcrho[1]}}
  }  
\end{mathpar}

\begin{lemma}
Let \LCTypeT be a type.
  If there exists a typing context $\Theta;\Gamma$ such that $\Theta;\Gamma\vdash \LCTypeT::\LCKindK$ then there exists a unique type value~$\LCTypeV$ such that $\LCTypeT\equiv \LCTypeV$.
\end{lemma}
\iftr
\begin{proof}
The existence of $\LCTypeV$ follows from induction on $\Theta;\Gamma\vdash \LCTypeT::\LCKindK$ and its uniqueness from induction on $\LCTypeT\equiv \LCTypeV$.
\end{proof}
\fi

\begin{lemma}[Type restriction]\label{thm:KindRes}
Let \LCTypeT be a type.
  If there exists a typing context $\Theta;\Gamma$ such that $\Theta;\Gamma\vdash \LCTypeT::\LCKindWithout{\LCKindK}{\LCProcrho}$ then $(\roles(\LCTypeT)\cup\ftv(\LCTypeT))\cap\LCProcrho=\emptyset$
\end{lemma}
\iftr
\begin{proof}
Follows from kinding rules.
\end{proof}
\fi

\begin{theorem}[Kindable types]
  Let $M$ be a choreography and \LCTypeT be a type such that $\Theta;\Gamma \vdash M : \LCTypeT$.
  Then  $\Theta;\Gamma \vdash \LCTypeT :: \LCKindAst$.
\end{theorem}
\iftr
\begin{proof}
Follows from induction on the derivation of $\Theta;\Gamma \vdash M : \LCTypeT$ and the kinding rules.
\end{proof}
\fi

We also find that types have the same kinds as their equivalent type values.
Due to $\beta$-expansion, a kindable type can be equivalent to an unkindable type, but not an unkindable type value.

\begin{theorem}[Kind Preservation]\label{thm:KindPres}
  Let \LCTypeT be a type.
  If there exists a typing context $\Theta;\Gamma$ such that $\Theta;\Gamma\vdash \LCTypeT::\LCKindK$, then $\Theta;\Gamma\vdash \LCTypeV::\LCKindK$ for any type value~$\LCTypeV$ such that $\LCTypeT\equiv \LCTypeV$.
\end{theorem}
\iftr
\begin{proof}
  Follows from the kinding and type equivalence rules.
  We take advantage of the fact that if we use the rule $\LCapp{(\LCTypefun{X}[\LCKindK]{\LCTypeT[1]})}{\LCTypeT[2]} \equiv \SingleSubst{\LCTypeT[1]}{X}{\LCTypeT[2]}$ to create an unkindable $\LCTypeT[2]\equiv \LCTypeT[1]$ with an extra application, then we must also use the same rule to remove this new type application before we get to a base type.
\end{proof}
\fi

\begin{example}
  We return to the delegation example from Section~\ref{sec:introduction} and try to type it.
  As $B$ appears free in the type of a value, $F$, being communicated between $\LCProcSeller$ and $\LCProcSeller[2]$, $B$ must actually have the restricted kind $\LCKindWithout{\LCProc}{\{\LCProcSeller,\LCProcSeller[2]\}}$.
  The choreography therefore gets the type
  $$
    \begin{array}{l}
      \LCForall{B}[\LCKindWithout{\LCProc}{\{\LCProcSeller,\LCProcSeller[2]\}}]{}\\
      \qquad \LCArr{\LCString{B}}{\{\LCProcSeller,\LCProcSeller[2]\}}{(\LCArr{(\LCArr{\LCNat[B]}{\LCEmptyProcSet}{\LCBool{B}})}{\{\LCProcSeller,\LCProcSeller[2]\}}{\LCUnit[B]})}
    \end{array}
    $$
This type shows both the input, output, and involved roles of the choreography.
\end{example}
\subsection{Operational Semantics}
\label{sec:semantics}

\begin{figure}
  \begin{mathpar}
    \inferrule*[left=AppTAbs]{\LCTypeT\equiv\LCTypeV}{
      \LCApp{(\LCFun{X}[\LCKindK]{M})}{\LCTypeT} \to_D \SingleSubst{M}{X}{\LCTypeV}
    }\and
    \inferrule*[left=MTApp1]{
      M_1 \to_D M_2
    }{
      \LCApp{M_1}{\LCTypeT} \to_D \LCApp{M_2}{\LCTypeT}
    }\\
    \inferrule*[left=Def]{ }{f \to_D D(f)}\\
    \inferrule*[left=Sel]{ }{\LCselect{\LCProcP}{\LCProcQ}{\ell}{M} \to_D M} \and
    \inferrule*[left=Com]{ }{
      \LCcom{\LCTypeT}{\LCProcP}{\LCProcQ}{V} \to_D \SingleSubst{V}{\LCProcP}{\LCProcQ}
    }
  \end{mathpar}
  \caption{Semantics of \chorlam (Selected Rules)}
  \label{fig:semantics}
\end{figure}

Finally, we consider the operational semantics of \chorlam.
These are mostly a standard call-by-value reduction semantics for a typed $\lambda$~calculus.
However, the reduction semantics must also carry a set~$D$ of function definitions.
Only a few rules are unusual or must be modified; those can be found in Figure~\ref{fig:semantics}.
\iftr You can find the rest of the rules in Appendix~\ref{sec:full-oper-sem}.\fi

The rules~\rulen{AppTAbs} and~\rulen{MTApp1} come from System~F$\omega$.
The rule~\rulen{AppTAbs} is similar to ordinary CBV $\beta$ reduction, but tells us how to reduce a \emph{type} abstraction applied to a \emph{type} value, but with the caveat that if we do not have a type value we must use type equivalence to get one before reducing.
The rule~\rulen{MTApp1} tells us that we can reduce a type function applied to any argument.

The rule~\rulen{Def} allows us to reduce function names by looking up their definition in the set~$D$.

Finally, we have the rules for communication.
The rule~\rulen{Sel} says that \LCKWselect acts as a no-op, as we stated earlier.
While this may seem redundant, such terms are vital for projection, as we will see in the next section.
More importantly, the \rulen{Com}~rule says tells us how we represent communication at the choreography level: via substitution of roles.
This also helps explain some of the restrictions in \rulen{Tcom}.
Since we replace \emph{all} mentions of \LCProcP with \LCProcQ in $V$, we cannot allow other mentions of \LCProcP in the type transformer of $V$.
Otherwise, there could be some mentions of $P$ which should not be replaced during communication, which we do not model. Unlike when typing $\LCcom{\LCTypeT}{\LCProcP}{\LCProcQ}{V}$, when executing a communication we know (since we only consider choreographies without free variables) that any type variables in $\LCTypeT$ or $V$ have already been instantiated and as such do we do not need to consider how to substitute variables which may later be instantiated to $\LCProcP$ or $\LCProcQ$.

It may be surprising to learn that our semantics are simply call-by-value reduction semantics, especially for those readers familiar with choreographies.
After all, choreographies are supposed to represent concurrent programs, and so multiple redices should be available at any time.
Indeed, previous works on choreographic programming~\citep[e.g.][]{HirschG22,Cruz-FilipeM17,CarboneM13} provided a semantics with \emph{out-of-order execution}, so that the operational semantics of the choreographies matched with all possible reductions in the concurrent interpretation.
We use these simpler semantics, without out-of-order execution, instead.
In exchange, our result in Section~\ref{sec:epp-correct} will be weaker: we only promise that any value which the choreography can reduce to, so can the concurrent interpretation.

To see why we chose to obtain this weaker result, consider the choreography
$$\LCapp{f}{\LCpair{(\LCcom{\LCTypefun{X}[\LCProc]{\LCNat[X]}}{\LCProcQ[1]}{\LCProcQ[2]}{(\LCLocalVal{3}[\LCProcQ[1]])})}{(\LCLocalVal{4}[\LCProcP])}}$$
Here we have a function $f$ which needs to be instantiated with a distributed pair. \LCProcP is ready to feed its part of the argument into $f$ and start computing the result, while \LCProcQ[1] and \LCProcQ[2] are still working on computing their part of the argument.
There are two ways we could interpret \chorlam concurrently: we can synchronize when all processes enter a function \emph{or} we can allow \LCProcP to enter the function early.
We take the second, more practical, route.
However, this means it is not possible to reflect at least one evaluation order into the semantics of the choreography without banning distributed values or allowing us to somehow call a single value in multiple steps.
This insight led to us adopting the weaker guarantee discussed above.

As is standard for call-by-value $\lambda$-calculi, we are able to show that our type system is \emph{sound} with respect to our operational semantics, as expressed in the following two theorems:
\begin{theorem}[Type Preservation]
  \label{thm:TypePres}
  Let $M$ be a choreography and $D$ a function mapping containing every function in $M$.
  If there exists a typing context $\Theta;\Gamma$ such that $\Theta;\Gamma\vdash M:\LCTypeT$ and $\Theta;\Gamma\vdash D$, then $\Theta;\Gamma\vdash M':\LCTypeT$ for any $M'$ such that $M\rightarrow_{D} M'$.
\end{theorem}
\iftr
\begin{proof}
  Follows from the typing and semantic rules and Theorem~\ref{thm:KindPres}.
\end{proof}
\fi
\begin{theorem}[Progress]\label{thm:Progress}
  Let $M$ be a closed choreography and $D$ a function mapping containing every function in $M$.
  If there exists a typing context $\Theta;\Gamma$ such that $\Theta;\Gamma\vdash M:\LCTypeT$ and $\Theta;\Gamma\vdash D$, then either $M=V$ or there exists $M'$ such that $M\rightarrow_{D} M'$.
\end{theorem}
\iftr
\begin{proof}
  Follows from the typing and semantic rules.
\end{proof}
\fi

\section{Endpoint Projection}
\label{sec:epp}
\newcommand{\procs}{\textcolor[cmyk]{0,.35,.96,.26}{\mathcal{P}}}
We now proceed to the most-important result for any choreographic programming language: \emph{endpoint projection}.
Endpoint projection gives a concurrent interpretation to our language~\chorlam by translating it to a parallel composition of programs, one for each process.
In order to define endpoint projection, though, we must define our process language, which we refer to as a \emph{local} language.
The syntax of the local language can be found in Figure~\ref{fig:netsyntax}.
There you can also find the syntax of local transition labels and network transition labels, both of which will be described when we describe the operational semantics of networks.

\begin{figure}
  \begin{syntax}
    \abstractCategory[Variables]{x,y,\ldots}
    \abstractCategory[Type Variables]{X,Y,\ldots}
    \abstractCategory[Process Names]{\LCProcP}
    \category[Local Transition Labels]{\mu}
      \alternative{\tau}
      \alternative{\LCProcP}
      \alternative{\LCsend{\LCProcP}{V~V'}}
      \alternative{\LCrecv{\LCProcP}{V'~V}}\\
      \alternative{\LCchoice{\LCProcP}{\ell}{}}
      \alternative{\LCoffr{\LCProcP}{\ell}}
    \category[Network Transition Labels]{\mu}
      \alternative{\tau_{\procs}}
    \category[Process labels]{\procs}
      \alternative{\LCProcP}
      \alternative{\LCProcP,\LCProcQ}
    \category[Types]{\LCTypeT}
      \alternative{\LCTypeV}
      \alternative{\LCapp{\LCTypeT[1]}{\LCTypeT[2]}}
      \alternative{\LCAmIType{\LCTypeV}{\LCTypeT[1]}{\LCTypeT[2]}}
      \alternative{\NWArr{\LCTypeT[1]}{\LCTypeT[2]}}\\
      \alternative{\LCSum{\LCTypeT[1]}{\LCTypeT[2]}}
      \alternative{\LCProd{\LCTypeT[1]}{\LCTypeT[2]}}
      \alternative{\LCForall{X}{\LCTypeT}}
      \alternative{\LCTypefun{X}{\LCTypeT}}
    \category[Type Values]{\LCTypeV}
      \alternative{X}
      \alternative{\LCUnit}
      \alternative{\LCNat}
      \alternative{\NWArr{\LCTypeV[1]}{\LCTypeV[2]}}
      \alternative{\LCProcP}
      \alternative{\LCBot}\\
      \alternative{\LCSum{\LCTypeV[1]}{\LCTypeV[2]}}
      \alternative{\LCProd{\LCTypeV[1]}{\LCTypeV[2]}}
      \alternative{\LCForall{X}{\LCTypeV}}
      \alternative{\LCTypefun{X}{\LCTypeV}}
    \category[Expressions]{M,N,\dots}
      \alternative{x}
      \alternative{\LCUnitVal}
      \alternative{n}
      \alternative{\LCfun{x}[\LCTypeT]{M}}
      \alternative{\LCFun{X}{M}}\\
      \alternative{\LCapp{M_1}{M_2}}
      \alternative{\LCApp{M}{\LCTypeT}}
      \alternative{\LCinl{\LCTypeT}{M}}
      \alternative{\LCinr{\LCTypeT}{M}}\\
      \alternative{\LCcase{M}{x}{M_1}{y}{M_2}}\\
      \alternative{\LCpair{M_1}{M_2}}
      \alternative{\LCfst{M}}
      \alternative{\LCsnd{M}}\\
      \alternative{\LCsendV{\LCTypeV}}
      \alternative{\LCrecvV{\LCTypeV}}\\
      \alternative{\LCoffr{\LCTypeV}{\{\ell_1 :M_1,\dots, \ell_n :M_n\}}}
      \alternative{\LCchoice{\LCTypeV}{\ell}{M}}\\
      \alternative{\rolesub{\LCTypeV[1]}{\LCTypeV[2]}}
      \alternative{f}
      \alternative{\LCAmI{\LCTypeV}{M_1}{M_2}}
    \category[Values]{V}
      \alternative{x}
      \alternative{\LCUnitVal}
      \alternative{n}
      \alternative{\LCBotVal}
      \alternative{\LCfun{x}[\LCTypeT]{M}}
      \alternative{\LCFun{X}{M}}\\
      \alternative{\LCinl{\LCTypeT}{V}}
      \alternative{\LCinr{\LCTypeT}{V}}
      \alternative{\LCpair{V_1}{V_2}}\\
      \alternative{\LCsendV{\LCTypeV}}
      \alternative{\LCrecvV{\LCTypeV}}
      \alternative{\rolesub{\LCTypeV[1]}{\LCTypeV[2]}}
  \end{syntax}
  
  \caption{Local Language Syntax}
  \label{fig:netsyntax}
\end{figure}

As in \chorlam, our local language inherits much of its structure from System~F$\omega$.
In particular, we have products, sums, functions, universal quantification, and $\LCTypelambda$ types, along with their corresponding terms.
In fact, some types look more like standard System~F$\omega$ than \chorlam:
function types do not need a set of processes which may participate in the function, and base types no longer need a location.

However, not everything is familiar; we have introduced new terms and new types.
The terms~$\LCsendV{\LCTypeV}$ and~$\LCrecvV{\LCTypeV}$ allow terms to send and receive values, respectively.
We also split \LCKWselect terms into two terms: an \emph{offer} term $\LCoffr{\LCTypeV}{\{\ell_1 :L_1,\dots, \ell_n :L_n\}}$ which allows $\LCTypeV$ to choose how this term will evolve.
We represent such choices using \emph{choice} terms of the form $\LCchoice{\LCTypeV}{\ell}{L}$.
This term informs the process represented by~\LCTypeV that it should reduce to its subterm labeled by~$\ell$, and then itself reduces to the term~$L$.
While these are unusual pieces of a polymorphic language like System~F$\omega$, they are familiar from process languages like $\pi$~calculus.
We also add undefined types and terms, written~\LCBot and~\LCBotVal, respectively.
These represent terms which are ill-defined; we use them to represent data which does not exist on some process~\LCProcP, but which needs to be written structurally in \LCProcP's program.

We also include a more-unusual feature: \emph{explicit substitutions} of processes.
The term~$\rolesub{\LCTypeV[1]}{\LCTypeV[2]}$ is a function which, when applied, replaces the role denoted by $\LCTypeV[1]$ with that denoted by $\LCTypeV[2]$ in its argument.
This function allows us to represent the view of communication according to third parties: the roles simply change, without any mechanism necessary.
For instance, imagine that \LCProcAlice wants to tell \LCProcBob to communicate an integer to \LCProcCathy.
She can do this by sending \LCProcBob the function~$\LCcomV{\LCTypefun{X}[\LCProc]{\LCNat[X]}}{\LCProcAlice}{\LCProcCathy}$.
In \chorlam, this corresponds to the choreography
$$\LCcom{\LCTypefun{X}[\LCProc]{\LCArr{\LCNat[X]}{\LCEmptyProcSet}{\LCNat[\LCProcCathy]}}}{\LCProcAlice}{\LCProcBob}{\left(\LCcomV{\LCTypefun{X}[\LCProc]{\LCNat[X]}}{\LCProcAlice}{\LCProcCathy}\right)}$$
In order to project this choreography, we need to be able to project the communication function above even when it is not applied to any arguments.
This is where we use explicit substitutions: we project the communication function to $\rolesub{\LCProcAlice}{\LCProcBob}$.

Finally, we introduce our unique feature: \LCKWAmI terms and their corresponding type.
These represent the ability of a process to know its own identity, and to take actions based on that knowledge. Process polymorphism requires an instantiation of a process variable at process $\LCProcP$ to be accompanied by a conditional determining whether the variable has been instantiated as $\LCProcP$ or as some other process $\LCProcP$ may interact with. 
In particular, the term~$\LCAmI{\LCTypeV}{M_1}{M_2}$ reduces to $M_1$ if the term is run by the process denoted by~\LCTypeV, and $M_2$ otherwise.
Since $M_1$ and $M_2$ may have different types, we provide types of the form~$\LCAmIType{\LCTypeV}{\LCTypeT[1]}{\LCTypeT[2]}$, which represent either the type $\LCTypeT[1]$ (if typing a term on the process denoted by~\LCTypeV) or $\LCTypeT[2]$ (otherwise).
These terms form a backbone of endpoint projection for \chorlam: every $\LCLambda$~term binding a process gets translated to include an \LCKWAmI term.
For instance, consider projecting the choreography $$\LCFun{X}[\LCProc]{\LCcom{\LCTypefun{X'}[\LCProc]{\LCNat[X']}}{\LCProcQ}{X}{\LCLocalVal{4}[\LCProcQ]}}$$
to some process \LCProcP.
Depending on the argument to which this function is applied, \LCProcP should behave very differently:
if it is applied to \LCProcP itself, it should receive something from \LCProcQ.
However, if it's applied to any other term, it should do nothing.
We therefore project the choreography above to the following program for \LCProcP:
$$\LCFun{X}{\LCAmI{X}{\LCrecv{\LCProcQ}{\LCBotVal}}{\LCBotVal}}$$
Note that the \LCKWAmI construct is necessary for process polymorphism in general, unless process variables cannot be instantiated to the process they are located at. It, and the combinatorial explosion caused by having multiple process abstractions, is not caused by the choreographic language but instead the choreographic language hides it and lets programmers avoid explicitly describing both sides of the \LCKWAmI separately.

Note that we do not have a kinding system for local programs.
In fact, we do not check the types of local programs at all.
However, because types have \emph{computational} content, we need to project them as well.
In order to preserve that computational content, we again use an equivalence of types which corresponds to $\beta,\eta$-equivalence.
However, in order to accommodate \LCKWAmIType~types, we must index that equivalence with a process.
Then, we have two rules regarding \LCKWAmIType~types:
\begin{mathpar}
  \inferrule*[left=IAm]
  { }{
    \NWEquiv{\LCProcP}{\LCAmIType{\LCProcP}{\LCTypeT[1]}{\LCTypeT[2]}}{\LCTypeT[1]}
  }\and
  \inferrule*[left=IAmNot]
  {
    \LCProcP \neq \LCProcQ
  }{
    \NWEquiv{\LCProcP}{\LCAmIType{\LCProcQ}{\LCTypeT[1]}{\LCTypeT[2]}}{\LCTypeT[2]}
  }
\end{mathpar}

Now that we have seen the syntax of the programs which run on each process, we can look at whole networks:
\begin{definition}
  A network $\mathcal{N}$ is a finite map from a set of processes to local programs.
  We often write $\LocalProg{\LCProcP[1]}{L_1} \mid \cdots \mid \LocalProg{\LCProcP[n]}{L_n}$ for the network where process $\LCProcP[i]$ has behaviour $L_i$.
\end{definition}

The parallel composition of two networks $\mathcal{N}$ and $\mathcal{N}'$ with disjoint domains,
$\mathcal{N} \mid \mathcal{N}'$, simply assigns to each process its behaviour in the network defining it.
Any network is equivalent to a parallel composition of networks with singleton domain, as suggested by the syntax above.

We now consider the operational semantics of local programs and networks.
These are given via labelled-transition systems; the syntax of both sorts of label can be found in Figure~\ref{fig:netsyntax}.
Selected rules for both operational semantics can be found in Figure~\ref{fig:NetSem}.
\iftr The full rules can be found in Appendix~\ref{sec:full-network-local-os}. \fi
As before, transitions are indexed by a set~$\DP$ of function definitions.
Function variables reduce by looking up their definition in $\DP$.
Since this transition involves no communication, it is labelled with the empty transition,~$\tau$.

\begin{figure}[h]
  \noindent\textbf{Local Language:}
  \begin{mathpar}
    \inferrule*[left=NDef]{}{
      f \xrightarrow{\tau}_\DP \DP(f)
    }\\
    \inferrule*[left=NBot]{}{
      \LCapp{\LCBotVal}{\LCBotVal} \xrightarrow{\tau}_\DP \LCBotVal
    }\and
    
    \inferrule*[left=NBott]{}{
      \LCApp{\LCBotVal}{\LCBot} \xrightarrow{\tau}_\DP \LCBotVal
    }\and
    \inferrule*[left=NSub]{}{
      \LCapp{\rolesub{\LCProcP}{\LCProcQ}}{V} \xrightarrow{\tau}_\DP \SingleSubst{V}{\LCProcP}{\LCProcQ}
    }\\
    \inferrule*[left=NAmIR]{}{
      \LCAmI{\LCProcP}{L_1}{L_2} \xrightarrow{\LCProcP}_{\DP} L_1
    }\and
    \inferrule*[left=NAmIL]{
      \LCProcQ\neq\LCProcP
    }{
      \LCAmI{\LCProcP}{L_1}{L_2} \xrightarrow{\LCProcQ}_{\DP} L_2
    }\\
    \inferrule*[left=NCho]{}{
      \LCchoice{\LCProcP}{\ell}{L}\xrightarrow{\LCchoice{\LCProcP}{\ell}{}}_{\DP} L
    }\and
    \inferrule*[left=Noff]{}{
      \LCoffr{\LCProcP}{\{\ell_1: L_1,\dots,\ell_n: L_n\}}\xrightarrow{\LCoffr{\LCProcP}{\ell_i}}_{\DP} L_i
    }\and
    \inferrule*[left=NSend]{}{
      \LCsend{\LCProcP}{V_1}\xrightarrow{\LCsend{\LCProcP}{V_1~V_2}}_{\DP} V_2
    } \and
    \inferrule*[left=NRecv]{}{
      \LCrecv{\LCProcP}{V_1}\xrightarrow{\LCrecv{\LCProcP}{V_2~V_1}}_{\DP} V_2
    }
  \end{mathpar}
  \noindent\textbf{Networks:}
  \begin{mathpar}
    \inferrule*[left=NCom]{
      L_1\xrightarrow{\LCsend{\LCProcP}{V~(\SingleSubst{V'}{\LCProcQ}{\LCProcP})}}_{\DP} L'_1\\
      L_2\xrightarrow{\LCrecv{\LCProcQ}{(\SingleSubst{V}{\LCProcQ}{\LCProcP})~V'}}_{\DP} L'_2
    }{
      \LCProcQ[] [L_1] \mid \LCProcP[] [L_2]\xrightarrow{\tau_{\LCProcQ,\LCProcP}}_{\DP} \LCProcQ[] [L'_1] \mid \LCProcP[] [L'_2]
    }\\
    \inferrule*[left=NSel]{
      L_1\xrightarrow{\LCchoice{\LCProcP}{\ell}{}}_{\DP} L'_1\\
      L_2\xrightarrow{\LCoffr{\LCProcQ}{\ell}}_{\DP} L'_2
    }{
      {\LCProcQ} [L_1] \mid {\LCProcP} [L_2]\xrightarrow{\tau_{\LCProcQ,\LCProcP}}_{\DP} {\LCProcQ} [L'_1] \mid {\LCProcP} [L'_2]
    }\\
    \inferrule*[left=NProam]{
      L \xrightarrow{\LCProcP}_\DP L'
    }{
      {\LCProcP} [L]\xrightarrow{\tau_{\LCProcP}}_{\DP} {\LCProcP} [L']
    }\and
    \inferrule*[left=NPro]{
      L \xrightarrow{\tau}_\DP L'
    }{
      \ProgOf{\LCProcP}{L} \xrightarrow{\tau_{\LCProcP}}_\DP \ProgOf{\LCProcP}{L'}
    }\and
    \inferrule*[left=NPar]{
      \mathcal{N}_1 \xrightarrow{\tau_{\procs}}_{\DP} \mathcal{N}_2
    }{
      \mathcal{N}_1 \mid \mathcal{N}' \xrightarrow{\tau_{\procs}}_{\DP} \mathcal{N}_2 \mid \mathcal{N}'
    }\\
  \end{mathpar}
  \caption{Semantics of Networks (Selected Rules)}
  \label{fig:NetSem}
\end{figure}

Perhaps surprisingly, undefined arguments to functions do not immediately cause the application to be undefined.
To see why, think about choreographies of the form $\LCapp{(\LCfun{x}[\LCNat[\LCProcP]]{M})}{N}$ where some process~\mbox{$\LCProcQ \neq \LCProcP$} is involved in both $M$ and $N$.
We project this to an application on $Q$ of the form $\LCapp{(\LCfun{x}[\LCBot]{\epp{M}{\LCProcQ}})}{\epp{N}{\LCProcQ}}$.
Note that because we know that $N$ has type $\LCNat[\LCProcP]$, the projection~$\epp{N}{\LCProcQ}$ has type $\LCBot$ and eventually evaluates to $\LCBotVal$.
Thus, if $\LCapp{(\LCfun{x}[\LCBot]{\epp{M}{\LCProcQ}})}{\LCBotVal}$ immediately evaluated to $\LCBotVal$, the process~\LCProcQ could not participate in $M$, as they need to do!
We therefore allow this to evaluate to $\epp{M}{\LCProcQ}$.
However, when the function is also undefined, we evaluate this to $\LCBotVal$ with the empty label $\tau$, as you can see in the rules~\rulen{NBot} and~\rulen{NBott}

As mentioned earlier, the explicit substitutions~$\rolesub{\LCProcP}{\LCProcQ}$ are functions which, when applied, perform the requested substitution in the value to which they are applied.
This is implemented in the rule~\rulen{NSub}.

The \LCKWAmI terms are given meaning via the rules~\rulen{NAmIR} and~\rulen{NAmIL}.
The rule~\rulen{NAmIR} says that the term $\LCAmI{\LCProcP}{L_1}{L_2}$ can evaluate to $L_1$ with label \LCProcP, while the rule~\rulen{NAmIL} says that it can instead reduce to $L_2$ with label~\LCProcQ where $\LCProcQ \neq \LCProcP$.
We will see later that in the network semantics, we only allow transitions labeled with the process performing the transition.

Choice and offer terms reduce via the rules~\rulen{NCho} and~\rulen{Noff}.
The first, \rulen{Ncho}, tells us that a choice term simply reduces to its continuation with a transition label indicating the choice that has been made.
The second, \rulen{Noff}, tells us that an offer term can reduce to \emph{any} continuation, with a transition label indicating the label of the continuation it reduced to.
We will see later that the semantics of networks only allows the offer term to reduce to the continuation chosen by a matching choice term.

Finally, the \LCKWsend and \LCKWrecv terms are given meaning via \rulen{NSend} and \rulen{NRecv}, respectively.
However, these rules behave somewhat-differently than might be expected: rather than acting as a plain send and receive, they behave more like a swap of information.

In a plain send, the sender would not have any information after the send---perhaps the term would come with a continuation, but this would not be related to the send.
Moreover, the receiver would not provide any information, but merely receive the information from the sender.
However, when sending a choreography with process polymorphism, the sender may need to participate in the continuation, depending on how polymorphic functions are applied.
For instance, consider the following choreography, where \LCProcP sends a polymorphic function to \LCProcQ, and the resulting polymorphic function is applied to \LCProcP:
$$\LCApp{(\LCcom{\LCTypefun{Y}[\LCProc]{\LCForall{X}[\LCProc]{\LCNat[X]}}}{\LCProcP}{\LCProcQ}{(\LCFun{X}[\LCProc]{\LCcom{\LCTypefun{Y'}[\LCProc]{\LCNat[Y']}}{\LCProcP}{X}{(\LCLocalVal{5}[\LCProcP])}})})}{\LCProcP}$$
The polymorphic function that results from the \LCKWcom above is as follows: $$\LCFun{X}[\LCProc]{\left(\LCcom{\LCTypefun{Y'}[\LCProc]{\LCNat[Y']}}{\LCProcQ}{X}{(\LCLocalVal{5}[\LCProcQ])}\right)}$$
Applying this to \LCProcP leads to a program where \LCProcP receives from \LCProcQ.
Since \LCProcP needs to participate in this program, \LCProcP must have a program remaining after sending the polymorphic function to \LCProcQ.

While this explains why \LCKWsend terms cannot simply, for instance, return unit, it does not explain why \LCKWsend and \LCKWrecv terms \emph{swap} results.
To see this, consider what happens when a term is sent from a process~\LCProcP to another process~\LCProcQ.
We know from our type system that \LCProcQ is not mentioned in the type of the term being sent, and we know that after the send all mentions of \LCProcP are changed to mentions of \LCProcQ.
Hence, after the send, \LCProcP's version of the term should be the view of a process not involved in the term.
This is exactly what \LCProcQ's version of the term is \emph{before} the send.
Thus, {\LCKWsend}s and {\LCKWrecv}s behaving as swaps leads to the correct behaviour.

Now that we have discussed the semantics of local programs, we discuss the semantics of networks.
Each transition in the network semantics has a silent label indexed with the processes participating in that reduction: $\tau_{\procs}$, where $\procs$ consists of either one process name (for local actions at that process) or two process names (for interactions involving these two processes). We treat $\procs$ as a set, implicitly allowing for exchange.

For instance, the rule~\rulen{NCom} describes communication.
Here, one local term must reduce with a \LCKWsend label, while another reduces with a \LCKWrecv label.
These labels must match, in the sense that the value received by the \LCKWrecv must be the value sent by the \LCKWsend---though with the receiver in place of the sender---and vice-versa.
Then, a network in which the local terms are associated with the appropriate processes, \LCProcQ and \LCProcP, can reduce with the label $\tau_{\LCProcQ, \LCProcP}$.
Similarly, the rules~\rulen{NSel} reduces matching choice and select terms, resulting in the label $\tau_{\LCProcQ, \LCProcP}$.

While \rulen{NCom} and \rulen{NSel} describe communication, the rest of the rules describe how a single process's term can evolve over time in a network.
Particularly interesting is \rulen{NProam}, which says that a \LCKWAmI~term can reduce only according to the process it is associated with.
We can see here that the resulting label is $\tau_{\LCProcP}$, indicating that this reduction step only involves \LCProcP.

The rules \rulen{NPro} tells us how to lift steps with an empty label $\tau$.
Such steps make no assumptions about the network, and so such terms can be associated with any process~\LCProcP.
When such a reduction takes place in a network, we label the resulting transition $\tau_{\LCProcP}$.

Finally, the rule~\rulen{NPar} says that if one part of a network can reduce with a label $\tau_{\procs}$, then the entire network can reduce with that same label.
This allows the other rules, which assume minimal networks, to be applied in larger networks. In the future we will use $\to^*$ and $\to^+$ to denote respectively a sequence and a sequence of at least one action with arbitrary labels.
\fabrizio{We should introduce here the notation for unlabelled reductions, see my comment on the completeness theorem for EPP.}

\subsection{Projection}
\label{sec:projection}

We can now define the endpoint projection (EPP) of choreographies.
This describes a single process's view of the choreography; the concurrent interpretation of a choreography is given by composing the projection to every process in parallel.
Endpoint projection to a particular process~\LCProcP is defined as a recursive function over typing derivations $\Theta; \Gamma \vdash M : \LCTypeT$.
For readability, however, we write it as a recursive function over the term $M$, and use the notation $\type(N)$ to refer to the types assigned to any term $N$ in the implicit typing derivation.
Similarly, we use $\kind(\LCTypeT)$ to refer to the kind of a type~\LCTypeT in the implicit typing derivation.
We write $\epp{M}{\LCProcP}$ to denote the projection of the term $M$ (implicitly a typing derivation for $M$, proving that it has \emph{some} type) to the process~\LCProcP.

Intuitively, projection translates a choreography term to its corresponding local behavior.
For example, a communication action projects to a send (for the sender), a receive (for the receiver), a substitution (for the other processes in the type of the value being communicated) or an empty process (for the remaining processes).
However, this is more complicated for \LCKWcase statements.
For instance, consider the following choreography, which matches on a sum type which is either an integer on \LCProcAlice or nothing on \LCProcAlice.
If it is an integer, then \LCProcBob receives that integer from Alice and returns it.
Otherwise, \LCProcBob returns the default value of $42$.
\LCProcAlice informs \LCProcBob of which branch she has taken using \LCKWselect terms.
\makeatletter
$$
  \left(
    \begin{array}[t]{l}
      \vspace{-3em}\\
    \LClambda\mkern3mu z \mathrel{\@LC@KeywordColor{:}} \LCSum{(\LCNat[\LCProcAlice])}{(\LCUnit[\LCProcAlice])}\mkern-1mu\@LC@KeywordColor{.}\\
      \mkern10mu\LCcase*{z}
      {x}{
      \LCselect{\LCProcAlice}{\LCProcBob}{\textsf{Just}}{
      (\LCcom{\LCTypefun{X}{\LCNat[X]}}{\LCProcAlice}{\LCProcBob}{x})
      }
      }
      {y}{
      \LCselect{\LCProcAlice}{\LCProcBob}{\textsf{Nothing}}{
      (\LCLocalVal{42}[\LCProcBob])
      }
      }
  \end{array}
  \right)~\LCinl{\LCUnit[\LCProcAlice]}{(\LCLocalVal{3}[\LCProcAlice])}
$$
\makeatother
Imagine projecting this to \LCProcBob's point of view.
He does not have any of the information in the sum, so he cannot participate in choosing which branch of the \LCKWcase expression gets evaluated.
Instead, he has to wait for \LCProcAlice to tell him which branch he is in.
If we na\"{i}vely translate just the first branch of the case expression, \LCProcBob waits for \LCProcAlice to send him the label~\textsf{Just} and then waits for \LCProcAlice to send him an integer.
Similarly, in the second branch \LCProcBob waits for \LCProcAlice to send him the label~\textsf{Nothing} before returning the default value~$42$.
Somehow, we need to combine these so that \LCProcBob waits for either label, and then takes the corresponding action.

We do this by \emph{merging} \LCProcBob's local programs for each branch~\citep{CarboneHY12,Cruz-FilipeM17,HondaYC16}.
Merging is a \emph{partial} function which combines two compatible local programs, combining choice statements.
In other words, the key property of merging is:
\begin{multline*}
\Merge{\LCoffr{\LCProcP}{\{\ell_i:B_i\}_{i\in I}}}{\LCoffr{\LCProcP}{\{\ell_j:B'_j\}_{j\in J}}} = \\ \LCoffr{\LCProcP}{\left(\{\Merge{\ell_k:B_k}{B'_k}\}_{k\in I\cap J} \cup \{\ell_i:B_i\}_{i\in I\setminus J}\cup \{\ell_j:B'_j\}_{j\in J\setminus I}\right)}
\end{multline*}
Merging is defined homomorphically on other terms, though it is undefined on incompatible terms.
Thus, for example, $\Merge{\LCinl{\LCTypeT}{M}}{\LCinl{\LCTypeT}{N}} = \LCinl{\LCTypeT}{(\Merge{M}{N})}$, but $\Merge{\LCinl{\LCTypeT[1]}{M}}{\LCinr{\LCTypeT[2]}{N}}$ is undefined.

We can then use this to project the choreography above to \LCProcBob as:
\makeatletter
$$
  (
    \LClambda\mkern3mu z \mathrel{\@LC@KeywordColor{:}} \LCBot\mkern-1mu\@LC@KeywordColor{.}\\
      \LCoffr{\LCProcAlice} 
      \{\textsf{Just}: (\LCrecv{\LCProcAlice}{\LCBotVal}, \textsf{Nothing}:42\}
  )~\LCBotVal
$$
\makeatother
Where $\LCBotVal$ represents a part of the choreography executed by $\LCProcAlice$.

\begin{definition}
  The EPP of a choreography $M$ for process $\LCProcP$ is defined by the rules in Figures~\ref{fig:proj-term}, \ref{fig:proj-term-ctd} and~\ref{fig:proj-typ}.

  To project a network from a choreography, we therefore project the choreography for each process and combine the results in parallel:
  $\epp M{}=\prod_{\LCProcP\in \roles(M)} \LCProcP\left[\epp M{\LCProcP}\right]$.
\end{definition}

\begin{figure}
  \begin{mathpar}
    \epp{x}{\LCProcP} = \left\{
      \begin{array}{ll}
        \LCBotVal & \text{if}~\epp{\type(x)}{\LCProcP} = \LCBot\\
        x & \text{otherwise}
      \end{array}
    \right.
    \and
    \epp{f}{\LCProcP} = f
    \\
    \epp{\LCUnitVal[\LCTypeV]}{\LCProcP} = \left\{
      \begin{array}{ll}
        \LCUnitVal & \text{if}~\epp{\LCTypeV}{\LCProcP} = \LCProcP\\
        \LCBotVal & \text{otherwise}
      \end{array}
    \right.
    \and
    \epp{\LCLocalVal{n}[\LCTypeV]}{\LCProcP} = \left\{
      \begin{array}{ll}
        \LCLocalVal{n} & \text{if}~\epp{\LCTypeV}{\LCProcP} = \LCProcP\\
        \LCBotVal & \text{otherwise}
      \end{array}
    \right.
    \\
    \epp{\LCfun{x}[\LCTypeT]{M}}{\LCProcP} = \left\{
      \begin{array}{ll}
        \LCBotVal & \begin{array}{l}\text{if}~$\epp{M}{\LCProcP} = \LCBotVal$\\\text{and}~$\epp{\LCTypeT}{\LCProcP} = \LCBot$\end{array}\\
        \LCfun{x}[\epp{\LCTypeT}{\LCProcP}]{\epp{M}{\LCProcP}} & \text{otherwise}
      \end{array}
    \right.
    \and
    \epp{\LCapp{M}{N}}{\LCProcP} = \left\{
      \begin{array}{ll}
        \LCBotVal & \text{if $\epp{M}{\LCProcP} = \epp{N}{\LCProcP} = \LCBotVal$}\\
        \LCapp{\epp{M}{\LCProcP}}{\epp{N}{\LCProcP}} & \begin{array}{l}\!\!\!\text{if $\LCProcP \in \roles(\type(M))$}\\ ~~\text{or $\epp{M}{\LCProcP} \neq \LCBotVal \neq \epp{N}{\LCProcP}$}\end{array}\\
        \epp{M}{\LCProcP} & \text{if $\epp{N}{\LCProcP} = \LCBotVal$}\\
        \epp{N}{\LCProcP} & \text{otherwise}
      \end{array}
    \right.
    \\
    \epp{\LCFun{X}[\LCKindK]{M}}{\LCProcP} = \left\{
      \begin{array}{ll}
        \LCFun{X}{\LCAmI*{X}{\epp{\SingleSubst{M}{X}{\LCProcP}}{\LCProcP}}{\epp{M}{\LCProcP}}} &
        \text{if}~\LCKindK \in \{\LCProc,\,\LCKindWithout{\LCProc}{\LCProcrho}\}\\
        \LCBotVal & \begin{array}{l}\text{if}~\epp{M}{\LCProcP} = \LCBotVal \\~~\text{and } \LCKindK = \LCKindWithout{\LCKindKPrime}{\LCProcSet{\LCProcP} \cup \LCProcrho}\end{array}\\
        \LCFun{X}{\epp{M}{\LCProcP}} & \text{otherwise}
      \end{array}
    \right.
    \and
    \epp{\LCApp{M}{\LCTypeT}}{\LCProcP} = \left\{
      \begin{array}{ll}
        \LCBotVal & \text{if $\epp M{\LCProcP}=\epp \LCTypeT{\LCProcP}=\LCBotVal$} \\ 
        \epp{M}{\LCProcP} & \begin{array}{l}\!\!\!\text{if $\epp \LCTypeT{\LCProcP}=\LCBotVal$}\\~~\text{and } \kind(\LCTypeT)=\LCKindWithout{\LCKindK}{(\LCProcSet{\LCProcP}\cup\LCProcrho)}\end{array} \\
        \epp{\LCTypeT}{\LCProcP} & \text{if $\epp M{\LCProcP}=\LCBotVal$} \\ 
        \LCApp{\epp{M}{\LCProcP}}{\epp{\LCTypeT}{\LCProcP}} & \text{otherwise}
      \end{array}
    \right.
    \\
    \begin{array}{l}
      \epp{\LCinl{\LCTypeT}{M}}{\LCProcP} \\\hspace{1em}{} = \left\{
      \begin{array}{ll}
        \LCBotVal & \begin{array}{l}\text{if}~\epp{M}{\LCProcP} = \LCBotVal~\text{and}\\ ~~\kind(\LCTypeT)=\LCKindWithout{\LCKindK}{\LCProcrho}\end{array}\\
        \epp{M_1}{\LCProcP} & \text{if} \epp{\type(M)}{\LCProcP}=\LCBot\\
        \LCinl{\epp{\LCTypeT}{\LCProcP}}{\epp{M_1}{\LCProcP}} & \text{otherwise}
      \end{array}
      \right.
    \end{array}\hspace{-1.7em}
    \and
    \begin{array}{l}
    \epp{\LCinr{\LCTypeT}{M}}{\LCProcP} \\\hspace{1em} {} = \left\{
      \begin{array}{ll}
        \LCBotVal & \begin{array}{l}\text{if}~\epp{M}{\LCProcP} = \LCBotVal~\text{and}\\
                      ~~\kind(\LCTypeT)=\LCKindWithout{\LCKindK}{\LCProcrho}\end{array}\\
        \epp{M_1}{\LCProcP} & \text{if} \epp{\type(M)}{\LCProcP}=\LCBot\\
        \LCinr{\epp{\LCTypeT}{\LCProcP}}{\epp{M_1}{\LCProcP}} & \text{otherwise}
      \end{array}
      \right.
      \end{array}
  \end{mathpar}
  \caption{Projection of \chorlam Programs}
  \label{fig:proj-term}
\end{figure}

\begin{figure}
  \begin{mathpar}
      \begin{array}{l}
      \epp{\LCcase{M}{x}{N_1}{y}{N_2}}{\LCProcP} =\\\hspace{2em}\left\{
      \begin{array}{ll}
        \LCcase{\epp{M}{\LCProcP}}{x}{\epp{N_1}{\LCProcP}}{y}{\epp{N_2}{\LCProcP}} & \text{if}~P \in \roles(\type(M))\\
        \LCBotVal & \text{if}~\epp{M}{\LCProcP} = \epp{N_1}{\LCProcP} = \epp{N_2}{\LCProcP} = \LCBotVal\\
        \epp{M}{\LCProcP} & \text{if}~\epp{N_1}{\LCProcP} = \epp{N_2}{\LCProcP} = \LCBotVal\\
        \Merge{\epp{N_1}{\LCProcP}}{\epp{N_2}{\LCProcP}} & \text{if}~\epp{M}{\LCProcP} = \LCBotVal\\
        \LCapp{\left(\LCfun{z}[\LCBot]{\left(\Merge{\epp{N_1}{\LCProcP}}{\epp{N_2}{\LCProcP}}\right)}\right)}{\epp{M}{\LCProcP}}~\textit{($z$ fresh)} & \text{otherwise}
      \end{array}
      \right.
    \end{array}
    \and
    \epp{(M_1,M_2)}{\LCProcP} = \left\{
      \begin{array}{ll}
        \LCBotVal & \text{if}~\epp{M_1}{\LCProcP} = \epp{M_2}{\LCProcP} = \LCBotVal\\
        (\epp{M_1}{\LCProcP}, \epp{M_2}{\LCProcP}) & \text{otherwise}
      \end{array}
    \right.
    \and
    \epp{\LCfst{M}}{\LCProcP} = \left\{
      \begin{array}{ll}
        \LCBotVal & \text{if}~\epp{M}{\LCProcP} = \LCBotVal\\
        \epp{M_1}{\LCProcP} & \text{if} \epp{\type(M)}{\LCProcP}=\LCBot\\
        \LCfst{\epp{M_1}{\LCProcP}} & \text{otherwise}
      \end{array}
    \right.
    \and\hspace{-2em}
    \epp{\LCsnd{M}}{\LCProcP}  = \left\{
      \begin{array}{ll}
        \LCBotVal & \text{if}~\epp{M}{\LCProcP} = \LCBotVal\\
        \epp{M_1}{\LCProcP} & \text{if} \epp{\type(M)}{\LCProcP}=\LCBot\\
        \LCsnd{\epp{M_1}{\LCProcP}} & \text{otherwise}
      \end{array}
    \right.
    \\
    \begin{array}{l}
      \epp{\LCselect{\LCProcQ[1]}{\LCProcQ[2]}{\ell}{M}}{\LCProcP} =\\[1em]\hspace{2em} \left\{
      \begin{array}{ll}
        \LCchoice{\LCProcQ'}{\ell}{\epp{M}{\LCProcP}} & \text{if}~\LCProcP = \LCProcQ[1] \neq \LCProcQ[2]\\
        \LCoffr{S}{\{\ell : \epp{M}{\LCProcP}\}} & \text{if}~\LCProcP=\LCProcQ[2] \neq \LCProcQ[1]\\
        \epp{M}{\LCProcP} & \text{otherwise}
      \end{array}
      \right.
    \end{array}
    \and
    \begin{array}{l}
      \epp{\LCcomV{\LCTypeT}{\LCProcQ[1]}{\LCProcQ[2]}}{\LCProcP} =\\[1em]\hspace{2em} \left\{
      \begin{array}{ll}
        \LCfun{x}[\LCapp{\LCTypeT}{\LCProcP}]{x} & \text{if}~\LCProcP = \LCProcQ[1] = \LCProcQ[2]\\
        \LCsendV{\LCProcQ[2]} & \text{if}~\LCProcP = \LCProcQ[1] \neq \LCProcQ[2]\\
        \LCrecvV{\LCProcQ[1]} & \text{if}~\LCProcP = \LCProcQ[2] \neq \LCProcQ[1]\\
        \rolesub{\LCProcQ[1]}{\LCProcQ[2]} & \text{if}~\epp{\LCTypeT}{\LCProcP} \neq \LCBot\\
        \LCBotVal & \text{otherwise}
      \end{array}
      \right.
    \end{array}
  \end{mathpar}
  \caption{Projection of \chorlam Programs (ctd.)}
  \label{fig:proj-term-ctd}
\end{figure}

Intuitively, projecting a choreography to a process that is not involved in it returns a $\LCBotVal$.
More complex choreographies, though, may involve processes that are not shown in their type.
This explains the first clause for projecting an application: even if $\LCProcP$ does not appear in the type of $M$, it may participate in interactions inside $M$.
A similar observation applies to the projection of $\LCKWcase$, where merging is also used.

Selections and communications follow the intuition given above, with one interesting detail:
self-selections are ignored, and self-communications project to the identity function.
This is different from many other choreography calculi, where self-communications are not allowed---we do not want to impose this in \chorlam, since we have process polymorphism and therefore do not want to place unnecessary restrictions on which processes a choreography can be instantiated with.

Any process~\LCProcP must prepare two behaviours for a process abstraction $\LCFun{X}[\LCProc]{M}$:
one for when $X$ is instantiated with \LCProcP itself, and one for when $X$ is instantiated with another process.
To do this, we use \LCKWAmI terms, which allow \LCProcP to use its knowledge of its identity to select which behaviour takes place.
(This also holds when the kind of $X$ is restricted, as long as the base kind is \LCProc, though if \LCProcP is excluded from the type of $X$ and \LCProcP does not participate in $M$ then we simply project to \LCBotVal.)
However, type abstractions $\LCFun{X}[\LCKindK]{M}$ do not use \LCKWAmI terms if \LCKindK is not a kind of processes, since \LCProcP cannot instantiate $X$.

In general, projecting a type yields $\LCBot$ at any process not used in that type. We use the restrictions on kinds to avoid projecting type variables and type abstractions when we know we do not need to and project all process names to themselves, but otherwise the projection of type constructs is similar to that of corresponding process terms.

Finally, to execute a projected choreography, we need to project the set of definitions of choreographic functions to a set of definitions of local functions.
Since these functions are all parametrised over every involved process, this is as simple as projecting the definitions onto an arbitrarily chosen process name.
$$\epp{D}{} = \{f \mapsto \epp{D(f)}{\LCProcP} \mid f\in\mathsf{domain}(D) \}\}$$
Note that function names always get projected everywhere.
This means that if we have a function which does not terminate when applied to some value in any process, then it diverges when applied to that value in the choreography and in every other process.
\begin{example}
We will now show how to project the bookseller service example from Section~\ref{sec:introduction}. As in that example we use $\LClet{x}{M~}{M'}$ as syntactic sugar for $\LCfun{x}[\LCTypeT]{M'}~M$ for some \LCTypeT and $\LCif{M_1}{M_2}{M_3}$ as syntactic sugar for $\LCcase{M_1}{x}{M_2}{x}{M_3}$ for some $x\notin(\fv(M_2)\cup\fv(M_3))$. We project for $\LCProcSeller$ and get the following process:

\NewDocumentCommand{\titlevar}{}{\textsf{title}}
\NewDocumentCommand{\pricelookup}{}{\textsf{price\_lookup}}
\NewDocumentCommand{\inbudget}{}{\textsf{buyAtPrice?}}
$$
\LCFun*{B}{\LCAmI*{B}{
  \LCfun*{\titlevar}{
    \LCfun*{\inbudget}
      {
        \LClet*{x}{
          \LCapp{(\LCfun{y}{y})}{\titlevar}
        }{
          \LClet*{y}{
            \LCapp{(\LCfun{z}{z})}{(\LCapp{\pricelookup}{x})}
          }{
            \LCif*{\LCapp{\inbudget}{y}}
            {{\LCUnitVal}}
            {{\LCUnitVal}}
          }
        }
    }
  }}
  {\LCfun*{\titlevar}{
    \LCfun*{\inbudget}
      {
        \LClet*{x}{
          \LCrecv{B}{\LCBotVal}
        }{
          \LClet*{y}{
            \LCsend{B}{(\LCapp{\pricelookup}{x})}
          }{\LCoffr{B}{\{{\textsf{Buy}}:{\LCUnitVal},{\textsf{Quit}}:{\LCUnitVal}\}}
          }
        }
    }
  }
  }
}
$$

Here we can see that if the buyer $B$ turns out to be \LCProcSeller itself, then all the communications become identity functions, and the seller does not inform itself of its choice. Otherwise, we get a function which, after being instantiated with a buyer, receives two $\LCBotVal$s representing values existing at $B$.
It then waits for $B$ to send a title, returns the price of this title, and waits for $B$ to decide whether to buy or not.
It might seem strange to have a function parametric on two $\LCBotVal$s, but this example actually illustrates why we cannot necessarily remove the first two $\lambda$s from the process without causing a deadlock.
Consider that $\LClet{y}{
  \LCsend{B}{(\LCapp{\pricelookup}{x})}
}{\LCoffr{B}{\{{\textsf{Buy}}:{\LCUnitVal},{\textsf{Quit}}:{\LCUnitVal}\}}
}$
is syntactic sugar for $(\LCfun{y}{\LCoffr{B}{\{{\textsf{Buy}}:{\LCUnitVal},{\textsf{Quit}}:{\LCUnitVal}\}}})~(\LCsend{B}{(\LCapp{\pricelookup}{x})})$.
Here we need to have the abstraction on $y$ even though it gets instantiated as $\LCBotVal$ after \LCProcSeller sends the result of $\LCapp{\pricelookup}{x}$ to $B$.
If instead we only had $({\LCoffr{B}{\{{\textsf{Buy}}:{\LCUnitVal},{\textsf{Quit}}:{\LCUnitVal}\}}})~(\LCsend{B}{(\LCapp{\pricelookup}{x})}),$
the first part of the application would not be a value, and would be waiting for $B$ to choose between $\textsf{Buy}$ and $\textsf{Quit}$ while $B$ has the abstraction on $y$ and therefore considers the first part of the application a function which must wait to be instantiated.
$B$ therefore expects to receive the result of $\LCapp{\pricelookup}{x}$, and we get a deadlock in our system.
This is why we never want to project a value to a non-value term, and need to keep any abstractions guarding a part of the choreography involving \LCProcSeller.         
\end{example}

\begin{figure}
  \begin{mathpar}
    \epp{X}{\LCProcP} = \left\{
      \begin{array}{ll}
        \LCBotVal & \text{if}~\kind(X) = \LCKindWithout{\LCKindK}{(\LCProcSet{\LCProcP} \cup \LCProcrho)} \mathrel{\text{for}} \LCKindK \neq \LCProc\\
        X & \text{otherwise}
      \end{array}
    \right.
    \and
    \epp{\LCProcQ}{\LCProcP} = \LCProcQ
    \and
    \epp{\LCUnit[\LCProcQ]}{\LCProcP} = \left\{
      \begin{array}{ll}
        \LCUnit & \text{if}~\LCProcP = \LCProcQ\\
        \LCBot & \text{otherwise}
      \end{array}
    \right.
    \and
    \epp{\LCNat[\LCProcQ]}{\LCProcP} = \left\{
      \begin{array}{ll}
        \LCNat & \text{if}~\LCProcP = \LCProcQ\\
        \LCBot & \text{otherwise}
      \end{array}
    \right.
    \\
    \epp{\LCProd{\LCTypeT[1]}{\LCTypeT[2]}}{\LCProcP} = \left\{
      \begin{array}{ll}
      \LCBot & \text{if}~\epp{\LCTypeT[1]}{\LCProcP} = \epp{\LCTypeT[2]}{\LCProcP} = \LCBot\\
      \LCProd{\epp{\LCTypeT[1]}{\LCProcP}}{\epp{\LCTypeT[2]}{\LCProcP}} & \text{otherwise}
      \end{array}
    \right.
    \and
    \epp{\LCSum{\LCTypeT[1]}{\LCTypeT[2]}}{\LCProcP} = \left\{
      \begin{array}{ll}
        \LCBot & \text{if}~\epp{\LCTypeT[1]}{\LCProcP} = \epp{\LCTypeT[2]}{\LCProcP} = \LCBot\\
        \LCSum{\epp{\LCTypeT[1]}{\LCProcP}}{\epp{\LCTypeT[2]}{\LCProcP}} & \text{otherwise}
      \end{array}
    \right.
    \and 
    \epp{\LCArr{\LCTypeT[1]}{\LCProcrho}{\LCTypeT[2]}}{\LCProcP} = \left\{
      \begin{array}{ll}
        \NWArr{\epp{\LCTypeT[1]}{\LCProcP}}{\epp{\LCTypeT[2]}{\LCProcP}} & \text{if}~\LCProcP \in \LCProcrho \mathrel{\text{or}} \epp{\LCTypeT[1]}{\LCProcP} \neq \LCBot \neq \epp{\LCTypeT[2]}{\LCProcP}\\
        \LCBot & \text{otherwise}
      \end{array}
    \right.
    \and
    \begin{array}{l}
    \epp{\LCForall{X}[\LCKindK]{\LCTypeT}}{\LCProcP} =\\[1em]\hspace{2em} \left\{
      \begin{array}{ll}
        \LCBot & \text{if}~\epp{\LCTypeT}{\LCProcP} = \LCBot \mathrel{\text{and}} \LCKindK = \LCKindWithout{\LCKindKPrime}{(\LCProcSet{\LCProcP} \cup \LCProcrho)}\\
        \LCForall{X}{\LCAmIType{X}{\epp{\SingleSubst{\LCTypeT}{X}{\LCProcP}}{\LCProcP}}{\epp{\LCTypeT}{\LCProcP}}} & \text{if}~\LCKindK \in \{ \LCProc,\, \LCKindWithout{\LCProc}{\LCProcrho}\}\\
        \LCForall{X}{\epp{\LCTypeT}{\LCProcP}} & \text{otherwise}
      \end{array}
    \right.
    \end{array}
    \and
    \epp{\LCapp{\LCTypeT[1]}{\LCTypeT[2]}}{\LCProcP} = \left\{
      \begin{array}{ll}
        \LCBot & \text{if}~\epp{\LCTypeT[1]}{\LCProcP} = \epp{\LCTypeT[2]}{\LCProcP} = \LCBot\\
        \epp{\LCTypeT[1]}{\LCProcP} & \text{if}~\epp{\LCTypeT[2]}{\LCProcP} = \LCBot \mathrel{\text{and}} \kind(\LCTypeT[2]) = \LCKindWithout{\LCKindK}{(\LCProcSet{\LCProcP}\cup{\LCProcrho})}\\
        \epp{\LCTypeT[2]}{\LCProcP} & \text{if}~\epp{\LCTypeT[1]}{\LCProcP} = \LCBot\\
        \LCapp{\epp{\LCTypeT[1]}{\LCProcP}}{\epp{\LCTypeT[2]}{\LCProcP}} & \text{otherwise}
      \end{array}
    \right.
    \and
    \begin{array}{l}
    \epp{\LCTypefun{X}[\LCKindK]{\LCTypeT}}{\LCProcP} =\\[1em]\hspace{2em} \left\{
      \begin{array}{ll}
        \LCBot & \text{if}~\epp{\LCTypeT}{\LCProcP} = \LCBot \mathrel{\text{and}} \LCKindK = \LCKindWithout{\LCKindKPrime}{(\LCProcSet{\LCProcP} \cup \LCProcrho)}\\
        \LCTypefun{X}{\LCAmIType{X}{\epp{\SingleSubst{{\LCTypeT}}{X}{\LCProcP}}{\LCProcP}}{\epp{\LCTypeT}{\LCProcP}}} & \text{if}~\LCKindK \in \{\LCProc,\,\LCKindWithout{\LCProc}{\LCProcrho}\}\\
        \LCTypefun{X}{\epp{\LCTypeT}{\LCProcP}} & \text{otherwise}
      \end{array}
    \right.
    \end{array}
  \end{mathpar}
  \caption{Projection of \chorlam Types}
  \label{fig:proj-typ}
\end{figure}

\iftr\section{The Correctness of Endpoint Projection}\fi
\label{sec:epp-correct}
We now show that there is a close correspondence between the executions of choreographies and of
their projections.
Intuitively, this correspondence states that a choreography can execute an action if, and only if, its projection can execute the same action, and both transition to new terms in the same relation.
However, this is not completely true: if a choreography $M$ reduces by rule \rulen{CaseL}, then the result has fewer branches than the network obtained by performing the corresponding reduction in the projection of $C$.

In order to capture this we revert to the branching relation~\citep{Montesi22,Cruz-FilipeM17}, defined by $M \sqsupseteq N$ iff $\Merge{M}{N}=M$.
Intuitively, if $M\sqsupseteq N$, then $M$ offers the same and possibly more behaviours than $N$.
This notion extends to networks by defining $\mathcal{N}\sqsupseteq\mathcal{N}'$ to mean that, for any role $\LCProcP$, $\mathcal{N}(\LCProcP)\sqsupseteq \mathcal{N}'(\LCProcP)$.

Using this, we can show that the EPP of a choreography can do all that (completeness) and only what (soundness) the original choreography does. For traditional imperative choreographic languages, this correspondence takes the form of one action in the choreography corresponding to one action in the projected network. We instead have a correspondence between one action in the choreography and multiple actions in the network due to allowing choreographies to manipulate distributed values in one action such as in $\LCApp{\LCfun{x}[\LCProd{\LCNat[\LCProcBob]}{\LCNat[\LCProcAlice]}]{M}}{\LCpair{\LCLocalVal{3}[\LCProcBob]}{\LCLocalVal{3}[\LCProcAlice]}}$ where both $\LCProcBob$ and $\LCProcAlice$ independently take the first part of the pair.
\begin{theorem}[Completeness]\label{thm:ChorToNet}
  Given a closed choreography $M$, if $M\rightarrow_{D} M'$, $\Theta;\Gamma\vdash D$, $\Theta;\Gamma\vdash M: \LCTypeT$, and $\epp{M}{}$ is defined, then
  there exist networks $\mathcal{N}$ and $M''$ such that:
  $\epp{M}{}\rightarrow^+_{\epp{D}{}}\mathcal{N}$ and
  $\mathcal{N}\sqsupseteq\epp{M'}{}$.\fabrizio{We should introduce $\to_D$, $\to_d$, and their zero-or-more and one-or-more extensions ($*$ and $+$). Something on the lines of ``we simply write $\to$ when the label is not important''}
\end{theorem}
\iftr
  In the foregoing  we use $L$ to denote local expressions and $U$ to denote local values in order to make the proofs more readable, as we will be switching back and forth between layers a lot.

Before we can prove completeness, we need a few lemmas. First, we show that choreographic values always project to local values.
\begin{lemma}\label{thm:Induction}
For any choreographic value $V$ and process $\LCProcP$, if $\Theta;\Gamma\vdash V:\LCTypeT$ then $\llbracket V\rrbracket_{\LCProcP}$ is either a value or undefined.
\end{lemma}
\begin{proof}
Straightforward from the projection rules.
\end{proof}

We then prove the same for type values.
\begin{lemma}\label{thm:InductionT}
Given a type value $\LCTypeV[1]$, if $\Theta;\Gamma\vdash \LCTypeV[1]::\LCKindK$ then for any process $\LCProcP$ in $\roles(\LCTypeV[1])$, $\llbracket \LCTypeV[2]\rrbracket_{\LCProcP}=\LCTypeV[2]$.
\end{lemma}
\begin{proof}
Straightforward from the projection rules.
\end{proof}

We then show that type values are projected to \LCBot at uninvolved processes.
\begin{lemma}\label{thm:TypeUnit}
Given a type value $\LCTypeV\neq\LCProcP$, for any process $\LCProcQ\notin \roles(\LCTypeV)$, $\llbracket \LCTypeV\rrbracket_{\LCProcQ}=\LCBot$.
\end{lemma}
\begin{proof}
Straightforward from induction on $\LCTypeV$.
\end{proof}

And similarly, we show that choreographic values project to $\LCBotVal$ at processes not involved in their type.
\begin{lemma}\label{thm:ValUnit}
Given a value $V$, if $\Theta;\Gamma\vdash V:\LCTypeT$ then for any process $\LCProcP\notin \roles(\LCTypeT)$, we have $\llbracket V\rrbracket_{\LCProcP}=\LCBotVal$ or $\llbracket V\rrbracket_{\LCProcP}$ is undefined.
\end{lemma}
\begin{proof}
Follows from Lemmas~\ref{thm:Induction} and~\ref{thm:TypeUnit} and the projection rules.
\end{proof}

Finally, we show that equivalent types are projected to equivalent local types.
\begin{lemma}\label{thm:TypeComp}
Given a closed type $\LCTypeT$, if $\LCTypeT\equiv \LCTypeT'$ and $\Theta;\Gamma\vdash \LCTypeT::\LCKindK$, then
  for any process $\LCProcP$, $\epp{\LCTypeT}{\LCProcP}\equiv_{\LCProcP}\epp{\LCTypeT'}{\LCProcP}$.
\end{lemma}
\begin{proof}
We prove this by structural induction on $\LCTypeT\equiv \LCTypeT'$. All but one case follow by simple induction.

The one interesting case is if $T=\LCTypefun{X}{\LCKindK}{\LCTypeT[1]}~\LCTypeV$ and $\LCTypeT'=\LCTypeT[1][X:=\LCTypeV]$. Then (1) if $\LCKindK=\LCKindWithout{\LCKindK'}{(\{R\}\cup\LCProcrho)}$ and $\epp{T[1]}{\LCProcP}=\LCBotVal$, we have $\epp{T}{\LCProcP}=\epp{T'}{\LCProcP}=\LCBot$. (2) If $\LCKindK\in\{\LCProc,\LCKindWithout{\LCProc}{\LCProcrho}\}$ then $\epp{T}{\LCProcP}=(\LCForall{X}{\LCAmI{X}{\llbracket {\LCTypeT[1]} [X:={\LCProcP}]\rrbracket_{\LCProcP}}{\llbracket \LCTypeT[1] \rrbracket_{\LCProcP}}})~\epp{\LCTypeV}{\LCProcP}$ and $\epp{\LCTypeT'}{\LCProcP}=\epp{\LCTypeT[1][X:=\LCTypeV]}{\LCProcP}$. Since $T$ is a closed type and $\Theta\Gamma\vdash \LCTypeV::\LCKindK$, $\LCTypeV$ must be a process. If $\LCTypeV=\LCProcP$ then obviously $(\LCForall{X}{\LCAmI{X}{\llbracket {\LCTypeT[1]} [X:={\LCProcP}]\rrbracket_{\LCProcP}}{\llbracket \LCTypeT[1] \rrbracket_{\LCProcP}}})~\epp{\LCTypeV}{\LCProcP}\equiv_{\LCProcP} \epp{\LCTypeT[1][X:=\LCTypeV]}{\LCProcP}$. If $\LCTypeV\neq\LCProcP$ then $(\LCForall{X}{\LCAmI{X}{\llbracket {\LCTypeT[1]} [X:={\LCProcP}]\rrbracket_{\LCProcP}}{\llbracket \LCTypeT[1] \rrbracket_{\LCProcP}}})~\epp{\LCTypeV}{\LCProcP}\equiv_{\LCProcP} \epp{\LCTypeT[1]}{\LCProcP}[X:=\epp{\LCTypeV}{\LCProcP}]$, but since $\LCTypeV$ is a process $\LCProcQ$, $\epp{\LCTypeV}{\LCProcP}=\LCProcQ$ and $\epp{X}{\LCProcP}=X$, and therefore we get $\epp{\LCTypeT}{\LCProcP}\equiv_{\LCProcP}\epp{\LCTypeT'}{\LCProcP}$. 
And (3) otherwise we have $\epp{\LCTypeT}=\LCTypefun{X}{\epp{\LCTypeT[1]}{\LCProcP}}~\epp{\LCTypeV}{\LCProcP}$ and $\llbracket \LCTypeT'\rrbracket_{\LCProcP}=\llbracket \LCTypeT[1]\rrbracket_{\LCProcP}[X:=\llbracket \LCTypeV\rrbracket_{\LCProcP}]$. We therefore get $\llbracket \LCTypeT\rrbracket_{\LCProcP}\equiv_{\LCProcP}\llbracket \LCTypeT'\rrbracket_{\LCProcP}$.
\end{proof}

We are now ready to prove completeness.
\begin{proof}[Proof of Theorem~\ref{thm:ChorToNet}]
We prove this by structural induction on $M\rightarrow_{D} M'$.
\begin{itemize}
\item Assume $M=\LCfun{x}[\LCTypeT]{N}~V$ and $M'=N[x:=V]$. Then for any process ${\LCProcP}$ such that $\epp{N}{\LCProcP}\neq\LCBotVal$ or $\epp{\LCTypeT}{\LCProcP}\neq\LCBot$, we have $\llbracket M\rrbracket_{\LCProcP}=(\LCfun{ x}[\epp{\LCTypeT}{{\LCProcP}}]{\llbracket N\rrbracket_{\LCProcP}})~\llbracket V\rrbracket_{\LCProcP}$ and $\llbracket M'\rrbracket_{\LCProcP}=\llbracket N\rrbracket_{\LCProcP}[x:=\llbracket V\rrbracket_{\LCProcP}]$, and for any other ${\LCProcP}'$, we have ${\LCProcP}'\notin \roles(\type(V))$ and therefore $\llbracket M\rrbracket_{{\LCProcP}'}=\epp {M'}{{\LCProcP}'}=\LCBotVal$. We therefore get ${\LCProcP}[\llbracket M\rrbracket_{\LCProcP}]\xrightarrow{\tau}_{\epp{D}{}} \llbracket M'\rrbracket_{\LCProcP}$ for all ${\LCProcP}\in\roles(\type(\LCfun{x}[\LCTypeT]{N})$ and define $\mathcal{N}=\prod\limits_{{\LCProcP}\in\roles(\type(\LCfun{x}[\LCTypeT]{N}))}{\LCProcP}[\llbracket M'\rrbracket_{\LCProcP}]\mid\prod\limits_{{\LCProcP}'\notin\roles(\type((\LCfun{x}[\LCTypeT]{N}))} {\LCProcP}'[\LCBotVal]$ and the result follows.

\item Assume $M=\LCFun{X}[\LCKindK]N~\LCTypeT$, $\LCTypeT\equiv\LCTypeV$, and $M'=N[X:=\LCTypeV]$.
  Then if $K\in\{\LCProcP,\,\LCKindWithout{\LCProcP}{\LCProcrho}\}$, for any process ${\LCProcP}$, $\epp M {\LCProcP}=\LCFun{X}[\LCKindK]{\LCAmI{X}{\epp{N[t:={\LCProcP}]}{{\LCProcP}}}{\epp{N}{{\LCProcP}}}}~\epp{\LCTypeT}{{\LCProcP}}$ and the result follows  Lemmas~\ref{thm:InductionT} and~\ref{thm:TypeComp}, and Rules \rulen{NBabs}, \rulen{NIamr}, and \rulen{NIaml}.
  If $\LCKindK\notin\{\LCProc,\LCKindWithout{\LCProc}{\LCProcrho}\}$ then for any process ${\LCProcP}$ such that $\epp{N}{{\LCProcP}}=\LCBotVal$ and $\LCKindK=\LCKindWithout{\LCKindK'}{(\{{\LCProcP}\}\cup\LCProcrho)}$, we have $\epp{M}{{\LCProcP}}=\epp{M'}{{\LCProcP}}=\LCBot$, for any other ${\LCProcP}'$ 
 we have $\llbracket M\rrbracket_{{\LCProcP}'}=(\LCFun{X}{\llbracket N\rrbracket_{{\LCProcP}'})}~\llbracket \LCTypeT\rrbracket_{{\LCProcP}'}$ and $\llbracket M'\rrbracket_{{\LCProcP}'}=\llbracket N\rrbracket_{{\LCProcP}'}[X:=\llbracket \LCTypeV\rrbracket_{{\LCProcP}'}]$. We therefore get ${\LCProcP}[\llbracket M\rrbracket_{\LCProcP}]\rightarrow^*_{\epp{D}{}} \llbracket M'\rrbracket_{\LCProcP}$ for all ${\LCProcP}$ and the result follows.

\item Assume $M=N~M''$, $M'=N'~M''$, and $N\rightarrow_{D} N'$.
Then for any process ${\LCProcP}$ such that $\epp N{{\LCProcP}}=\epp{M''}{{\LCProcP}}=\LCBotVal$, by induction we have $\epp{N'}{{\LCProcP}}=\LCBotVal$, and therefore $\epp M{{\LCProcP}}=\epp{M'}{{\LCProcP}}=\LCBotVal$.
For any process ${\LCProcP}'$ such that ${\LCProcP}'\in \roles(\type(N))$ or $\epp N{{\LCProcP}'} \neq \LCBotVal \neq \epp{M''}{{\LCProcP}'}$, $\llbracket M\rrbracket_{{\LCProcP}'}= \llbracket N\rrbracket_{{\LCProcP}'}~\llbracket M''\rrbracket_{{\LCProcP}'}$ and $\llbracket M'\rrbracket_{{\LCProcP}'}=\llbracket N'\rrbracket_{{\LCProcP}'}~\llbracket M''\rrbracket_{{\LCProcP}'}$.  
For any other process ${\LCProcP}''$ such that $\epp{N}{{\LCProcP}''}=\LCBotVal$, by induction we get $\epp{N'}{{\LCProcP}''}=\LCBotVal$ and therefore $\epp M{{\LCProcP}''}=\epp{M'}{{\LCProcP}''}=\epp{M''}{{\LCProcP}''}$. 
For any other process ${\LCProcP}'''$ such that $\epp{M''}{{\LCProcP}'''}=\LCBotVal$, we get $\epp{M}{{\LCProcP}'''}=\epp{N}{{\LCProcP}'''}$ and $\epp{M'}{{\LCProcP}'''}=\epp{N'}{{\LCProcP}'''}$. 
And by induction $\llbracket N\rrbracket\rightarrow^*_{\epp{D}{}} \mathcal{N}_N$ $\mathcal{N}_N\sqsupseteq\epp{N'}{}$. 
For any process ${{\LCProcP}}$ we therefore get $\llbracket N\rrbracket_{\LCProcP}\xrightarrow{\mu_0}_{\epp{D}{}}\xrightarrow{\mu_1}_{\epp{D}{}}\dots L_{\LCProcP}$ for $L_{\LCProcP}\sqsupseteq \epp{N'}{{\LCProcP}}$ for some sequences of transitions $\xrightarrow{\mu_0}_{\epp{D}{}}\xrightarrow{\mu_1}_{\epp{D}{}}\dots$, and from the network semantics we get 
\[\llbracket M\rrbracket \rightarrow^* \begin{array}{l}
\prod\limits_{\epp N{{\LCProcP}}=\epp{M''}{{\LCProcP}}=\LCBotVal} {\LCProcP}[\LCBotVal] \mid \prod\limits_{{\LCProcP}'\in \roles(\type(N)) \text{ or } \epp N{{\LCProcP}'} \neq \LCBotVal \neq \epp{M''}{{\LCProcP}'}}{\LCProcP}'[L_{{\LCProcP}'}~\llbracket M''\rrbracket_{{\LCProcP}'}] \\ \mid \prod\limits_{\epp{M}{{\LCProcP}''}=\epp{M''}{{\LCProcP}''}} {\LCProcP}''[\epp{M''}{{\LCProcP}''}] \mid \prod\limits_{\epp{M}{{\LCProcP}'''}=\epp{N}{{\LCProcP}'''}} {\LCProcP}''[L_{{\LCProcP}''}] \\
\end{array}=\mathcal{N}\] and 
$  M' \rightarrow_{D} N'~M''$.
And since $\llbracket N\rrbracket\rightarrow^*_{\epp{D}{}} \mathcal{N}'$ and $\llbracket N'\rrbracket\rightarrow^*_{\epp{D}{}} \mathcal{N}'_N$, we know these sequences of transitions can synchronise when necessary, and if $\epp N{{\LCProcP}''''}\neq  \epp{N'}{{\LCProcP}''''}=\LCBotVal$ then we can do the extra application to get rid of this unit.

\item Assume $M=V~N$, $M'=V~N'$, and $N\rightarrow_{D} N'$. This is similar to the previous case, using Lemma~\ref{thm:Induction} to ensure every process can use Rule \rulen{NApp2}.

\item Assume $M=\LCcase{N}{x}{N'}{x'}{N''}$, $M'=\LCcase{M''}{x}{N'}{x}{N''}$, and $N\rightarrow_{D} M''$. Then for any process ${{\LCProcP}}$ such that ${\LCProcP}\in\roles(\type(N))$, we have $\llbracket M\rrbracket_{\LCProcP}=\LCcase{\llbracket N\rrbracket_{\LCProcP}}{x}{\llbracket N'\rrbracket_{\LCProcP}}{x'}{\llbracket N''\rrbracket_{\LCProcP}}$ and $\llbracket M'\rrbracket_{\LCProcP}=\LCcase{\llbracket M''\rrbracket_{\LCProcP}}{x}{\llbracket N'\rrbracket_{\LCProcP}}{x'}{\llbracket N''\rrbracket_{\LCProcP}}$. For any other process ${\LCProcP}'$ such that $\epp N{{\LCProcP}'}=\epp{N'}{{\LCProcP}'}=\epp{N''}{{\LCProcP}'}=\LCBotVal$, by induction we get $\epp{M''}{{\LCProcP}'}=\LCBotVal$, and therefore $\epp{M}{{\LCProcP}'}=\epp{M'}{{\LCProcP}'}=\LCBotVal$. For any other process ${\LCProcP}''$ such that $\epp{N}{{\LCProcP}''}=\LCBotVal$, we get $\epp{M}{{\LCProcP}''}=\epp{M'}{{\LCProcP}''}=\epp{N'}{{\LCProcP}''}\sqcup \epp{N''}{{\LCProcP}''}$. For any other processes ${\LCProcP}'''$ such that $\epp{N'}{{\LCProcP}'''}=\epp{N''}{{\LCProcP}'''}=\LCBotVal$, we have $\epp{M}{{\LCProcP}'''}=\epp{N}{{\LCProcP}'''}$ and $\epp{M'}{{\LCProcP}'''}=\epp{M''}{{\LCProcP}'''}$. For any other process ${\LCProcP}''''$, we have $\llbracket M\rrbracket_{{\LCProcP}''''}=(\lambda x:\LCBot.\llbracket N'\rrbracket_{{\LCProcP}''''}\sqcup\llbracket N''\rrbracket_{{\LCProcP}''''})~\llbracket N\rrbracket_{{\LCProcP}''''}$ and $\llbracket M'\rrbracket_{{\LCProcP}''''}=(\lambda x.\llbracket N'\rrbracket_{{\LCProcP}''''}\sqcup\llbracket N''\rrbracket_{{\LCProcP}''''})~\llbracket M''\rrbracket_{{\LCProcP}''''}$ for $x\notin \fv(N')\cup \fv(N'')$. The rest follows by simple induction similar to the second case.

\item Assume $M=\LCcase{\LCinl{\LCTypeT}{V}}{x}{N}{x'}{N'}$ and $M'=N[x:= V]$. Then for any process ${\LCProcP}\in\roles(\type(\LCinl{\LCTypeT}{V}))$, we have $\llbracket M\rrbracket_{\LCProcP}=\LCcase{\LCinl{\epp{\LCTypeT}{\LCProcP}}{\llbracket V\rrbracket_{\LCProcP}}}{x}{\llbracket N\rrbracket_{\LCProcP}}{x'}{\llbracket N'\rrbracket_{\LCProcP}}$ and $\llbracket M'\rrbracket_{\LCProcP}=\llbracket N[x:=\llbracket V\rrbracket_{\LCProcP}]\rrbracket_{\LCProcP}$. By Lemma~\ref{thm:ValUnit}, $\llbracket N[x:=\llbracket V\rrbracket_{\LCProcP}]\rrbracket_{\LCProcP}=\llbracket N\rrbracket_{\LCProcP}[x:=\llbracket V\rrbracket_{\LCProcP}]$. For any other process ${\LCProcP}'\notin\roles(\type(\LCinl{\LCTypeT}{V}))$, $\epp{\LCinl{V}{\LCTypeT}{{\LCProcP}'}}=\LCBotVal$, and therefore $\epp{M}{{\LCProcP}'}=\epp{N}{{\LCProcP}'}\sqcup \epp{N'}{{\LCProcP}'}\sqsupseteq \epp{N}{{\LCProcP}'}=\epp{M'}{{\LCProcP}'}$. The result follows.

\item Assume $M=\LCcase{\LCinr{\LCTypeT}{V}}{x}{N}{x'}{N'}$ and $M'=N'[x':= V]$. This case is similar to the previous.

\item Assume $M=\LCcom{\LCProcP}{\LCTypeT}{\LCProcQ}{V}$ and $M'=V[{\LCProcQ}:={\LCProcP}]$. Then if ${\LCProcQ}\neq {\LCProcP}$, $\llbracket M\rrbracket_{\LCProcP}=\LCrecvV{{\LCProcQ}}~\epp{V}{{\LCProcP}}$, $\llbracket M'\rrbracket_{\LCProcP}=\llbracket V[{\LCProcQ}:={\LCProcP}] \rrbracket_{\LCProcP}= \llbracket V \rrbracket_{\LCProcQ}[{\LCProcQ}:={\LCProcP}]$, $\llbracket M\rrbracket_{\LCProcQ}=\LCsendV{{\LCProcP}}~\llbracket V \rrbracket_{\LCProcQ}$, $\llbracket M' \rrbracket_{\LCProcQ}=\llbracket V[{\LCProcQ}:={\LCProcP}] \rrbracket_{\LCProcQ}= \llbracket V \rrbracket_{\LCProcP}[{\LCProcQ}:={\LCProcP}]$, and for any ${\LCProcP}'$ such that $\epp{\LCTypeT}{\LCProcP'}\neq\LCBot$, we have $\llbracket M\rrbracket_{{\LCProcP}'}=\rolesub{{\LCProcQ}}{{\LCProcP}}~\llbracket V\rrbracket_{{\LCProcP}'}$ and $\llbracket M'\rrbracket_{{\LCProcP}'}=\llbracket V[{\LCProcQ}:={\LCProcP}]\rrbracket_{{\LCProcP}'}=\llbracket V\rrbracket_{{\LCProcP}'}[{\LCProcQ}:={\LCProcP}]$, and for any other ${\LCProcP}''$, $\llbracket M\rrbracket_{{\LCProcP}''}=\llbracket M'\rrbracket_{{\LCProcP}''}=\LCBotVal$. We therefore get $\llbracket M\rrbracket_{\LCProcP}\xrightarrow{\LCrecvV{{\LCProcQ}} \llbracket V \rrbracket_{\LCProcQ}[{\LCProcQ}:={\LCProcP}]~\llbracket V \rrbracket_{\LCProcP}}_{\epp{D}{}} \llbracket M'\rrbracket_{\LCProcP}$, $\llbracket M\rrbracket_{\LCProcQ}\xrightarrow{\LCsendV{{\LCProcP}} \llbracket V \rrbracket_{\LCProcQ}~\llbracket V \rrbracket_{\LCProcP}[{\LCProcQ}:={\LCProcP}]}_{\epp{D}{}} \llbracket M'\rrbracket_{\LCProcQ}$, and $\llbracket M\rrbracket_{{\LCProcP}'}\xrightarrow{\tau}_{\epp{D}{}}\llbracket M'\rrbracket_{{\LCProcP}'}$. We define $\mathcal{N}=\mathcal{N}'=\llbracket M'\rrbracket$ and the result follows. If ${\LCProcQ}={\LCProcP}$, then $\llbracket M\rrbracket_{\LCProcP}=(\lambda x. x)~\llbracket V \rrbracket_{\LCProcP}$ and $\llbracket M'\rrbracket_{\LCProcP}=\llbracket V \rrbracket_{\LCProcP}$ and $\mathcal{N}=\mathcal{N}'=\llbracket M'\rrbracket$ and the result follows.

\item Assume $M=\LCselect{{\LCProcQ}}{{\LCProcP}}{\ell}{M'}$. Then $\llbracket M \rrbracket_{{\LCProcQ}}=\LCchoice{{\LCProcP}}{\ell}{\llbracket M' \rrbracket_{{\LCProcQ}}}$, $\llbracket M \rrbracket_{{\LCProcP}}=\LCoffr{\LCProcQ}{\{\ell:\llbracket M' \rrbracket_{{\LCProcP}}\}}$, and for any ${\LCProcP}'\notin \{{\LCProcQ},{\LCProcP}\}$, $\llbracket M \rrbracket_{{\LCProcP}'}=\llbracket M' \rrbracket_{{\LCProcP}'}$. We therefore get $\llbracket M \rrbracket\xrightarrow{\tau_{{\LCProcP},{\LCProcQ}}}_{\epp{D}{}} \llbracket M \rrbracket\setminus \{{\LCProcP},{\LCProcQ}\}\mid {\LCProcP}[\llbracket M' \rrbracket_{{\LCProcP}}]\mid {\LCProcQ}[\llbracket M' \rrbracket_{{\LCProcQ}}]$ and the result follows.

\item Assume $M=\LCpair{N}{N'}$, $N\rightarrow_{D} N''$, and $M'=\LCpair{N''}{N'}$. Then the result follows from simple induction.
\item Assume $M=\LCpair{V}{N}$, $N\rightarrow_{D} N'$, and $M'=\LCpair{V}{N'}$. Then the result follows from simple induction.

\item Assume $M=\LCfst{\LCpair{V}{V'}}$ and $M'=V$. Then for any process ${\LCProcP}$ such that $\epp{V}{\LCProcP}\neq\LCBotVal$ or $\epp{V}{\LCProcP}\neq\LCBotVal$, $\llbracket M \rrbracket_{\LCProcP}=\LCfst{\LCpair{\llbracket M' \rrbracket_{\LCProcP}}{\llbracket V' \rrbracket_{\LCProcP}}}$ and for any other process ${\LCProcP}'\notin \roles(\type(\LCpair{M'}{V'})$, we have $\llbracket M \rrbracket_{{\LCProcP}'}=\LCBotVal$ and $\llbracket M' \rrbracket_{{\LCProcP}'}=\LCBotVal$. We define $\mathcal{N}=\mathcal{N}'=\llbracket M'\rrbracket$ and the result follows.

\item Assume $M=\LCsnd{\LCpair{V}{V'}}$ and $M'=V'$. Then the case is similar to the previous.

\item Assume $M=f$ and $M'=D(f)$. Then the result follows from the definition of $\llbracket D\rrbracket$.\qedhere

\end{itemize}
\end{proof}
\fi
\begin{theorem}[Soundness]\label{thm:NetToChor}
Given a closed choreography $M$ and a function mapping $D$, if $\Theta;\Gamma\vdash M:\LCTypeT$, $\Theta;\Gamma\vdash D$, and
  $\epp M{}\rightarrow^*_{\epp D{}} \mathcal{N}$ for some network $\mathcal N$, then
  there exist a choreography $M'$ and a network $\mathcal N'$ such that: $M\rightarrow^*_{D} M'$,
  $\mathcal{N}\rightarrow^*\mathcal{N}'$, and $\mathcal{N'}\sqsupseteq \epp{M'}{}$.

\end{theorem}
\iftr
  
As with completeness, we need some ancillary lemmas before we can prove soundness. For this, we need a notion of removing processes from a network.
\begin{definition}
Given a network $\mathcal{N}=\prod\limits_{\LCProcP\in\LCProcrho} {\LCProcP}[L_{\LCProcP}]$, we have $\mathcal{N}\setminus \LCProcrho'= \prod\limits_{\LCProcP\in(\LCProcrho\setminus \LCProcrho')} {\LCProcP}[L_{\LCProcP}]$
\end{definition}
First we show that actions in a network do not affect the roles not mentioned in the transition label.
\begin{lemma}
For any process $\LCProcP$ and network $\mathcal{N}$, if $\mathcal{N}\xrightarrow{\tau_{\procs}}_{\DP} \mathcal{N}'$ and $\LCProcP\notin \procs$ then $\mathcal{N}(\LCProcP)=\mathcal{N}'(\LCProcP)$.
\end{lemma}
\begin{proof}
Straightforward from the network semantics.
\end{proof}

Then we show that removing processes from a network does not prevent it from performing actions involving different processes.
\begin{lemma}
For any set of processes $\LCProcrho$ and network $\mathcal{N}$, if $\mathcal{N}\xrightarrow{\tau_{\procs}} \mathcal{N}'$ and $\procs\cap \LCProcrho = \emptyset$ then $\mathcal{N}\setminus \LCProcrho \xrightarrow{\tau_{\procs}}_{\DP} \mathcal{N}'\setminus \LCProcrho$.
\end{lemma}
\begin{proof}
Straightforward from the network semantics.
\end{proof}

We finally show that if the projection of a choreographic type is equivalent to a local type value, then the original choreographic type is equivalent to a choreographic type value.
\begin{lemma}\label{thm:TypeSound}
Given a closed type $\LCTypeT[1]$ and process $\LCProcP$, if $\Theta;\Gamma\vdash \LCTypeT[1]::\LCKindK$ and
  $\epp{\LCTypeT[1]}{\LCProcP} \equiv_{\LCProcP} \LCTypeV$, then
  there exist a type $\LCTypeV'$ such that: $\LCTypeT[1] \equiv \LCTypeV'$ and $\epp{\LCTypeV'}{\LCProcP}=\LCTypeV$.
\end{lemma}
\begin{proof}
We prove this by structural induction on $\LCTypeT[1]$. All but one case follows from simple induction.
 
 Assume $\LCTypeT[1]=\LCTypeT[2]~\LCTypeT[3]$. Then if $\epp {\LCTypeT[2]}{\LCProcP}=\epp{\LCTypeT[2]}{\LCProcP}=\LCBot$, we have $\epp {\LCTypeT[1]}{\LCProcP}=\LCBotVal$ and therefore $\LCTypeV=\LCBot=\LCTypeV'$.
Otherwise, if $\epp{\LCTypeT[3]}{\LCProcP'}=\LCBot$ and $\kind(\LCTypeT[3])=\LCKindWithout{\LCKindK}{(\{\LCProcP\}\cup \LCProcrho)}$, we get $\epp{\LCTypeT[1]}{\LCProcP}=\epp{\LCTypeT[2]}{\LCProcP}$ and the result follows form induction.  
Otherwise if $\epp{\LCTypeT[2]}{\LCProcP}=\LCBot$, we get $\epp{\LCTypeT[1]}{\LCProcP}=\epp{\LCTypeT[3]}{\LCProcP}$ and the result follows from induction. 
Otherwise, we get $\llbracket \LCTypeT\rrbracket_{\LCProcP}= \llbracket \LCTypeT[2]\rrbracket_{\LCProcP}~\llbracket \LCTypeT[3]\rrbracket_{\LCProcP}$. By induction, $\llbracket \LCTypeT[2]\rrbracket_{\LCProcP}\equiv_{\LCProcP} \LCTypeV[2]$ and there exists $\LCTypeV[2]'$ such that $\LCTypeT[2] \equiv \LCTypeV[2]'$ and $\epp{\LCTypeV[2]'}{\LCProcP}=\LCTypeV[2]$ and $\llbracket \LCTypeT[3]\rrbracket_{\LCProcP}\equiv_{\LCProcP} \LCTypeV[3]$ and there exists $\LCTypeV[3]'$ such that $\LCTypeT[3] \equiv \LCTypeV[3]'$ and $\epp{\LCTypeV[3]'}{\LCProcP}=\LCTypeV[3]$. Because $\LCTypeT[1]$ is kindable, we have a kind $\LCKindK'$ such that $\Theta;\Gamma\vdash \LCTypeT[2]::\LCKindArrow{\LCKindK'}{\LCKindK}$ and $\Theta;\Gamma\vdash \LCTypeT[3]::{\LCKindK'}$ This means that $\LCTypeV[2]'=\LCTypefun{X}{\LCKindK'}{\LCTypeV[4]}$ and if $\LCKindK'\in\{\LCProc,\LCKindWithout{\LCProc}{\LCProcrho}\}$ then $\LCTypeV[2]=\LCTypefun{X}{\LCAmI{X}{\epp{{\LCTypeV[4]}[X:=\LCProcP]}{\LCProcP}}{\epp{\LCTypeV[4]}{\LCProcP}}}$, otherwise $\LCTypeV[2]=\LCTypefun{X}{\epp{\LCTypeV[4]}{\LCProcP}}$. We then get $\LCTypeV\equiv_{\LCProcP} \epp{{\LCTypeV[4]'}[X:={\LCTypeV[3]}]}{\LCProcP}$ and $\LCTypeV'\equiv {\LCTypeV[4]'}[X:={\LCTypeV[3]'}]$, and since $X$ and $\LCTypeV[3]'$ are both base types, so are $\epp{{\LCTypeV[4]'}[X:={\LCTypeV[3]}]}{\LCProcP}$ and ${\LCTypeV[4]'}[X:={\LCTypeV[3]'}]$.
\end{proof}

We are then ready to prove soundness.
\begin{proof}[Proof of Theorem~\ref{thm:NetToChor}]
We prove this by structural induction on $M$.
\begin{itemize}
	\item Assume $M=V$. Then for any process $\LCProcP$, $\llbracket M\rrbracket_{\LCProcP}=U$, and therefore $\llbracket M\rrbracket\not\xrightarrow{\tau_{\procs}}$. 
	\item Assume $M=N_1~N_2$. Then for any process ${\LCProcP}$ such that $\epp {N_1}{{\LCProcP}}=\epp{N_2}{{\LCProcP}}=\LCBotVal$, we have $\epp M{{\LCProcP}}=\LCBotVal$.
For any process ${\LCProcP}'$ such that ${\LCProcP}'\in \roles(\type(N_1))$ or $\epp {N_1}{{\LCProcP}'} \neq \LCBotVal \neq \epp{N_2}{{\LCProcP}'}$, $\llbracket M\rrbracket_{{\LCProcP}'}= \llbracket N_1\rrbracket_{{\LCProcP}'}~\llbracket N_2\rrbracket_{{\LCProcP}'}$.  
For any other process ${\LCProcP}''$ such that $\epp{N_2}{{\LCProcP}''}=\LCBotVal$, we get $\epp M{{\LCProcP}''}=\epp{N_1}{{\LCProcP}''}$. 
For any other process ${\LCProcP}'''$, we get $\epp{M}{{\LCProcP}'''}=\epp{N_2}{{\LCProcP}'''}$. We then have 2 cases.
		\begin{itemize}
		\item Assume $N_2=V$. Then $\llbracket N_2\rrbracket_{\LCProcP}=U$ by Theorem~\ref{thm:Induction}, and for any ${\LCProcP}'$ such that ${\LCProcP}'\notin \roles(\type(N_2))\subseteq \roles(\type(N_1))$, by Theorem~\ref{thm:ValUnit}, $\llbracket N_2\rrbracket_{{\LCProcP}'}=\LCBotVal$ and therefore $\epp{M}{{\LCProcP}'}=\epp{N_1}{{\LCProcP}'}$, and we have 5 cases.
		\begin{itemize}
			\item Assume $N_1=\LCfun{x}[\LCTypeT]{N_3}$. Then for any process ${\LCProcP}$ such that $\epp{N_3}{{\LCProcP}}\neq \LCBotVal$ or $\epp{\LCTypeT}{{\LCProcP}}\neq\LCBot$, $\llbracket N_1\rrbracket_{\LCProcP}=\LCfun{x}[\epp{\LCTypeT}{{\LCProcP}}]{\llbracket N_3\rrbracket_{\LCProcP}}$. And for any other process, $\llbracket N_1\rrbracket_{\LCProcP}=\LCBotVal$. The only transition available at any process, would then use Rule~\rulen{NAbsApp}. 
			
			This means there exists ${\LCProcP}''$ such that $\mathbf{{\LCProcP}}={\LCProcP}''$. We then get $\llbracket M \rrbracket\xrightarrow{\tau_{\LCProcP}}\llbracket M \rrbracket\setminus \{{\LCProcP}''\}\mid {\LCProcP}''[\llbracket N_3\rrbracket_{{\LCProcP}''}[x:=\llbracket N_2\rrbracket_{{\LCProcP}''}]]$. We say that $M'=N_3[x:=N_2]$ and the result follows from using Rule~\rulen{NAbsApp} in every process ${\LCProcP}$ such that $\epp{M}{{\LCProcP}}\neq\LCBotVal$ and induction.
			
			\item Assume $N_1=\LCcomV{\LCProcQ}{\LCTypeT}{\LCProcP}$. Then if ${\LCProcQ}\neq {\LCProcP}$, $\llbracket M\rrbracket_{\LCProcQ}=\LCsendV{{\LCProcP}}~\llbracket N_2\rrbracket_{\LCProcQ}$, $\llbracket M\rrbracket_{\LCProcP}=\LCrecvV{{\LCProcP}}~\LCBotVal$, for any ${\LCProcP}'\in \roles(\LCTypeT)$, $\llbracket M\rrbracket_{{\LCProcP}'}=\rolesub{{\LCProcQ}}{{\LCProcP}}~\llbracket V\rrbracket_{{\LCProcP}'}$, and for any other process ${\LCProcP}''$, $\epp{N_1}{{\LCProcP}''}=\LCBotVal=\llbracket M\rrbracket_{{\LCProcP}''}$. And if ${\LCProcQ}={\LCProcP}$ then $\llbracket N_1\rrbracket_{\LCProcP}=\lambda x. x$.
			
			If $\procs={\LCProcQ},{\LCProcP}$ then $\mathcal{N}=\llbracket M\rrbracket\setminus \{{\LCProcQ},{\LCProcP}\}\mid {\LCProcQ}[\epp{N_2}{{\LCProcP}}[{\LCProcQ}:={\LCProcP}]]\mid {\LCProcP}[\llbracket N_2\rrbracket_{\LCProcQ}[{\LCProcQ}:={\LCProcP}]]$. Because $\llbracket N_2\rrbracket_{\LCProcP}=\LCBotVal$ and $\llbracket N_2\rrbracket_{\LCProcQ}=U$, $N_2=V$. Therefore $M\xrightarrow{\procs}_{D} V[{\LCProcQ}:={\LCProcP}]$ and for any ${\LCProcP}'\in \roles(\LCTypeT)$, by Rule~\rulen{NSub}, $\mathcal{N}({\LCProcP}')\xrightarrow{\tau}_{\epp{D}{}} \epp{V[{\LCProcQ}:={\LCProcP}]}{{\LCProcP}'}$ and the result follows from induction.

			If $\procs={\LCProcP}$ then either ${\LCProcQ}={\LCProcP}$ or $\epp{N_1}{{\LCProcP}}=\rolesub{{\LCProcQ}}{{\LCProcP}}$. If ${\LCProcQ}={\LCProcP}$ then $\mathcal{N}=\llbracket M\rrbracket \setminus \{{\LCProcP}\} \mid {\LCProcP}[\llbracket N_2\rrbracket_{\LCProcP}]$ and the rest is similar to above. If $\epp{N_1}{{\LCProcP}}=\rolesub{{\LCProcQ}}{{\LCProcP}}$ then the case is similar to one of the other two.			

			\item Otherwise, $N_1\neq V$ and either $\mathbf{{\LCProcP}}={\LCProcP}$ or $\mathbf{{\LCProcP}}={\LCProcP},{\LCProcQ}$.

			If $\procs={\LCProcP}$ then either $\llbracket N_1\rrbracket_{\LCProcP}\xrightarrow{\tau} L$ and ${\LCProcP}\in\roles(\type(N_1))$, $\mathcal{N}=\llbracket M\rrbracket\setminus \{{\LCProcP}\} \mid {\LCProcP}[L~\llbracket N_2\rrbracket_{\LCProcP}]$. We therefore have $\llbracket N_1\rrbracket\xrightarrow{\tau_{\LCProcP}} \llbracket N_1\rrbracket\setminus \{{\LCProcP}\}\mid {\LCProcP}[L]$, and by induction, $N_1\rightarrow^*_{D} N_1'$ such that $\llbracket N_1\rrbracket\setminus \{{\LCProcP}\}\mid {\LCProcP}[L]\rightarrow^*\mathcal{N}_1\sqsupseteq \llbracket N_1'\rrbracket$. Since all these transitions can be propagated past $N_2$ in the network and $\llbracket N_2\rrbracket_{{\LCProcP}'}$ in any process ${\LCProcP}'$ involved, we get the result for $M'=N_1'~N_2$.

			If $\procs={\LCProcP},{\LCProcQ}$ then the case is similar.
		\end{itemize}
		\item If $N_2\neq V$ then we have 2 cases.
		\begin{itemize}
		\item If $\procs={\LCProcP}$ then either $\llbracket N_1\rrbracket_{\LCProcP}\xrightarrow{\tau} L$ or $\epp{N_1}{{\LCProcP}}=U$ and $\llbracket N_2\rrbracket_{\LCProcP}\xrightarrow{\tau} L$ and the case is similar to the previous.

		\item If $\procs={\LCProcQ},{\LCProcP}$ then there exists $U$ such that either $\llbracket N_1\rrbracket_{\LCProcQ}\xrightarrow{\LCsendV{{\LCProcP}}~U} L_{\LCProcQ}$ or $\llbracket N_2\rrbracket_{\LCProcQ}\xrightarrow{\LCsendV{{\LCProcP}}~U} L_{\LCProcQ}$ and $\llbracket N_1\rrbracket_{\LCProcP}\xrightarrow{\LCrecvV{{\LCProcQ}}~U[{\LCProcQ}:={\LCProcP}]} L_{\LCProcP}$ or $\llbracket N_2\rrbracket_{\LCProcP}\xrightarrow{\LCrecvV{{\LCProcQ}}~U[{\LCProcQ}:={\LCProcP}]} L_{\LCProcP}$.

		If $\llbracket N_1\rrbracket_{\LCProcQ}\xrightarrow{\LCsendV{{\LCProcP}}~U} L_{\LCProcQ}$ then $\llbracket N_1\rrbracket_{\LCProcQ}\neq U'$ and therefore $\llbracket N_1\rrbracket_{\LCProcP}\xrightarrow{\LCrecvV{{\LCProcQ}}~U[{\LCProcQ}:={\LCProcP}]} L_{\LCProcP}$ and the case is similar to the previous. If $\llbracket N_2\rrbracket_{\LCProcQ}\xrightarrow{\LCsendV{{\LCProcP}}~U} L_{\LCProcQ}$ then $\llbracket N_1\rrbracket_{\LCProcQ}= U'$, and therefore $\llbracket N_2\rrbracket_{\LCProcP}\xrightarrow{\LCrecvV{{\LCProcQ}}~U[{\LCProcQ}:={\LCProcP}]} L_{\LCProcP}$ and the case is similar to the previous.
		\end{itemize}
	\end{itemize}
	\item Assume $M=N~\LCTypeT$. Then for any process ${\LCProcP}$ such that $\epp {N}{{\LCProcP}}=\epp{\LCTypeT}{{\LCProcP}}=\LCBotVal$, we have $\epp M{{\LCProcP}}=\LCBotVal$.
For any process ${\LCProcP}'$ such that $\epp {\LCTypeT}{{\LCProcP}'} =\LCBot$ and $\kind(\LCTypeT)=\LCKindWithout{\LCKindK}{(\{{\LCProcP}\}\cup\LCProcrho)}$, $\llbracket M\rrbracket_{{\LCProcP}'}= \llbracket N\rrbracket_{{\LCProcP}'}$.  
For any other process ${\LCProcP}''$ such that $\epp{\LCTypeT}{{\LCProcP}''}=\LCBotVal$, we get $\epp M{{\LCProcP}''}=\epp{N}{{\LCProcP}''}$. 
For any other process ${\LCProcP}'''$, we get $\epp{M}{{\LCProcP}'''}=\epp{N}{{\LCProcP}'''}~\epp{\LCTypeT}{{\LCProcP}'''}$. This case is similar to the previous unless $N=\LCTypefun{t}[K]{N'}$ and $\LCTypeT=B$.

If $N=\LCFun{X}[\LCKindK]{N'}$ and $\LCTypeT\equiv\LCTypeV$ then we have two cases. Either $K\in\{\LCProc,\LCKindWithout{\LCProc}{\LCProcrho}\}$ or not. If $K\in\{\LCProc,\LCKindWithout{\LCProc}{\LCProcrho}\}$ then for any ${\LCProcP}'$, $\epp{M}{{\LCProcP}'}=\LCapp{\LCFun{X}{\LCAmI X {\epp{N'[t:={\LCProcP}']}{\LCProcP}} {\epp{N'}{{\LCProcP}'}}}}{\epp{\LCTypeV}{{\LCProcP}'}}$. As $\epp{\LCTypeV}{{\LCProcP}'}={{\LCProcP}}$ for some ${\LCProcP}$, the only available transition is using Rule~\rulen{NBabs}, and we therefore get $\procs={{\LCProcP}''}$ for some ${\LCProcP}''$ and $\mathcal{N}=\epp{M}{}\setminus \{{\LCProcP}''\}\mid {\LCProcP}''[\LCAmI {\LCProcP} {\epp{N'[X:={\LCProcP}'']}{{\LCProcP}''}} {\epp{N'}{{\LCProcP}''}}]$. We then define $M'=N'[X:=\LCTypeV] $ and see that the result follows form using Rules~\rulen{NIamr} and~\rulen{NProam} on ${\LCProcP}''$ if ${\LCProcP}''={\LCProcP}$ and otherwise using Rules~\rulen{NIaml} and~\rulen{NProam}, at at all other processes using Rule~\rulen{NBAbs} and then either Rules~\rulen{NIamr} and~\rulen{NProam} or Rules~\rulen{NIaml} and~\rulen{NProam} and the result follows from induction.

If $\LCKindK\notin\{\LCProc,\LCKindWithout{\LCProc}{\LCProcrho}\}$ then the case is similar to $N_1=\LCfun{X}[\LCTypeT]{N_3}$ above.

\item Assume $M=\LCfst{N}$. Then either $N\neq V$ and the result follows from induction, or $N=\LCpair{V}{V'}$ and for any process ${\LCProcP}\in \roles(\type(\LCpair{V}{V'}))$, $\llbracket M \rrbracket_{\LCProcP}=\LCfst{\LCpair{\llbracket V \rrbracket_{\LCProcP}}{\llbracket V' \rrbracket_{\LCProcP}}}$ and for any other process ${\LCProcP}'\notin \roles(\type(\LCpair{V}{V'})$, by Theorem~\ref{thm:ValUnit} we have $\llbracket M \rrbracket_{{\LCProcP}'}=\epp{N}{{\LCProcP}'}=\LCBotVal$, and therefore $\epp{M}{{\LCProcP}'}\not\rightarrow$.

			If $\procs={\LCProcP}\in\roles(\type(\LCpair{V}{V'}))$ then $\mathcal{N}= \llbracket M \rrbracket \setminus \{{\LCProcP}\} \mid {\LCProcP}[\llbracket V \rrbracket_{\LCProcP}]$ and $M\xrightarrow{\mathbf{{\LCProcP}}}_{D} V$. The result follows by use of Rule~\rulen{NProj1} and Theorem~\ref{thm:ValUnit} and induction.

			\item Assume $N_1=\LCsnd{N_2}$. This case is similar to the previous.
\item Assume $M=\LCpair{M_1}{M_2}$. Then the result follows from simple induction.
	\item Assume $M=\LCcase{N}{x}{N'}{x'}{N''}$. Then for any process ${\LCProcP}$ such that ${\LCProcP}\in\roles(\type(N))$, we have $\llbracket M\rrbracket_{\LCProcP}=\LCcase{\llbracket N\rrbracket_{\LCProcP}}{x}{\llbracket N'\rrbracket_{\LCProcP}}{x'}{\llbracket N''\rrbracket_{\LCProcP}}$. For any other process ${\LCProcP}'$ such that $\epp N{{\LCProcP}'}=\epp{N'}{{\LCProcP}'}=\epp{N''}{{\LCProcP}'}=\LCBotVal$, $\epp{M}{{\LCProcP}'}=\LCBotVal$. For any other process ${\LCProcP}''$ such that $\epp{N}{{\LCProcP}''}=\LCBotVal$, we get $\epp{M}{{\LCProcP}''}=\epp{N'}{{\LCProcP}''}\sqcup \epp{N''}{{\LCProcP}''}$. For any other processes ${\LCProcP}'''$ such that $\epp{N'}{{\LCProcP}'''}=\epp{N''}{{\LCProcP}'''}=\LCBotVal$, we have $\epp{M}{{\LCProcP}'''}=\epp{N}{{\LCProcP}'''}$. For any other process ${\LCProcP}''''$, we have $\llbracket M\rrbracket_{{\LCProcP}''''}=(\lambda x:\LCBot.\llbracket N'\rrbracket_{{\LCProcP}''''}\sqcup \llbracket N''\rrbracket_{{\LCProcP}''''})~\llbracket N\rrbracket_{{\LCProcP}''''}$. We have two cases.
	\begin{itemize}
		\item Assume $\procs={\LCProcP}\in \roles(\type(N))$. Then we have three cases.
		\begin{itemize}
			\item Assume $N=\LCinl{\LCTypeT}{V}$. Then $\llbracket N\rrbracket_{\LCProcP}=\LCinl{\epp{\LCTypeT}{\LCProcP}}{\llbracket V\rrbracket_{\LCProcP}}$ and $\mathcal{N}=\llbracket M\rrbracket\setminus \{{\LCProcP}\} \mid {\LCProcP}[\llbracket N'[x:= \llbracket V\rrbracket_{\LCProcP}]\rrbracket_{\LCProcP}]$. We define $M'=N'$ and since $\llbracket N'\rrbracket_{{\LCProcP}'}\sqsupseteq \llbracket N'\rrbracket_{{\LCProcP}'}\sqcup\llbracket N''\rrbracket_{{\LCProcP}'}$ the result follows from using Rules~\rulen{NAbsApp} and~\rulen{NCasel} and induction.
			\item Assume $N=\LCinr{\LCTypeT}{V}$. Then the case is similar to the previous.
			\item Otherwise, we use Rule~\rulen{NCase} and we have a transition $\llbracket N \rrbracket_{{\LCProcP}}\xrightarrow{\tau} L$ such that \[\mathcal{N}=\llbracket M\rrbracket\setminus \{{\LCProcP}\} \mid {\LCProcP}[\LCcase{L}{x}{\llbracket N'\rrbracket_{\LCProcP}}{x'}{\llbracket N''\rrbracket_{\LCProcP}}]\] and the result follows from induction similar to the last application case.
		\end{itemize}
		\item Assume $\procs={\LCProcQ},{\LCProcP}$. Then the logic is similar to the third subcases of the previous case.
	\end{itemize}
	\item Assume $M=\LCselect{{\LCProcQ}}{{\LCProcP}}{\ell}{N}$. This is similar to the $N_1=\LCcomV{\LCProcQ}{\LCTypeT}{\LCProcP}$ case above.
	\item Assume $M=f$. Then for any process ${\LCProcP}$, $\epp{M}{{\LCProcP}}=f$. We therefore have some process \LCProcP such that $\procs={\LCProcP}$ and $\mathcal{N}=(\llbracket M\rrbracket\setminus {\LCProcP}) \mid {\LCProcP}[\llbracket D \rrbracket(f)]$. We then define the required choreography $M'=D(f)$ and network $\mathcal{N}'=\llbracket M'\rrbracket$ and the result follows.\qedhere
\end{itemize}
\end{proof}
\fi

From Theorems~\ref{thm:TypePres}, \ref{thm:Progress}, \ref{thm:ChorToNet}, and \ref{thm:NetToChor}, we get the following corollary, which states that a network derived from a well-typed closed choreography can continue to reduce until all roles contain only local values.
\begin{corollary}\label{thm:NetReduc}
  Given a closed choreography $M$ and a function environment $D$ containing all the function names of
  $M$, if $\Theta;\Gamma\vdash M:T$ and $\Theta;\Gamma\vdash D$, then: whenever
  $\epp M{}\rightarrow^*_{\epp D{}}\mathcal{N}$ for some network~$\mathcal{N}$, either there exists $\procs$ such that
  $\mathcal{N}\xrightarrow{\tau_{\procs}}_{\epp D{}}\mathcal{N}'$ or
  $\mathcal{N}=\prod\limits_{\LCProcP\in \roles(M)} {\LCProcP}[V_{\LCProcP}]$.
\end{corollary}


\section{Related Work}
\label{sec:related}

\subsection{Choreographies}
\label{sec:choreographies}
Choreographies are inspired by the ``Alice and Bob'' notation for security protocols by~\citet{NS78}, which included a term for expressing a communication from a participant to another.
The same idea inspired later several graphical notations for modelling interactions among processes, including sequence diagrams and message sequence charts~\citep{uml,msc}.

The first sophisticated languages for expressing choreographies were invented to describe interactions between web services.
The Web Services Choreography Description Language (WS-CDL) by the W3C~\citep{W3C04} is a choreographic language which describes the expected observable interactions between web services from a global point of view~\citep{ZongyanXCH07}.
\citet{CarboneHY12} later formalized endpoint projection for a theory of choreographies based on WS-CDL, and in particular introduced the merging operator (which we adjusted and extended to our setting).
This inspired more work on choreographies and projection and eventually the birth of choreographic programming---where choreographies are programs compiled to executable code---and the first choreographic programming language, Chor~\citep{Montesi13}.
As choreographic programming languages became more complex, \citet{Cruz-FilipeM17} developed a \emph{core calculus of choreographies} (CC). \citet{Montesi22} revisited and generalised CC in his text on foundations of choreographic languages. \citet{CruzFilipeMP21a,CruzFilipeMP21b}\fabrizio{Broken cit, maybe because it can't distinguish the two?..} then formalized this new version and its properties in the Coq theorem prover~\citep{Coq04}. Later, \citet{PohjolaGSN22} developed a certified end-to-end compiler from another variation of CC to CakeML by using the HOL theorem prover.

One of the primary design goals of all of choreographic programming languages is \emph{deadlock-freedom by design}~\citep{CarboneM13}---the operational correspondence between the choreography and the distributed network ensures deadlock-freedom for the network.
\chorlam achieves this goal.
\fabrizio{Removed the commented text here because actually even the cited paper uses multiple steps.}

The first choreographic language with limited process polymorphism was Procedural Choreographies~(PC)~\citep{Cruz-FilipeM17c}.
In PC, a choreography comes with an environment of predefined \emph{procedures} parametric on process names which may be called by the choreography.
These procedures have a number of limitations compared to the process polymorphism of \chorlam: they cannot contain any free processes, they cannot be partially instantiated, and they are lower-order---that is, they must be defined in the environment rather than as part of a larger choreography.
These limitations allow the projection function to know how the procedure will be instantiated, whereas in \chorlam we may need to compute the processes involved first. This has major implications for projection: in PC, it is easy to tell when projecting a procedure call which processes are involved and therefore need a local copy of the call.
However, \chorlam's  full process polymorphism allows the function and process names to each be enclosed in a context.
While this allows greater flexibility for programmers, it forces us to project a process-polymorphic functions to every process and let each process determine at runtime whether it is involved.

Recently, there has been a fair amount of interest in higher-order programming for choreographies~\citep{GiallorenzoMP20,HirschG22,CruzFilipeGLMP21}.
The first higher-order choreographic programming language, Choral~\citep{GiallorenzoMP20}, is an object-oriented choreographic language compiled to Java code.
Thus, Choral choreographies can depend on other choreographies, allowing programmers to reuse code. Choral was also the first choreographic language to treat $\LCcomV{\LCTypeT}{\LCProcP}{\LCProcQ}$ as a first-class function.

While Choral gave a practical implementation of higher-order choreographies, it did not establish their theoretical properties.
Two different---but independently-developed---works filled this gap, including Chor$\lambda$, the basis of \chorlam.
Chor$\lambda$ is a functional choreographic calculus based on the $\lambda$-calculus.
In this work, we extended Chor$\lambda$ with type and process polymorphism and the ability to send non-local values such as choreographies.
Chor$\lambda$, and hence \chorlam, provides a core language for higher-order choreographies, thus allowing us to establish their properties.
Since the original Chor$\lambda$ has parametric procedures like PC and Choral, it lacks \chorlam's property that a choreography diverging in one process must diverge in every process.
This forces Chor$\lambda$ to have both a complex notion of out-of-order execution and a more lax correspondence between actions in the network and the choreography.

The other work establishing the theoretical properties of higher-order choreographic programming is Pirouette~\citep{HirschG22}, which is also a functional choreographic programming language based on simply-typed $\lambda$~calculus.
Unlike Chor$\lambda$ (and thus \chorlam), Pirouette does not allow processes to send messages written in Pirouette.
Instead, it takes inspiration from lower-order choreographic programming languages in which (the computations to produce) messages are written in their own separate language.
Like the choreographic languages in~\citep{Montesi22,CruzFilipeMP21a}, Pirouette's design is parametrized by the language for writing messages. Thus, Pirouette can describe communication patterns between processes that draw from a large swath of languages for their local computations.
Nevertheless, this design means that Pirouette fundamentally cannot allow programs to send choreographic functions, unlike \chorlam.

Moreover, unlike Chor$\lambda$ and \chorlam, Pirouette forces every process to synchronize when applying a function.
This conceit allows Pirouette to establish a bisimulation relation with its network programming language, a result formalized in Coq.
This result allows a traditional---and verified---proof of deadlock-freedom by construction.
However, this constant synchronization would be a bottleneck in real-world systems; \chorlam's choice to obtain a weaker---but strong-enough---connection between the languages allows it to avoid this high cost.

\subsection{Concurrent Functional Programming}
\label{sec:conc-funct-progr}
Functional concurrent programming has a long history, starting with attempts to parallelize executions of functional programs~\citep{Burton87}.
The first language for functional programming with communications on channels was Facile~\citep{GiacaloneMP89} which, unlike later choreographic languages, had an abstraction over \emph{process IDs} very similar to process polymorphism.
A more recent work, which more-closely resembles choreographic programming, is Links~\citep{CooperLWY06}, with the RPC calculus~\citep{CooperW09} as its core language.
Links and the RPC calculus, like choreographies, allow a programmer to describe programs where different parts of the computation takes place at different locations and then compile it to separate code for each location.
Interestingly, though Links has explicit communication, in the RPC calculus the top level does not, and communications are created only in the projection when projecting a function located at a different process.
Moreover, the RPC calculus does not feature multiple threads of computation; instead, on communication the single thread of computation moves to a new location while other locations block.
The RPC calculus was later extended with location polymorphism, very similar to our and Facile's process polymorphism~\citep{ChoiCFL20}. However, as the RPC calculus only deals with systems of 2 processes, a client and a server, they project a process abstraction as a pair, and then the location as picking the correct part of this pair. This solution creates a simpler network language but is not suitable for a framework with an arbitrary number of participants such as \chorlam.
Moreover, the RPC calculus---like \chorlam but unlike traditional choreographic languages---does not have out-of-order execution at the top level.

Session types were applied to a concurrent functional calculus with asynchronous communication by \citet{GayV10}.
However, their session types do not give them a deadlock-freedom result.
Instead, they get a \emph{runtime safety} result: the only time a thread cannot perform an action is if it has terminated, is trying to read from an empty buffer, or is trying to start a new channel with another thread which isn't ready yet.

ML5~\citep{LicataH10,MurphyCH07} is a functional concurrent programming language based on the semantics of modal logic.
However, instead of the \LCKWsend and \LCKWrecv terms of choreographic languages, they have a primitive get[$w$]~$M$, which makes another process $w$ evaluate $M$ and return the result.
Since $M$ may include other gets, this construct gives ML5 something resembling \chorlam's ability to send a full choreography.
However,  the result of evaluating this ``choreography'' must be at the receiver and then returned to the sender.

\emph{Multitier} programming languages, like ScalaLoci~\citep{WeisenburgerS20}, offer another paradigm for describing global views of distributed systems.
Like Choral, ScalaLoci is built on top of an existing object-oriented language: in this case, Scala.
In ScalaLoci and other \emph{tierless} languages, an object describes a whole system containing multiple processes and functions.
Differently from choreographic programming, multitier programming does not allow for modelling the intended flow of communications. Rather, communication happens implicitly through remote function calls and the concrete protocol to be followed is largely left to be decided by a runtime middleware.
For a more detailed comparison of choreographic and multitier languages, see the work of \citet{GiallorenzoMPRS21}.


\section{Conclusion}
\label{sec:conclusion}

In this paper, we presented \chorlam, the first higher-order choreographic programming language with process polymorphism.
\chorlam has a type and kind system based on System~F$\omega$, but extended such that process names are types of kind~\LCProc.
Moreover, we showed how to obtain a concurrent interpretation of \chorlam programs in a process language by using a new construct corresponding by the ability of a process to know its identity. We found that this construct was necessary if process variables are able to be instantiated as the process they are located at, but using a choreographic language abstracts from this necessity.
Our explorations of process polymorphism also allowed \chorlam to describe a communication of a non-local value from \LCProcP to \LCProcQ as sending the part of the message owned by \LCProcP to \LCProcQ. These non-local values include full choreographies, creating a simple and flexible way to describe delegation by communicating a distributed function describing the delegated task. This innovation required a new notion of communication as an exchange in which the delegator rather than being left with an empty value after sending a choreography is left with a function which will allow it to potentially take part in the delegated task, e.g. by receiving a result at the end.

Process polymorphism fills much of the gap between previous works on the theory of higher-order choreographies and practical languages.
However, there is still more work to do.
For instance, currently \chorlam does not support recursive types.
In order to support recursive types, we would need to either make endpoint projection capable of projecting to a possibly-nonterminating description of a process, or limit recursive types ability to make type computations fail to terminate.

Furthermore, one can imagine allowing processes to send types and process names as well as values.
This would, for example, allow us to program a server to wait to receive the name of a client which it will have to interact with.
Since this form of delegation is common in practice, understanding how to provide this capability in a choreographic language, while retaining the guarantees of choreographic programming, would enable programmers to apply their usual programming patterns to choreographic code.

We project local type despite lacking a typing system for local processes. Our unusual network communication semantics have made it difficult to define local typing rules for $\LCKWsend$s and $\LCKWrecv$s, and we therefore leave local typing (or alternatively type erasure) as future work.

Finally, while certain, instant, and synchronous communication is convenient for theoretical study, such assumptions do not match real-world distributed systems.
\citet{Cruz-FilipeM17b} modeled asynchronous communication in choreographies via runtime terms representing messages in transit.
We could adapt this method to \chorlam by having the communication primitive reduce in two steps: first to a runtime term and then to the delivered value.
However, this extension would be nontrivial, since it is not clear how to represent messages in transit when those messages are non-local values such as choreographies.

While these gaps between theory and practice persist, process polymorphism in \chorlam brings us much closer to realistic choreographic languages for distributed systems.
Choreographic programs promise to provide easier and cleaner concurrent and distributed programming with better guarantees.
With higher-order choreographic programming and process polymorphism, the fulfilment of that promise is nearly within reach.



\bibliography{main}

\appendix
\section{Full \chorlam Typing Rules}
\label{sec:full-typing-rules}
\begin{mathparpagebreakable}
  \inferrule*[left=Tunit]{
    \Theta;\Gamma\vdash \LCTypeV:: \LCProc
  }{
    \Theta;\Gamma\vdash \LCUnitVal[\LCTypeV]:\LCUnit[\LCTypeV]
  }\and
  \inferrule*[left=Tint]{
    \Theta;\Gamma\vdash \LCTypeV :: \LCProc
  }{
    \Theta;\Gamma\vdash \LCLocalVal{n}[\LCTypeV]:\LCNat[\LCTypeV]
  } \and
  \inferrule*[left=Tapp]{
    \Theta;\Gamma\vdash N:\LCArr{\LCTypeT[1]}{\LCProcrho}{\LCTypeT[2]}\\
    \Theta;\Gamma\vdash M:\LCTypeT[1]
  }{
    \Theta;\Gamma\vdash N~M:\LCTypeT[2]
  } \and
  \inferrule*[left=Tabs]{
    \Theta;\Gamma\vdash \LCTypeT[1]::\LCKindAst \\
    \Theta;\Gamma'\vdash \LCTypeV:: \LCProc \text{ for all } \LCTypeV\in\LCProcrho\\
    \Theta\cap(\LCProcrho\cup \roles(\LCTypeT[1]) \cup \roles(\LCTypeT[2])\cup\ftv(\LCTypeT[1])\cup\ftv(\LCTypeT[2]));\Gamma,x:\LCTypeT[1]\vdash M:\LCTypeT[2] \\
  }{
    \Theta;\Gamma\vdash\LCfun{x}[\LCTypeT[1]]{M}:\LCArr{\LCTypeT[1]}{\LCProcrho}{\LCTypeT[2]}
  } \and
  \inferrule*[left=Tsel]{
    \Theta;\Gamma\vdash \LCTypeV[1] :: \LCProc\\
    \Theta;\Gamma\vdash \LCTypeV[2] :: \LCProc\\
    \Theta;\Gamma\vdash M : \LCTypeT
  }{
    \Theta;\Gamma\vdash \LCselect{\LCTypeV[1]}{\LCTypeV[2]}{\ell}{M}:\LCTypeT
  } \and
  \inferrule*[left=Tcom]{
    \Theta;\Gamma\vdash \LCTypeT:: \LCKindArrow{\LCProc}{\LCKindAst}\\
    \Theta;\Gamma\vdash \LCTypeV[1]:: \LCKindWithout{\LCProc}{(\roles(\LCTypeT)\cup \ftv(\LCTypeT)})\\
    \Theta;\Gamma\vdash \LCTypeV[2]::\LCKindWithout{\LCProc}{(\roles(\LCTypeT)\cup \ftv(\LCTypeT)})
  }{
    \Theta;\Gamma \vdash \LCcomV{\LCTypeT}{\LCTypeV[1]}{\LCTypeV[2]}:(\LCArr{\LCTypeT~\LCTypeV[1]}{\LCEmptyProcSet}{\LCTypeT~\LCTypeV[2]})
  } \and
  \inferrule*[left=TappT]{
    \Theta;\Gamma\vdash M:\LCForall{X}[\LCKindK]{\LCTypeT[1]}\\
    \Theta;\Gamma \vdash \LCTypeT[2]::\LCKindK
  }{
    \Theta;\Gamma\vdash \LCApp{M}{\LCTypeT[2]} : \SingleSubst{\LCTypeT[]}{X}{\LCTypeT[2]}
  } \and
  \inferrule*[left=TabsT]{
    \Theta';\Gamma', X :: \LCKindK \vdash M : \LCTypeT\\
    \text{if}\;\exists \LCKindKPrime, \LCProcrho.\,\LCKindK = \LCKindWithout{\LCKindKPrime}{\LCProcrho}
    \mathrel{\text{then}}\Gamma' = \addkind{(\Gamma + X)}{\LCProcrho}{X}
    \mathrel{\text{else}}\Gamma' = \Gamma + X\\
    \text{if}\;\LCKindK = \LCProc \mathrel{\text{or}} \exists \LCProcrho.\, \LCKindK = \LCKindWithout{\LCProc}{\LCProcrho}
    \mathrel{\text{then}} \Theta' = \Theta, X
    \mathrel{\text{else}} \Theta' = \Theta
  }{
    \Theta;\Gamma\vdash\LCFun{X}[\LCKindK]{M}:\LCForall{X}[\LCKindK]{\LCTypeT}
  }\and
  \inferrule*[left=Teq]{
    \Theta;\Gamma\vdash M:\LCTypeT[1]\\
    \LCTypeT[1] \equiv \LCTypeT[2]\\
    \Theta;\Gamma \vdash \LCTypeT[2]::\LCKindAst
  }{
    \Theta;\Gamma\vdash M:\LCTypeT[2]
  } \and
  \inferrule*[left=Tdefs]{
    \forall f\in \mathsf{domain}(D).\, f \mathrel{:} \LCTypeT \in \Gamma \land \emptyset;\Gamma\vdash D(f) \mathrel{:} \LCTypeT
  }{
    \Theta;\Gamma\vdash D
  } \and
  \inferrule*[left=Tvar]
  {x:\LCTypeT\in \Gamma }
  {\Theta;\Sigma;\Gamma\vdash x:\LCTypeT} \and
  \inferrule*[left=Tcase]
  {\Gamma\vdash N:\LCSum{\LCTypeT[1]}{\LCTypeT[2]} \\
    \Theta;\Gamma,x:T_1\vdash M':\LCTypeT \\
    \Theta;\Gamma,x':T_2\vdash M'':\LCTypeT} 
  {\Theta;\Gamma\vdash \LCcase{N}{x}{M'}{x'}{M''}:\LCTypeT} \and
  \inferrule*[left=Tfun]
  {f:\LCTypeT\in \Gamma }
  {\Theta;\Gamma\vdash f:\LCTypeT} \and
  \inferrule*[left=Tpair]
  {\Theta;\Gamma\vdash M:\LCTypeT[1] \\
    \Theta;\Gamma\vdash N:\LCTypeT[2]} 
  {\Theta;\Gamma\vdash \LCpair{M}{N}: \LCProd{\LCTypeT[1]}{\LCTypeT[2]}} \and
  \inferrule*[left=Tproj1]
  {\Theta;\Gamma\vdash M:\LCProd{\LCTypeT[1]}{\LCTypeT[2]}}
  {\Theta;\Gamma\vdash \LCfst{M}:\LCTypeT[1]}\and
  \inferrule*[left=Tproj2]
  {\Theta;\Gamma\vdash M:\LCProd{\LCTypeT[1]}{\LCTypeT[2]}}
  {\Theta;\Gamma\vdash \LCsnd{M}:\LCTypeT[2]} \and
  \inferrule*[left=Tinl]
  {\Theta;\Sigma;\Gamma\vdash M:\LCTypeT[1]}
  {\Theta;\Sigma;\Gamma\vdash \LCinl{M}{\LCTypeT[2]}:\LCSum{\LCTypeT[1]}{\LCTypeT[2]}} \and
  \inferrule*[left=Tinr]
  {\Theta;\Sigma;\Gamma\vdash M:\LCTypeT[2]}
  {\Theta;\Sigma;\Gamma\vdash \LCinr{M}{\LCTypeT[2]}:\LCSum{\LCTypeT[1]}{\LCTypeT[2]}} 
\end{mathparpagebreakable}

\section{Full \chorlam Operational Semantics}
\label{sec:full-oper-sem}
\begin{mathparpagebreakable}
  \inferrule*[left=AppAbs]{ }{
    \LCapp{(\LCfun{x}[\LCTypeT]{M})}{V} \to_{D} \SingleSubst{M}{x}{V}
  }\\
  \inferrule*[left=App1]{
    M_1 \to_D M_2
  }{
    \LCapp{M_1}{N} \to_D \LCApp{M_2}{N}
  }\and
  \inferrule*[left=App2]{
    N_1 \to_D N_2
  }{
    \LCapp{V}{N_1} \to_D \LCapp{V}{N_2}
  } \\
  \inferrule*[left=AppTAbs]{\LCTypeT\equiv\LCTypeV}{
    \LCApp{(\LCFun{X}[\LCKindK]{M})}{\LCTypeT} \to_D \SingleSubst{M}{X}{\LCTypeV}
  }\and
  \inferrule*[left=MTApp1]{
    M_1 \to_D M_2
  }{
    \LCApp{M_1}{\LCTypeT} \to_D \LCApp{M_2}{\LCTypeT}
  }\\
  \inferrule*[left=MTApp2]{
    \LCTypeT[1] \to_D \LCTypeT[2]
  }{
    \LCApp{V}{\LCTypeT[1]} \to_D \LCApp{V}{\LCTypeT[2]}
  }\\
  \inferrule*[left=Inl]{
    M_1 \to_D M_2
  }{
    \LCinl{\LCTypeT}{M_1} \to_D \LCinl{\LCTypeT}{M_2}
  }\and
  \inferrule*[left=Inl]{
    M_1 \to_D M_2
  }{
    \LCinr{\LCTypeT}{M_1} \to_D \LCinr{\LCTypeT}{M_2}
  }\and
  \inferrule*[left=Case]{
    N_1 \to_D N_2
  }{
    \LCcase{N_1}{x}{M_1}{y}{M_2} \to_D \LCcase{N_2}{x}{M_1}{y}{M_2}
  }\and
  \inferrule*[left=CaseL]{
  }{
    \LCcase{\LCinl{\LCTypeT}{V}}{x}{M_1}{y}{M_2} \to_D \SingleSubst{M_1}{x}{V}
  }\and
  \inferrule*[left=CaseR]{
  }{
    \LCcase{\LCinr{\LCTypeT}{V}}{x}{M_1}{y}{M_2} \to_D \SingleSubst{M_1}{x}{V}
  }\\
  \inferrule*[left=Pair1]{
    M_1 \to_D M_2
  }{
    \LCpair{M_1}{N} \to_D \LCpair{M_2}{N}
  }\and
  \inferrule*[left=Pair2]{
    N_1 \to_D N_2
  }{
    \LCpair{V}{N_1} \to_D \LCpair{V}{N_2}
  }\and
  \inferrule*[left=Fst]{
    M_1 \to_D M_2
  }{
    \LCfst{M_1} \to_D \LCfst{M_2}
  }\and
  \inferrule*[left=Snd]{
    M_1 \to_D M_2
  }{
    \LCsnd{M_1} \to_D \LCsnd{M_2}
  }\\
  \inferrule*[left=Proj1]{ }{
    \LCapp{\LCfst}{\LCpair{V_1}{V_2}} \to_D V_1
  }\and
  \inferrule*[left=Proj2]{ }{
    \LCapp{\LCsnd}{\LCpair{V_1}{V_2}} \to_D V_2
  }\\
  \inferrule*[left=Def]{ }{f \to_D D(f)}\\
  \inferrule*[left=Sel]{ }{\LCselect{\LCProcP}{\LCProcQ}{\ell}{M} \to_D M} \and
  \inferrule*[left=Com]{ }{
    \LCcom{\LCTypeT}{\LCProcP}{\LCProcQ}{V} \to_D \SingleSubst{V}{\LCProcP}{\LCProcQ}
  }
\end{mathparpagebreakable}

\section{Full Network and Local Program Operational Semantics}
\label{sec:full-network-local-os}
\begin{mathparpagebreakable}
  \inferrule*[left=NSend]{}{
    \LCsend{\LCProcP}{V}\xrightarrow{\LCsend{\LCProcP}{V~V'}}_{\DP} V'
  } \and
  \inferrule*[left=NRecv]{}{
    \LCrecv{\LCProcP}{V}\xrightarrow{\LCrecv{\LCProcP}{V'~V}}_{\DP} V'
  }\and
  \inferrule*[left=NCom]{
    L_1\xrightarrow{\LCsend{\LCProcP}{V~V'}[\LCProcQ := \LCProcP]}_{\DP} M'_1\\
    L_2\xrightarrow{\LCrecv{\LCProcQ}{V[\LCProcQ := \LCProcP]~V'}}_{\DP} M'_2
  }{
    \LCProcQ[] [M_1] \mid \LCProcP[] [M_2]\xrightarrow{\tau_{\LCProcQ,\LCProcP}}_{\DP} \LCProcQ[] [M'_1] \mid \LCProcP[] [M'_2]
  }\and
  \inferrule*[left=NCho]{}{
    \LCchoice{\LCProcP}{\ell}{M}\xrightarrow{\LCchoice{\LCProcP}{\ell}{}}_{\DP} M
  }\and
  \inferrule*[left=Noff]{}{
    \LCoffr{\LCProcP}{\{\ell_1: M_1,\dots,\ell_n: M_n\}}\xrightarrow{\LCoffr{\LCProcP}{\ell_i}}_{\DP} M_i
  }\and
  \inferrule*[left=NSel]{
    L_1\xrightarrow{\LCchoice{\LCProcP}{\ell}{}}_{\DP} M'_1\\
    L_2\xrightarrow{\LCoffr{\LCProcQ}{\ell}}_{\DP} M'_2
  }{
    {\LCProcQ} [M_1] \mid {\LCProcP} [M_2]\xrightarrow{\tau_{\LCProcQ,\LCProcP}}_{\DP} {\LCProcQ} [M'_1] \mid {\LCProcP} [M'_2]
  }\and
  \inferrule*[left=NAmIR]{}{
    \LCAmI{\LCProcP}{M}{M'} \xrightarrow{\LCProcP}_{\DP} M
  }\and
  \inferrule*[left=NAmIL]{
    \LCProcQ\neq\LCProcP
  }{
    \LCAmI{\LCProcP}{M}{M'} \xrightarrow{\LCProcQ}_{\DP} M'
  }\and
  \inferrule*[left=NAbsApp]{}{
    \LCapp{(\LCfun{x}[\LCTypeT]{M})}{V} \xrightarrow{\tau}_\DP \SingleSubst{M}{x}{V}
  }\and
  \inferrule*[left=NSub]{}{
    \LCapp{\rolesub{\LCProcP}{\LCProcQ}}{V} \xrightarrow{\tau}_\DP \SingleSubst{V}{\LCProcP}{\LCProcQ}
  }\and
  \inferrule*[left=NProam]{
    M \xrightarrow{\LCProcP}_\DP M'
  }{
    {\LCProcP} [M]\xrightarrow{\tau_{\LCProcP}}_{\DP} {\LCProcP} [M']
  }\and
  \inferrule*[left=NBAbs]{
    \LCTypeT\equiv_{\LCProcP}\LCTypeV
  }{
    \LCApp{(\LCFun{X}{M})}{\LCTypeT} \xrightarrow{\LCProcP}_{\DP} \SingleSubst{M}{X}{\LCTypeV}
  }\\
  \inferrule*[left=NApp1]{
    M_1 \xrightarrow{\mu}_\DP M_2
  }{
    \LCapp{M_1}{M'} \xrightarrow{\mu}_\DP \LCApp{M_2}{M'}
  }\and
  \inferrule*[left=NApp2]{
    M\xrightarrow{\mu}_\DP M'
  }{
    \LCapp{V}{M}\xrightarrow{\mu}_\DP \LCApp{V}{M'}
  }\and
  \inferrule*[left=NTApp1]{
    M \xrightarrow{\mu}_\DP M'
  }{
    \LCApp{M}{\LCTypeT} \xrightarrow{\mu}_\DP \LCApp{M}{\LCTypeT}
  }\and
  \inferrule*[left=NPro]{
    M \xrightarrow{\tau}_\DP M'
  }{
    \ProgOf{\LCProcP}{M} \xrightarrow{\tau_{\LCProcP}}_\DP \ProgOf{\LCProcP}{M'}
  }\and
  \inferrule*[left=NInl]{
    M \xrightarrow{\mu}_\DP M'
  }{
    \LCinl{\LCTypeT}{M} \xrightarrow{\mu}_\DP \LCinl{\LCTypeT}{M'}
  }\and
  \inferrule*[left=NInr]{
    M \xrightarrow{\mu}_\DP M'
  }{
    \LCinr{\LCTypeT}{M} \xrightarrow{\mu}_\DP \LCinr{\LCTypeT}{M'}
  }\and
  \inferrule*[left=NCase]{
    M \xrightarrow{\mu}_\DP M'
  }{
    \LCcase{M}{x}{M_1}{y}{L_2}\xrightarrow{\mu}_\DP \LCcase{M'}{x}{M_1}{y}{M_2}
  }\and
  \inferrule*[left=NCaseL]{}{
    \LCcase{\LCinl{\LCTypeT}{V}}{x}{M_1}{y}{M_2} \xrightarrow{\tau}_\DP \SingleSubst{M_1}{x}{V}
  }\and
  \inferrule*[left=NCaseR]{}{
    \LCcase{\LCinr{\LCTypeT}{V}}{x}{M_1}{y}{M_2} \xrightarrow{\tau}_\DP \SingleSubst{M_2}{x}{V}
  }\and
  \inferrule*[left=NPair1]{
    M_1 \xrightarrow{\mu}_{\DP} M_1'
  }{
    \LCpair{M_1}{M_2} \xrightarrow{\mu}_\DP \LCpair{M_1'}{M_2}
  }\and
  \inferrule*[left=NPair2]{
    M_2 \xrightarrow{\mu}_{\DP} M_2'
  }{
    \LCpair{M_1}{M_2} \xrightarrow{\mu}_\DP \LCpair{M_1}{M_2'}
  }\and
  \inferrule*[left=Fst]{
    M_1 \to_D M_2
  }{
    \LCfst{M_1} \to_D \LCfst{M_2}
  }\and
  \inferrule*[left=Snd]{
    M_1 \to_D M_2
  }{
    \LCsnd{M_1} \to_D \LCsnd{M_2}
  }\and
  \inferrule*[left=NProj1]{}{
    \LCapp{\LCfst}{\LCpair{V_1}{V_2}} \xrightarrow{\tau}_\DP V_1
  }\and
  \inferrule*[left=NProj2]{}{
    \LCapp{\LCsnd}{\LCpair{V_1}{V_2}} \xrightarrow{\tau}_\DP V_2
  }\and
  \inferrule*[left=NDef]{}{
    f \xrightarrow{\tau}_\DP \DP(f)
  }\and
  \inferrule*[left=NBot]{}{
    \LCapp{\LCBotVal}{\LCBotVal} \xrightarrow{\tau}_\DP \LCBotVal
  }\and
  \inferrule*[left=NBott]{}{
    \LCApp{\LCBotVal}{\LCBot} \xrightarrow{\tau}_\DP \LCBotVal
  }\and
  \inferrule*[left=NPar]{
    \mathcal{N}_1 \xrightarrow{\tau_{\LCProcP}}_{\DP} \mathcal{N}_2
  }{
    \mathcal{N}_1 \mid \mathcal{N}' \xrightarrow{\tau_{\LCProcP}}_{\DP} \mathcal{N}_2 \mid \mathcal{N}'
  }
\end{mathparpagebreakable}


\end{document}